%% file: ms.tex
\documentclass[a4paper, 10pt]{article}

\usepackage{color,multicol,tcolorbox,mathrsfs,xifthen,authblk,float,placeins,abstract,caption}
\usepackage{amsmath,amssymb}
\usepackage[a4paper, margin=1.81cm]{geometry}
\usepackage[left]{lineno}
\usepackage[superscript]{cite}

\begin{document}

\renewcommand{\familydefault}{\sfdefault}
\sffamily

\title{Deriving a genetic regulatory network from an optimization principle}

\author[a,b]{Thomas R. Sokolowski}
\author[c,d]{Thomas Gregor}
\author[c,e]{William Bialek}
\author[a]{Ga\v sper Tka\v cik}

\affil[a]{Institute of Science and Technology Austria, AT-3400 Klosterneuburg, Austria}
\affil[b]{Frankfurt Institute for Advanced Studies, DE-60438 Frankfurt am Main, Germany}
\affil[c]{Joseph Henry Laboratory of Physics \& Lewis-Sigler Institute for Integrative Genomics, Princeton University, Princeton, NJ 08544, USA}
\affil[d]{Department of Stem Cell and Developmental Biology, UMR3738, Institut Pasteur, 25 rue du Docteur Roux, FR-75015 Paris, France}
\affil[e]{Center for Studies in Physics and Biology, Rockefeller University, New York NY 10065}

\maketitle

\renewcommand{\abstractnamefont}{\large\bfseries}
\renewcommand{\abstracttextfont}{\large}

\vspace{0.5cm}
\begin{abstract}

\input{content/abstract-summary}

\end{abstract}
\vfill
\pagebreak

%\linenumbers

%%%%%%%%%%%%%%%%%%%%
%%% MAIN CONTENT %%%
%%%%%%%%%%%%%%%%%%%%
\newcommand{\FigOneWidth}{0.5\linewidth}
\newcommand{\IncludeTheBox}{1} % set to 1 for including box within the main text (works for PNAS style, not for Springer/Nature style)
\newcommand{\DiscussionName}{Conclusions} % how to name the Discussion section
\newcommand{\MySection}[1]{\section*{#1}} % sections heading style
\newcommand{\MyDropcap}[1]{#1} % toggle first capital letter in PNAS

%%% INPUT MAIN TEXT %%%

\input{content/main-content}

\section*{Supplementary information}
This article is accompanied by a supplementary document with additional text and figures describing the methods and supplementary results in more detail.

\section*{Correspondence}
Correspondence and requests for materials should be directed by e-mail to: \texttt{gtkacik@ist.ac.at}.

\clearpage

\section*{Acknowledgments}
\input{content/main-acknow}

\section*{Author Contributions}
T.R.S. and G.T. designed and carried out the research. T.R.S. derived the embryo model, implemented the code and performed the data analysis. T.G. contributed experimental data essential for deriving and parametrizing the embryo model. G.T. and W.B. concepted the research approach and contributed important revisions. All authors together wrote the paper.

\section*{Data Availability}
The data that support the findings of this study are available from the corresponding author upon reasonable request.

\section*{Code Availability}
The custom source code used for creating and analysing the data presented in this study is available from the corresponding author upon reasonable request.

\section*{Competing Interests}
The authors declare no competing interests.

%\vspace{1cm}

% Bibliography
\bibliographystyle{sn-nature.bst}
\bibliography{ms}

\end{document}

\maketitle

\renewcommand{\listfigurename}{List of Supplementary Figures}
\vfill
\tableofcontents
\listoffigures
\vfill

\clearpage
\section{Spatial-stochastic gene expression model}
\subsection{Principal ansatz}
\label{secAnsatz}
We model the stochastic gene expression dynamics of the gap gene system using a set
of coupled stochastic differential equations describing the time evolution of gene products
of all gap gene species $\alpha$ present in the system in all nuclear volumes $i$.
These dynamics are governed by a production term that depends on the local levels of maternal
inputs and gap gene proteins in volume $i$, a degradation term, a diffusion term describing
spatial exchange with neighboring volumes, and several noise sources that we specify in detail later.
For clarity here we first derive a model that conceptually considers protein production only and skips over the intermediate mRNA production step.
Later, in Sec.~\ref{secMRNAmodel}, we construct an extended model that incorporates mRNA in an approximate fashion
and conserves the principal structure of the ansatz below,
such that all derivations made in this section also apply to the extended model.

Let us denote the normalized protein copy number of gene $\alpha$ in volume $i$ by $g_i^\alpha$
and non-normalized protein copy numbers by $G_i^\alpha$;
we will explain the choice of the normalization factor $\Nmax$ further below.
The dynamics of proteins of gene species $\alpha$ in volume $i$ then are governed by:
\begin{align}
 \partial_t g_i^\alpha &= \frac{r_\alpha(\gset_i, \cset_i)}{\Nmax} - \frac{1}{\tau_\alpha} g_i^\alpha 
		        - h_\alpha \sum_{\nni} \left( g_i^\alpha - g_{\nni}^\alpha \right) + \sum_k \GammaNoise_{ik}^\alpha \eta_k
 \label{eqAnsatz}
\end{align}
where the production function $r_\alpha(\gset_i, \cset_i)$ depends on the set of all maternal and gap gene protein levels,
$\cset_i = (c_i^1, c_i^2, c_i^3)$ and $\gset_i = (g_i^1, g_i^2, g_i^3, g_i^4)$, respectively, in the volume;
note that the ansatz above remains valid for arbitrary numbers of gap genes and maternal inputs.
$\tau_\alpha$ is a (gene-specific) product lifetime, while $h_\alpha \equiv D_\alpha/\Delta^2$ denotes a gene-specific ``hopping rate'' 
for spatial exchange of gene products between neighboring nuclei;
herein $D_\alpha$ is the diffusion coefficient and $\Delta$ the distance between neighboring nuclear volumes, assumed to be the same for all nuclei (uniform spacing).
The sum of the diffusion term runs over all $2d$ nearest neighbors $\nni$ of $i$, whose number varies with the dimension $d$ of the nuclear lattice;
we considered a cylindrical lattice in $d=2$ and a uniformly spaced one-dimensional lattice ($d=1$),
but keep $d$ arbitrary in the following derivations.
The $k$-sum runs over all noise sources in the system that affect the copy number of gene $\alpha$ in volume $i$. 
In this sum, the noise powers $\GammaNoise_{ik}^\alpha$ are understood to be normalized via $\Nmax$ as well,
and will be specified later, in Sec.~\ref{secNoisePowers}.
All noise sources $\eta_k$ are assumed to be zero-mean Gaussian white noise, i.e.
\begin{align}
 \langle \eta_k(t) \rangle &= 0	\nn\\
 \langle \eta_k(t) \eta_k(t+dt) \rangle &= \delta(dt)
\end{align}

By default we assume that all genes can be expressed with the same maximal mean copy number $\Nmax$, i.e., each nucleus can potentially produce $\Nmax$ gene products in stationary state on average, when all stochastic fluctuations are averaged out.
Our model straightforwardly extends to the more general case in which the maximal mean copy numbers 
of the zygotically expressed species $\zeta$ can be different.
We can then take the largest mean copy number as a reference:
\begin{align}
 \Nmax \equiv \max_\zeta \la G_\zeta \ra
\end{align}
All other copy numbers then will be smaller or equal to $\Nmax$, and we can account for this by introducing additional, species-specific
production rates in the production function
\begin{align}
 \frac{r_\alpha(\gset_i, \cset_i)}{\Nmax} &= \frac{1}{\Nmax} \left[ \rb^\alpha + (\rmax^\alpha - \rb^\alpha) f_\alpha(\gset_i, \cset_i) \right]	\nn\\
				      &= \frac{1}{\tau} \left[ \frac{\rb^\alpha}{\rmax} + \frac{\rmax^\alpha - \rb^\alpha}{\rmax} f_\alpha(\gset_i, \cset_i) \right]	\nn\\
				      &\equiv \frac{1}{\tau} \left[ \rbhat^\alpha + (\rmaxhat^\alpha - \rbhat^\alpha) f_\alpha(\gset_i, \cset_i) \right]
\end{align}
where we denote the production rate and lifetime of the species with the largest copy number $\Nmax$ by $\rmax$ and $\tau$, respectively, i.e. $\Nmax\equiv\rmax\tau$.
Rate $\rb^\alpha$ is a small basal expression rate, while $\rbhat^\alpha$ denotes the same rate measured in units of $\rmax$.
Note that the ``normalized'' rate $\rmaxhat^\alpha \equiv \rmax^\alpha / \rmax$ then can be smaller or larger than 1,
because the lifetimes $\tau$ and $\tau_\alpha$ are not taken into account in the normalization. 
% We can, however, expect that the factors $\rmaxhat^\alpha$ will not differ much if the gene products belong to a similar family
% and are produced via the same machinery (under equal polymerase/ribosome resource constraints).

The regulation function $f_\alpha(\gset_i, \cset_i)$ also can be imposed later,
such that in the following calculations we will leave the specific form of $f_\alpha$ open for tractability.
In our model we opted for a regulatory function based on the MWC model, described in detail in \ref{secRegFun}.
In the following we will also operate with the function
\begin{align}
 F_\alpha(\gset_i, \cset_i) \equiv \frac{1}{\tau} \left[ \rbhat^\alpha + (\rmaxhat^\alpha - \rbhat^\alpha) f_\alpha(\gset_i, \cset_i) \right] - \frac{1}{\tau_\alpha} g_i^\alpha
\end{align}
grouping the production and degradation terms for $g_i^\alpha$.

We will now proceed with the calculation of the means and covariances of the spatially coupled gene expression model,
needed for later computation of the positional information contained in the expression pattern.

\subsection{Mean expression levels}
\label{secMeans}
The stationary mean expression levels $\gbar_i^\alpha \equiv \la \gbar_i^\alpha \ra$ in principle could be obtained from Eq.~(\ref{eqAnsatz}) 
by setting the time derivatives on the left-hand side to zero and averaging over the right-hand side.
Practically, however, this is only feasible if the production term $\la r_\alpha(\gset_i,\cset_i) \ra$ and the regulatory function $f_\alpha(\gset_i,\cset_i)$ in particular could be approximated by (the leading terms of) its Taylor-expansion around $\gbarm_i$ [see Eq.~(\ref{eqRegFunTaylor}) below]. 
Then, since $\forall k: \la \eta_k \ra = 0$, this would yield the following coupled system
\begin{align}
  & \partial_t \gbar_i^\alpha = \frac{r_\alpha(\gbarm_i, \cset_i)}{\Nmax} - \frac{1}{\tau_\alpha} \gbar_i^\alpha
		                - h_\alpha \sum_{\nni} \left( \gbar_i^\alpha - \gbar_{\nni}^\alpha \right) \equiv 0 \quad\Leftrightarrow	\nn\\
  & \left( 2d h_\alpha + \frac{1}{\tau_\alpha} \right) \gbar_i^\alpha  = \frac{r_\alpha(\gbarm_i, \cset_i)}{\Nmax} + h_\alpha \sum_\nni \gbar_\nni^\alpha \quad\Leftrightarrow	\nn\\
  & \gbar_i^\alpha = \frac{\tau_\alpha}{\tau} \frac{\rbhat^\alpha + (\rmaxhat^\alpha - \rbhat^\alpha) f_\alpha(\gbarm_i, \cset_i)}{1 + 2d\cdot h_\alpha\tau_\alpha} 
    + \frac{h_\alpha\tau_\alpha}{1 + 2d\cdot h_\alpha\tau_\alpha} \sum_\nni \gbar_\nni^\alpha \nn\\
\label{eqMeans}
\end{align}
where $d$ denotes the dimensionality of the spatial lattice.
Here the prefactors have the same form and interpretation as in the single-input/single-output system studied earlier by us \cite{Sokolowski2015},
and reflect the fact that with increasing diffusive coupling (increasing $h_\alpha$) the expression level in volume $i$
is increasingly determined by the spatial average of the neighboring expression levels (second term on the right),
such that neighboring levels increasingly equilibrate in spite of (potentially) different local production rates (first term on the right).
%In the limit $h_\alpha \rightarrow \infty$ this results in the same constant expression level in the whole spatial system.

%Eq.~\ref{eqMeans} defines a coupled equation system relating the stationary means in volume $i$ with the stationary means in the nearest-neighbor volumes $n$.
In practice the coupled system defined above is  hard to solve because the regulatory function $f_\alpha(\gset_i,\cset_i)$ (yet to impose) can be highly nonlinear.
Moreover, it is not a priori clear around which expression levels  one should linearize, as these levels arise as a consequence of the non-linear dynamics. We therefore opted for numerical forward-integration of the means, which do not require linearizing $f_\alpha(\gset_i,\cset_i)$ and can be carried out with sufficient computational efficiency for our needs; the details are described in Sec.~\ref{secMeansIntegration} along with the other numerical methods that we used for solving and optimizing the spatial-stochastic model derived here.

\subsection{Covariances}
\label{secCovs}
Assuming that we have determined the full set of mean expression levels in the system,
let us now compute the steady-state covariances
between the product copy number $g_i^\alpha$ of gene $\alpha$ in volume $i$ and the product copy number $g_j^\beta$ of gene $\beta$ in volume $j$,
defined as:
\begin{align}
C_{ij}^{\alpha\beta} \equiv \la \dg_i^\alpha \dg_j^\beta \ra
\end{align}
Here the fluctuations $\dg_i^\alpha$ are assumed to be linear and symmetric around the mean, i.e. $g_i^\alpha = \bar g_i^\alpha + \dg_i^\alpha$.
Note the symmetry $C_{ij}^{\alpha\beta}=C_{ji}^{\beta\alpha}$.

We compute $C_{ij}^{\alpha\beta}$ using the same formula as in our earlier work on a single-input/single-output system \cite{Sokolowski2015},
which correlates the fluctuations of local expresssion level $g_i^\alpha$ with the ``forces'' (production, degradation and diffusion) 
driving expression level $g_j^\beta$ and vice versa, and all noise sources affecting the respective expression levels:
\begin{align}
 \partial_t \la \dg_i^\alpha \dg_j^\beta \ra &= \underset{\FTerm_\beta}{\underbrace{ \la \dg_i^\alpha F_\beta(\gset_j,\cset_j)\vphantom{^\beta} \ra }} 
					      + \underset{\FTerm_\alpha}{\underbrace{ \la \dg_j^\beta F_\alpha(\gset_i,\cset_i) \ra }}	\nn\\
					     &\quad + \underset{\DTerm_\beta}{\underbrace{ h_\beta \la \dg_i^\alpha \sum_{n_j} \left( g_{n_j}^\beta - g_j^\beta \right) \ra }}
					            + \underset{\DTerm_\alpha}{\underbrace{ h_\alpha \la \dg_j^\beta \sum_{n_i} \left( g_{n_i}^\alpha - g_i^\alpha \right) \ra }}	\nn\\
					     &\quad + \underset{\GTerm}{\underbrace{ \sum_k \la \GammaNoise_{ik}^\alpha \GammaNoise_{jk}^\beta \ra }}
\label{eqODECovar}
\end{align}
We will now compute the different terms denoted by capital blackletters separately.
Following our usual strategy, we will specify and compute the noise term $\GTerm$ only at the end.

Using simple algebraic steps, we can write out $\FTerm_\alpha$ as:
\begin{align}
 \FTerm_\beta &= \frac{\rmaxhat^\beta-\rbhat^\beta}{\tau} \la \dg_i^\alpha f_\beta(\gset_j,\cset_j) \ra - \frac{1}{\tau_\beta} \la \dg_i^\alpha \dg_j^\beta \ra
\end{align}
To proceed further, we have to compute how fluctuations in all the involved gene copy numbers $\delta\gset_j$ propagate through the regulatory function $f_\beta$.
Here, in contrast to the computation of the mean expression levels (Sec.~\ref{secMeans}), we  pursue a linearization ansatz assuming the fluctuations to be small compared to the allowed expression range.
We can expand around the mean $\gbarm_j$, retaining terms up to first order, as follows:
\begin{align}
 f_\beta(\gset_j,\cset_j) &= f_\beta(\gbarm_j + \delta\gset_j,\cset_j)
		           \simeq f_\beta(\gbarm_j,\cset_j) + \sum_\zeta \left(\frac{\partial f_\beta}{\partial g_j^\zeta}\right)_{\gbarm_j,\cset_j} \dg_j^\zeta
\label{eqRegFunTaylor}
\end{align}
Herein $\zeta$ loops over all zygotic (gap gene) species in the system.
From this it follows that
\begin{align}
 \la \dg_i^\alpha f_\beta(\gset_j,\cset_j) \ra &= \sum_\zeta \left(\frac{\partial f_\beta}{\partial g_j^\zeta}\right)_{\gbarm_j,\cset_j} \la \dg_i^\alpha \dg_j^\zeta \ra
					        \equiv \sum_\zeta f_\beta'^{\zeta}(\gbarm_j,\cset_j) C_{ij}^{\alpha\zeta}
\end{align}
where we introduce the short notation $f_\beta^{'\zeta}$ for indicating that the derivative of the regulatory function $f_\beta$ is taken with respect
to zygotic species $\zeta$. %(which can be equal to $\beta$ in case of self-regulation).
Note that here we allow for $\zeta=\beta$, which corresponds to self-regulation of species $\beta$.
With simple intermediate steps we obtain
\begin{align}
 \FTerm_\beta  &= \frac{\rmaxhat^\beta - \rbhat^\beta}{\tau} \sum_\zeta f_\beta'^{\zeta}(\gbarm_j,\cset_j) C_{ij}^{\alpha\zeta}   - \frac{1}{\tau_\beta} C_{ij}^{\alpha\beta}	\nn\\
 \FTerm_\alpha &= \frac{\rmaxhat^\alpha - \rbhat^\alpha}{\tau} \sum_\zeta f_\alpha'^{\zeta}(\gbarm_i,\cset_i) C_{ji}^{\beta\zeta} - \frac{1}{\tau_\alpha} C_{ij}^{\alpha\beta}
\end{align}
where the calculation for $\FTerm_\alpha$ is completely analogous, and in the last step we have used $C_{ji}^{\beta\alpha}=C_{ij}^{\alpha\beta}$.

For the diffusive coupling terms $\DTerm_\beta$ and $\DTerm_\alpha$, after taking the average, we get:
\begin{align}
 \DTerm_\beta  &= h_\beta  \sum_{n_j} C_{i n_j}^{\alpha\beta} - 2d h_\beta  C_{ij}^{\alpha\beta}	\nn\\
 \DTerm_\alpha &= h_\alpha \sum_{n_i} C_{j n_i}^{\beta\alpha} - 2d h_\alpha C_{ij}^{\alpha\beta}
\end{align}
Here, $d$ again is the lattice dimension.

Putting all terms together, setting $\partial_t [...] = 0$ in Eq.~(\ref{eqODECovar}), and collecting parts that only contain 
covariance $C_{ij}^{\alpha\beta}$ on one side, we obtain the following:
\newcommand{\vphOneOverTauBeta}{\vphantom{\frac{1}{\tau_\beta}}}
\begin{align}
\lefteqn{
 \Bigg[  \underset{\text{from }\FTerm_\alpha}{\underbrace{ \frac{1}{\tau_\alpha} \vphOneOverTauBeta}}
       + \underset{\text{from }\FTerm_\beta}{ \underbrace{ \frac{1}{\tau_\beta}  \vphOneOverTauBeta }}
       + \underset{\text{from }\DTerm_\alpha\text{ and }\DTerm_\beta}{ \underbrace{ 2d(h_\alpha + h_\beta) \vphOneOverTauBeta }}
 \Bigg] C_{ij}^{\alpha\beta} =
 } \nn\\
 &\quad   \frac{\rmaxhat^\beta -\rbhat^\beta}{\tau} \sum_\zeta f_\beta'^{\zeta}(\gbarm_j,\cset_j) C_{ij}^{\alpha\zeta} 
	+ \frac{\rmaxhat^\alpha-\rbhat^\alpha}{\tau} \sum_\zeta f_\alpha'^{\zeta}(\gbarm_i,\cset_i) C_{ji}^{\beta\zeta}	\nn\\
 &\quad + h_\beta  \sum_{n_j} C_{i n_j}^{\alpha\beta} + h_\alpha \sum_{n_i} C_{j n_i}^{\beta\alpha} + \GTerm
\end{align}
With the definitions
\begin{align}
 T_{\alpha\beta} &\equiv 2 \left[ \frac{1}{\tau_\alpha} + \frac{1}{\tau_\beta} + 2d (h_\alpha + h_\beta) \right]^{-1}~,	\\
 \Lambda^2_\alpha &\equiv h_\alpha T_{\alpha\beta}~, \quad\quad\quad \Lambda^2_\beta \equiv h_\beta T_{\alpha\beta}~,
\end{align}
which can be interpreted as an effective averaging time and effective diffusive coupling lengths, respectively,
we can rewrite the equation above in the more compact final form:
\begin{align}
 C_{ij}^{\alpha\beta} &= \frac{T_{\alpha\beta}}{2} \Bigg[ \frac{\rmaxhat^\beta-\rbhat^\beta}{\tau} \sum_\zeta f_\beta'^{\zeta}(\gbarm_j,\cset_j) C_{ij}^{\alpha\zeta}	\nn\\
		      &\quad\quad\quad\quad\quad + \frac{\rmaxhat^\alpha-\rbhat^\alpha}{\tau} \sum_\zeta f_\alpha'^{\zeta}(\gbarm_i,\cset_i) C_{ji}^{\beta\zeta} \Bigg]	\nn\\
		      &\quad~ + \frac{1}{2} \Bigg[ \Lambda^2_\beta  \sum_{n_j} C_{i n_j}^{\alpha\beta} + \Lambda^2_\alpha \sum_{n_i} C_{j n_i}^{\beta\alpha} \Bigg] + \frac{T_{\alpha\beta}}{2} \GTerm
\label{eqCovarGeneral}
\end{align}
This defines a linear system for the whole set of inter-gene/inter-position covariances (or entries of the four-dimensional covariance matrix of the system) $C_{ij}^{\alpha\beta}$,
which can be solved after specifying the noise term $\GTerm$ and imposing suitable boundary conditions.
The equation above has the same structure as in the single-input/single-output model studied in our earlier work \cite{Sokolowski2015}, 
with additional terms introduced by the dependencies of the regulatory functions on the additional gap genes in the system.
Accordingly, by setting $\alpha=\beta$ and imposing that the regulatory functions only depend on a single maternal input $c_i$,
implying $\forall\zeta:~f_\alpha'^{\zeta}(\gbarm_j,\cset_j)=0=f_\beta'^{\zeta}(\gbarm_j,\cset_j)$,
we recover the formula for the single gene system (see Sec.~\ref{secSingleGeneLimit} below).
We will now consider this and other special cases of the general formula above, 
including a simplified case which reproduces a known formula from our earlier work as a consistency test.
We will then proceed by imposing a ``short-correlation assumption'', which assumes that significant covariance correlations 
only arise between nearby neighbor volumes and markedly simplifies the (numerical) computation of the covariance matrix.

\subsubsection{Local inter-gene covariance}
By setting $i=j$ in Eq.~(\ref{eqCovarGeneral}) we obtain the formula for covariance between $g_\alpha$ and $g_\beta$ in nuclear volume i:
\begin{align}
 C_{ii}^{\alpha\beta} &= \frac{T_{\alpha\beta}}{2} \Bigg[ \frac{\rmaxhat^\beta-\rbhat^\beta}{\tau} \sum_\zeta f_\beta'^{\zeta}(\gbarm_i,\cset_i) C_{ii}^{\alpha\zeta}	\nn\\
		      &\quad\quad\quad\quad\quad + \frac{\rmaxhat^\alpha-\rbhat^\alpha}{\tau} \sum_\zeta f_\alpha'^{\zeta}(\gbarm_i,\cset_i) C_{ii}^{\beta\zeta} \Bigg]	\nn\\
		      & \quad~ + \frac{1}{2} \sum_{n_i} \Bigg[ \Lambda^2_\beta  C_{i n_i}^{\alpha\beta} + \Lambda^2_\alpha C_{i n_i}^{\beta\alpha} \Bigg] 
			       + \frac{T_{\alpha\beta}}{2} \sum_k \la \GammaNoise_{ik}^\alpha \GammaNoise_{ik}^\beta \ra
\label{eqCovarLocal}
\end{align}
Assuming equal parameters for all genes (which is {\it not} the same as setting $\alpha=\beta$), this simplifies further to:
\begin{align}
 C_{ii}^{\alpha\beta} &= \frac{T}{2} \frac{1-\rbhat}{\tau} \sum_\zeta f'^{\zeta}(\gbarm_i,\cset_i) \Big( C_{ii}^{\alpha\zeta} + C_{ii}^{\beta\zeta} \Big)	\nn\\
		      & \quad~ + \frac{\Lambda^2}{2} \sum_{n_i} \Big( C_{i n_i}^{\alpha\beta} + C_{i n_i}^{\beta\alpha} \Big) 
			       + \frac{T}{2} \sum_k \la \GammaNoise_{ik}^\alpha \GammaNoise_{ik}^\beta \ra
\end{align}
Obviously, in this case $\rmaxhat^\alpha = \rmaxhat^\beta \equiv 1$ because $\rmax^\alpha = \rmax^\beta \equiv \rmax$ is the same for all genes.

\subsubsection{Local variance}
If we locally consider even the same gene, i.e. set $\alpha=\beta$ in Eq.~(\ref{eqCovarLocal}), we get the formula for the local variance $\sigma^2_{i,\alpha}$ 
of expression level $g_\alpha$:
\begin{align}
 \sigma^2_{i,\alpha} \equiv C_{ii}^{\alpha\alpha} 
		      &= T_{\alpha\alpha} \frac{\rmaxhat^\alpha-\rbhat^\alpha}{\tau_\alpha} \sum_\zeta f_\alpha'^{\zeta}(\gbarm_i,\cset_i) C_{ii}^{\alpha\zeta}	\nn\\
		      & \quad~ + \Lambda^2_\alpha \sum_{n_i} C_{i n_i}^{\alpha\alpha}
			       + \frac{T_{\alpha\alpha}}{2} \sum_k \la \big[\GammaNoise_{ik}^\alpha\big]^2 \ra
\end{align}
If the parameters for all genes again are equal, we can also directly write this as:
\begin{align}
 \sigma^2_{i,\alpha} &= T \frac{1-\rbhat}{\tau} \sum_\zeta f'^{\zeta}(\gbarm_i,\cset_i) C_{ii}^{\alpha\zeta}	\nn\\
		     & \quad~ + \Lambda^2 \sum_{n_i} C_{i n_i}^{\alpha\alpha}
			       + \frac{T}{2} \sum_k \la \big[\GammaNoise_{ik}^\alpha\big]^2 \ra
\label{eqLocalVarEqGenes}
\end{align}

\subsubsection{Inter-volume covariance for the same gene}
\label{secSingleGeneLimit}
This special case is obtained by setting $\alpha=\beta$ in Eq.~(\ref{eqCovarGeneral}) directly:
\begin{align}
\lefteqn{C_{ij}^{\alpha\alpha} =}	\nn\\
	&\quad \frac{T_{\alpha\alpha}}{2} \frac{\rmaxhat^\alpha-\rbhat^\alpha}{\tau} 
	      \sum_\zeta \Big( f_\alpha'^{\zeta}(\gbarm_j,\cset_j) C_{ij}^{\alpha\zeta} + f_\alpha'^{\zeta}(\gbarm_i,\cset_i) C_{ji}^{\alpha\zeta} \Big)	\nn\\
	&\quad\quad~ + \frac{\Lambda^2_\alpha}{2} \Bigg[ \sum_{n_j} C_{i n_j}^{\alpha\alpha} + \sum_{n_i} C_{j n_i}^{\alpha\alpha} \Bigg] + \frac{T_{\alpha\alpha}}{2} \sum_k \la \GammaNoise_{ik}^\alpha \GammaNoise_{jk}^\alpha \ra
\label{eqCovarInterVolume}
\end{align}
This formula again reduces to the single-input/single-output formula from \cite{Sokolowski2015}
when setting $\forall\zeta:~f_\alpha'^{\zeta}=0$ (limit of no mutual and no self-regulation).

\subsubsection{A single self-regulating gene without diffusive coupling}
As a consistency test, we can check whether the extended spatial framework derived here correctly reproduces the formula 
for the variance of the self-interacting gene without spatial coupling, derived earlier in \cite{Tkacik2012}.

If there is only a single species $g_\alpha \equiv g$ which is self-regulating, Eq.~(\ref{eqLocalVarEqGenes})
for the local variance reduces to:
\begin{align}
\sigma^2_{i,\alpha} &= T \frac{1-\rbhat}{\tau} \left(\frac{\partial f_\alpha}{\partial g_\alpha}\right)_{\gbar^\alpha_i,\cset_i} 
							  \underset{\sigma^2_{i,\alpha}}{\underbrace{C_{ii}^{\alpha\alpha}}}	\nn\\
		    & \quad~ + \Lambda^2 \sum_{n_i} C_{i n_i}^{\alpha\alpha} + \frac{T}{2} \sum_k \la \big[\GammaNoise_{ik}^\alpha\big]^2 \ra
\end{align}
For the last part of this section, we will drop the index $\alpha$.
In the limiting case without diffusion we have
\begin{align}
 \Lambda^2 &= 0~, \nn\\
 T &= \tau~.
\end{align}
This leads to
\begin{align}
\sigma^2_{i} &= (1-\rbhat) \left(\frac{\partial f}{\partial g}\right)_{\gbar_i,\cset_i} \sigma^2_{i} + \frac{\tau}{2} \sum_k \la \GammaNoise_{ik}^2 \ra		\nn\\
\Leftrightarrow \quad\quad \sigma^2_{i} &= \frac{\tau}{1-(1-\rbhat) \left(\frac{\partial f}{\partial g}\right)_{\gbar_i,\cset_i} } \frac{1}{2} \sum_k \la \GammaNoise_{ik}^2 \ra	\nn\\
				      &= \frac{1}{\frac{1}{\tau}- \frac{\rmax-\rb}{\Nmax} \left(\frac{\partial f}{\partial g}\right)_{\gbar_i,\cset_i} } \frac{1}{2} \sum_k \la \GammaNoise_{ik}^2 \ra	\nn\\
				      &\overset{(\rb=0)}{=} \frac{1}{\frac{1}{\tau}- \frac{\rmax}{\Omega} \left(\frac{\partial f}{\partial \left( G/\Omega \right)}\right)_{\bar{G}_i,\cset_i} } \frac{1}{2} \sum_k \la \GammaNoise_{ik}^2 \ra
\label{eqVarSelfReg}
\end{align}
where $\Omega$ is the volume.
This  recovers the formula obtained from the fluctuation-dissipation approach in \cite{Tkacik2012}, 
in which the variance is attenuated by a ``Langevin damping force''.

\subsubsection{Short-correlations assumption}
\label{secSCA}
In the short-correlations assumption we assume that only nearest-neighbor volumes have significant correlations.
We thus set
\begin{align}
 \forall \alpha \forall \beta : \quad C_{ij}^{\alpha\beta} &= 0 \quad \text{if } j\notin N(i)\cup \lbrace i \rbrace
\end{align}
where $N(i)$ is the set of indices of the nearest-neighbor volumes of $i$.

In formula (\ref{eqCovarGeneral}), this simplifies the diffusive coupling terms originating from parts $\DTerm_\alpha$ and $\DTerm_\beta$.
In particular, the terms $C^{\alpha\beta}_{i n_j}$, which denote correlations between volume $i$ and the neighbors of $j$ (!)
now only have nonzero values when $i$ is a nearest neighbor volume of $j$, i.e. for $n_j = i$, while all other terms under the sum vanish.
This correspondingly holds for the terms $C^{\beta\alpha}_{j n_i}$, such that only $C^{\beta\alpha}_{j j} \geq 0$ when $n_i = j$.
We thus obtain the markedly simpler formula
\begin{align}
 C_{ij}^{\alpha\beta} &= \frac{T_{\alpha\beta}}{2} \Bigg[ \frac{\rmaxhat^\beta-\rbhat^\beta}{\tau} \sum_\zeta f_\beta'^{\zeta}(\gbarm_j,c_j) C_{ij}^{\alpha\zeta}	\nn\\
		      &\quad\quad\quad\quad\quad + \frac{\rmaxhat^\alpha-\rbhat^\alpha}{\tau} \sum_\zeta f_\alpha'^{\zeta}(\gbarm_i,\cset_i) C_{j i}^{\beta\zeta} \Bigg]	\nn\\
		      &\quad~ + \frac{1}{2} \Bigg[ \Lambda^2_\beta C_{ii}^{\alpha\beta} + \Lambda^2_\alpha C_{j j}^{\beta\alpha} \Bigg] + \frac{T_{\alpha\beta}}{2} \sum_k \left\langle \GammaNoise_{ik}^\alpha \GammaNoise_{j k}^\beta \right\rangle
\label{eqCovarSCA_NN}
\end{align}
which now is restricted to $j \in N(i)$, i.e. volume $j$ being a nearest neighbor volume of volume $i$ (while $C_{ij}^{\alpha\beta} = 0$ by covention if this is not the case).

Note that in the case $i=j$ the terms in the coupling sums do {\emph not} vanish, because then the terms $C^{\beta\alpha}_{j n_i} = C^{\beta\alpha}_{i n_i}$
and $C^{\beta\alpha}_{j n_i} = C^{\beta\alpha}_{i n_i}$ all are nonzero again.
However, in this case we obtain the same formula as in Eq.~(\ref{eqCovarLocal}).
Eqs.~(\ref{eqCovarSCA_NN}) and (\ref{eqCovarLocal}) together define a complete equation system 
for all nearest-neighbor covariances $C_{ij}^{\alpha\beta}$ and local (inter-gene) covariances $C_{ii}^{\alpha\beta}$ that contains
significantly less unknowns than the system for general $C_{ij}^{\alpha\beta}$ (where $j$ can also denote a non-neighboring volume) defined by Eq.~(\ref{eqCovarGeneral}).

%\pagebreak
\subsection{Noise powers}
\label{secNoisePowers}
%Under the assumption that the system is not in the multi-stable regime yet, the noise powers will be entirely composed of contributions that we know how to treat.
Building on our earlier work \cite{Sokolowski2015}, our model comprises three noise sources affecting the expression of the gap genes in the nuclei: 
(1.) ``input noise'' from diffusive arrival of transcription factors (TFs);
(2.) ``output noise'' (birth-death noise) coming from the production and degradation of gap gene products following gene activation;
(3.) ``diffusion noise'' from the stochastic hopping of gene products between neighboring volumes.

The diffusion noise not only contributes to the summed noise sources affecting the copy number of gene $\alpha$ in volume $i$, $\Big\langle \left[\NNoise^\alpha_{ii}\right]^2 \Big\rangle$;
it also induces anticorrelated fluctuations in neighboring volumes that we have to include in the nearest-neighbor noise correlations $\Big\langle \left[\NNoise^\alpha_{i\nu}\right]^2 \Big\rangle \equiv \Big\langle \GammaNoise_{ik}^\alpha \GammaNoise_{\nu k}^\alpha \Big\rangle = - \Big\langle \left[\DNoise^\alpha_{[i \rightarrow \nu]}\right]^2 \Big\rangle - \Big\langle \left[\DNoise^\alpha_{[\nu \rightarrow i]}\right]^2 \Big\rangle$, where $\nu$ indicates one of the nearest neighbor volumes of $i$.
Here and throughout we will assume that diffusive processes of different protein species are uncorrelated, i.e. the corresponding term $\Big\langle \GammaNoise_{ik}^\alpha \GammaNoise_{\nu k}^\beta \Big\rangle \equiv 0$.
We calculate the detailed form of the diffusion noise contributions further below.

We will now first compute the normalized ``local'' noise term $\Big\langle \left[\NNoise^\alpha_{ii}\right]^2 \Big\rangle$, which in more detail we can write as follows:
\begin{align}
 \la \left[\NNoise^\alpha_{ii}\right]^2 \ra &\equiv \underset{\text{input noise}}{\underbrace{\la \left[\JNoise^\alpha_{i}\right]^2 \ra}}
						  + \underset{\text{output noise}}{\underbrace{\la \left[\ONoise^\alpha_{i}\right]^2 \ra}}
					          + \underset{\text{diffusion/hopping noise}}{\underbrace{
						    \sum_\nu \la \left[\DNoise^\alpha_{[i \rightarrow \nu]}\right]^2 + \left[\DNoise^\alpha_{[\nu \rightarrow i]} \right]^2 \ra
						  }}
\end{align}
Let us first consider the input noise sources.
Here we assume that the combined diffusive search and binding processes responsible for this noise will be uncorrelated among the different genes.
This ``crosstalk-free'' situation is expected to hold when the copy numbers of all transcription factors are comparably low,
and the accessibility of a TF-specific binding site does not depend on the presence of a different TF type.
Then we can simply add up the input noise contributions contributed by all the different TFs that regulate gene $\alpha$:
\begin{align}
\lefteqn{ \la \left[\JNoise^\alpha_i\right]^2 \ra = 
\underset{\equiv \sum_\kappa \JNoise_c^\kappa}{\underbrace{ \frac{1}{\Nmax^2} (\rmax^\alpha - \rb^\alpha)^2 \sum_\kappa \left[ \left( \frac{\partial f_\alpha}{\partial c^\kappa} \right)^2 \frac{2 c^\kappa ~\Phi_\kappa^\alpha}{D_\kappa l_\alpha} \right]_{c_i^\kappa} }} }	\nn\\
& + \underset{\equiv \sum_\zeta \JNoise_g^\zeta}{\underbrace{
  \frac{1}{\Nmax^2} (\rmax^\alpha - \rb^\alpha)^2 \sum_\zeta 
  \left[ \left( \frac{\partial f_\alpha}{\partial (G^\zeta / \Omega)} \right)^2 \frac{2 (G^\zeta/\Omega) \Phi_\zeta^\alpha}{D_\zeta l_\alpha} \right]_{\la G_i^\zeta \ra}
  }}
\end{align}
where $G_j^\zeta$ are absolute copy numbers of the zygotic species, 
$\Omega$ the reaction (nuclear) volume, and $c_i^\kappa$ the concentration of maternal input species $\kappa$,
and $l_\alpha$ the spatial size of the TF binding sites, which we assume to be the same for all TF species and simply denote by $l$.
$D_\kappa$ and $D_\zeta$ denote the internal diffusion constants of the maternal and zygotic TF species, which we assume to be equal as well
for all TF species, denoting it by $D_\text{int}$ from now on.
$\Phi_\kappa^\alpha$ and $\Phi_\zeta^\alpha$ are functions of the internal structure of the regulatory region of gene $\alpha$,
which may depend on the average occupancies.
While throughout this work we set all $\Phi_{\kappa(\zeta)}^\alpha \equiv 1$, we included these factors in the derivations for potential future applications of the model.

The first terms for the input noise from maternal inputs can (for a particular $\kappa$) be simplified further as 
\begin{align}
 \JNoise_c^\kappa
 &= \frac{1}{\Nmax\tau} \frac{(\rmax^\alpha - \rb^\alpha)^2}{\rmax^2} \left[ \left( \frac{\partial f_\alpha}{\partial c^\kappa} \right)^2 \frac{2 c^\kappa \Nmax}{D_\text{int} l \tau} \Phi_\kappa^\alpha \right]_{c_i^\kappa}	\nn\\
 &= \frac{1}{\Nmax\tau} (\rmaxhat^\alpha - \rbhat^\alpha)^2 \left[ \left( \frac{\partial f_\alpha}{\partial c^\kappa} \right)^2 2c^\kappa c_0 \Phi_\kappa^\alpha \right]_{c_i^\kappa}	\nn\\
 \label{eqInputNoiseMaternal}
\end{align}
where $c_0\equiv \Nmax / D_\text{int} l\tau$ defines a natural concentration scale, as in our previous work \cite{Tkacik2009,Walczak2010,Tkacik2012,Sokolowski2015}.

For the other ($G_i^\zeta$-dependent) terms, we can similarly write, for each $\zeta$ separately:
\begin{align}
 \JNoise_g^\zeta
 &= \frac{1}{\Nmax^2} (\rmax^\alpha - \rb^\alpha)^2 
    \left[ \left( \frac{\partial f_\alpha}{\partial \left(\frac{G^\zeta}{\Nmax}\right)} \right)^2 \frac{\Omega}{\Nmax} \frac{2 \left(\frac{G^\zeta}{\Nmax}\right) \Phi_\zeta^\alpha}{D_\text{int} l} \right]_{\la G_i^\zeta \ra}	\nn\\
 &= \frac{1}{\Nmax\tau} \frac{(\rmax^\alpha - \rb^\alpha)^2}{\rmax^2} 
 \left[ \left( \frac{\partial f_\alpha}{\partial g^\zeta} \right)^2 \frac{\Omega}{\Nmax} \frac{2 g^\zeta\Nmax}{D_\text{int} l \tau} \Phi_\zeta^\alpha \right]_{\gbar_i^\zeta}	\nn\\
%  &= \frac{1}{\Nmax\tau} \frac{1}{\gammaMax} (\rmaxhat^\alpha - \rbhat^\alpha)^2 \left[ \left( \frac{\partial f_\alpha}{\partial g_i^\zeta} \right)^2 \frac{2 g_i^\zeta\Nmax}{D_\text{int} l \tau} \Phi_\zeta^\alpha \right]_{\gbar_i^\zeta}	\nn\\
 &= \frac{1}{\Nmax\tau} \frac{1}{\gammaMax} (\rmaxhat^\alpha - \rbhat^\alpha)^2 \left[ \left( \frac{\partial f_\alpha}{\partial g^\zeta} \right)^2 2 g^\zeta c_0 \Phi_\zeta^\alpha \right]_{\gbar_i^\zeta}	\nn\\
 \label{eqInputNoiseZygotic}
\end{align}
where $\gammaMax\equiv \Nmax/\Omega$ is the maximal mean concentration.

%\pagebreak
The output noise is simple birth-death shot noise, and given by the sum of the average birth and death rates:
\begin{align}
 \lefteqn{ \la \left[\ONoise^\alpha_i\right]^2 \ra  } \nn\\
 &= \frac{1}{\Nmax^2} \left[ \left( \rb^\alpha + (\rmax^\alpha - \rb^\alpha) f_\alpha(\gbarm_i,\cset_i)\right) + \frac{1}{\tau_\alpha} \la G_i^\alpha \ra \right]			\nn\\
 &= \frac{1}{\Nmax} \left[ \frac{1}{\tau}\left( \hat\rb^\alpha + (\rmaxhat^\alpha - \hat\rb^\alpha) f_\alpha(\gbarm_i,\cset_i)\right) + \frac{1}{\tau_\alpha} \gbar_i^\alpha \right]	\nn\\
\end{align}

Finally, for the diffusion noise sources we have:
\begin{align}
 \la \left[\DNoise^\alpha_{[i\rightarrow\nu]}\right]^2 \ra &= \frac{1}{\Nmax^2} h_\alpha \la G_i^\alpha \ra = \frac{1}{\Nmax} h_\alpha \gbar_i^\alpha	\nn\\
 \la \left[\DNoise^\alpha_{[\nu\rightarrow i]}\right]^2 \ra &= \frac{1}{\Nmax^2} h_\alpha \la G_\nu^\alpha \ra = \frac{1}{\Nmax} h_\alpha \gbar_\nu^\alpha
\end{align}
These terms not only appear in $\Big\langle \left[\NNoise^\alpha_{ii}\right]^2 \Big\rangle$;
in their negated form, they also constitute the nearest-neighbor noise covariance $\Big\langle \left[\NNoise^\alpha_{i\nu}\right]^2 \Big\rangle$,
listed in the final summary of all noise contributions below.

\pagebreak
Taken together, assuming equal internal TF diffusion constants and binding size lengths, the noise terms read
% \fbox{
% \begin{minipage}{0.46\textwidth}
\newcommand{\NoiseHspace}{-2EM}
\begin{align}
\lefteqn{ \la \left[\NNoise^\alpha_{ii}\right]^2 \ra = \frac{1}{\Nmax} \Bigg\lbrace }	\nn\\
&\quad\quad \frac{1}{\tau}\left[ \hat\rb^\alpha + (\rmaxhat^\alpha - \hat\rb^\alpha) f_\alpha(\gbarm_i,\cset_i)\right] + \frac{1}{\tau_\alpha} \gbar_i^\alpha
&\hspace{\NoiseHspace}\text{``output noise''}	\nn\\
&\quad + \frac{(\rmaxhat^\alpha - \rbhat^\alpha)^2}{\tau} \sum_\kappa \left[ \left( \frac{\partial f_\alpha}{\partial c^\kappa} \right)^2 2 c^\kappa c_0 \Phi_\kappa^\alpha \right]_{c_i^\kappa}
&\hspace{\NoiseHspace}\text{``maternal input noise''}	\nn\\
&\quad + \frac{1}{\gammaMax} \frac{(\rmaxhat^\alpha - \rbhat^\alpha)^2}{\tau} \sum_\zeta \left[ \left( \frac{\partial f_\alpha}{\partial g^\zeta} \right)^2 2 g^\zeta c_0 \Phi_\zeta^\alpha \right]_{\gbar_i^\zeta}
&\hspace{\NoiseHspace}\text{``zygotic input noise''}	\nn\\
&\quad + 2d \cdot h_\alpha \gbar_i^\alpha + \sum_\nu h_\alpha \gbar_\nu^\alpha	~\Bigg\rbrace
&\hspace{\NoiseHspace}\text{``diffusion noise''}	\\
&&\nn\\
& \la \left[\NNoise^\alpha_{i\nu}\right]^2 \ra = -\frac{h_\alpha}{\Nmax} \left( \gbar_i^\alpha + \gbar_\nu^\alpha \right)
&\hspace{\NoiseHspace}\text{``diffusion noise''}
\end{align}
% \vspace{0.5EX}
% \end{minipage}}
where $\kappa$ and $\zeta$, respectively, run over all maternal input and gap gene output species in the system,
and $\nu \in N(i)$ over all nearest neighbors of volume $i$.

\subsection{Regulatory function}
\label{secRegFun}
Until now we could carry out all calculations without specifying the nonlinear regulatory function $f_\alpha(\gset_i, \cset_i)$
that converts the local maternal input and gap protein levels into an average activity level of the regulated gap gene.
While multiple choices are possible here, such as \Hill{}-type regulatory functions, in this work we opted for the 
\Monod{}-\Wyman{}-\Changeux{} (MWC) regulation model, which allows exploration of a richer set of solutions.
Our parametrization of the model is based on the approximation used in \cite{Walczak2010},
which we briefly recapitulate here once more for completeness.

The MWC model was originally introduced for allosteric enzymes that can bind ligands with different affinities in two different conformational states named $T$ (``tense'') and $R$ (``relaxed''). In the absence of any ligand, the occupancy of the respective states is determined by a single equilibrium constant $L = [T] / [R]$. The ligands can bind to the enzyme in both states, but with different affinities, characterized by different dissociation constants $K_T$ and $K_R$. Since the affinity of the ligand to binding sites in the $R$ state is higher, the equilibrium is shifted towards the $R$ state upon ligand binding. One notable limitation of the MWC model is that it requires all binding subunits to be in the same state; as such, the whole enzyme switches state at once, partial switching of subunits is not accounted for. With these assumptions, formulas for two quantities can be derived: the (mean) fractional occupancy of (all) ligand binding sites, $\bar Y$, and the fraction of enzymes in the $R$ state, $\bar R$:
\begin{align}
 \bar Y &= \frac{\xi(1+\xi)^{n-1} + LC\xi(1+C\xi)^{n-1}}
                {(1+\xi)^n + L(1+C\xi)^n}   \\
 \bar R &= \frac{(1+\xi)^n}{(1+\xi)^n + L(1+C\xi)^n}
 \label{eq:MWC_R_occupancy}
\end{align}
Here $C=K_R/K_T$ is the state affinity ratio, $n$ is the number of (identical) subunits that can bind the ligand, and $\xi = [X]/K_R$ is the ligand concentration $[X]$ normalized to the affinity of the $R$ state.

This model can be employed as a phenomenological model of gene regulation, as in previous published works. This model is phenomenological in that the gene regulation is likely not a process at thermal equilibrium; nevertheless, the model provides a convenient and versatile parametrization of the regulatory functions, as we will show in what follows. Applied to gene regulation, a gene can exist in ``off'' or ``on'' expression states (mimicking the balance of the $T$ and $R$ states in the original model application to haemoglobin), and the binding of transcription factors (TFs) (which take the role of ligands in the original model) shifts the effective balance between the two gene expression states. 
For a specific gene $\alpha$, let us denote the constant describing the ratio between the ``on'' and ``off'' states by $L_\alpha$, the TF concentration by $c$, and the dissociation constants for TF binding in the ``on'' and ``off'' states by $K_{\rm on}^{\alpha 0}$ and $K_{\rm off}^{\alpha 0}$, respectively;
the indices $\alpha 0$ are used to indicate that there is only one regulator protein species, which we extend to the case of multiple regulators further below.
Then we can rewrite Eq.~(\ref{eq:MWC_R_occupancy}), which now will describe the activity $f_\alpha(c)$ of the gene as a function of the TF concentration $c$, as follows:
\begin{align}
 f_\alpha(c) &= \frac{ \left(1 + \frac{c}{K_{\rm on}^{\alpha 0}} \right)^{H^{\alpha 0}_{\rm reg}} }
		     { L_\alpha \left(1 + \frac{c}{K_{\rm off}^{\alpha 0}} \right)^{H^{\alpha 0}_{\rm reg}} + \left(1 + \frac{c}{K_{\rm on}^{\alpha 0}} \right)^{H^{\alpha 0}_{\rm reg}} }
\end{align}
Here the parameter $n$ (which could be interpreted as the number of binding sites for the TF) determines the strength of regulatory action of input $c$ on species $\alpha$, denoted by $H^{\alpha 0}_{\rm reg}$.
Notice that the constant $C$ in Eq.~(\ref{eq:MWC_R_occupancy}) now is absorbed into the different denominators dividing the TF concentration $c$.
For further clarity, we can also rewrite the formula as follows
\begin{align}
 f_\alpha(c) &= \frac{ e^{-\beta \varepsilon_{\rm on}}  \left(1 + \frac{c}{K_{\rm on}^{\alpha 0}} \right)^{H^{\alpha 0}_{\rm reg}} }
		     { e^{-\beta \varepsilon_{\rm off}} \left(1 + \frac{c}{K_{\rm off}^{\alpha 0}} \right)^{H^{\alpha 0}_{\rm reg}} 
		     + e^{-\beta \varepsilon_{\rm on}} \left(1 + \frac{c}{K_{\rm on}^{\alpha 0}} \right)^{H^{\alpha 0}_{\rm reg}}  }	
\end{align}
where $L_\alpha=e^{\beta\Delta\varepsilon} \equiv  e^{\beta(\varepsilon_{\rm on}-\varepsilon_{\rm off})}$ now corresponds to the ratio of the Boltzmann-like weights of the (unregulated/unbound) ``on'' and ``off'' states. 
As it will become clear later, the difference in the exponential corresponds to an effective ``base energy'' appearing in the denominator of $f_\alpha(c)$, controlling to the probability of spontaneous (``basal'') activity.

In reference \cite{Walczak2010} the constant $L_\alpha$ is written as $L_\alpha = \tilde{L}_\alpha^{H^{\alpha 0}_{\rm reg}}$, which makes it easier to combine it with the other terms.
Let us now consider the limit in which spontaneous activity is improbable (compared to regulated activity), i.e. $\varepsilon_{\rm on} \gg \varepsilon_{\rm off}$ or $L_\alpha \gg 1$,
and in which binding to the off-state is improbable, too, $K_{\rm off}^{\alpha 0} \gg K_{\rm on}^{\alpha 0}$.
Then we can ignore the term describing the probability of the bound-but-inactive state and write:
\begin{align}
 f_\alpha(c) &\simeq \frac{ \left(1 + \frac{c}{K_{\rm on}^{\alpha 0}} \right)^{H^{\alpha 0}_{\rm reg}} }
		     { \tilde{L}_\alpha^{H^{\alpha 0}_{\rm reg}} + \left(1 + \frac{c}{K_{\rm on}^{\alpha 0}} \right)^{H^{\alpha 0}_{\rm reg}} }
	      = \frac{ 1 }{ 1+ \tilde{L}_\alpha^{H^{\alpha 0}_{\rm reg}} \left/ \left(1 + \frac{c}{K_{\rm on}^{\alpha 0}} \right)^{H^{\alpha 0}_{\rm reg}} \right.}			\nn\\
	     &= \frac{ 1 }{ 1+ \exp \left[ -H^{\alpha 0}_{\rm reg} \ln \left( \frac{ 1 + c/K_{\rm on}^{\alpha 0} }{ 1 + \tilde c/K_{\rm on}^{\alpha 0} } \right) \right] }
	      = \frac{ 1 }{ 1+ \exp \left[ -H^{\alpha 0}_{\rm reg} \ln \left( 1 + \frac{c}{K_{\rm on}^{\alpha 0}} \right) + F_\alpha^{(0)} \right] }
\end{align}
Here, $\tilde c \equiv (\tilde{L}_\alpha-1) K_{\rm on}^{\alpha 0}$ has the interpretation of an effective half-activation (half-regulation) threshold concentration,
while we see that $F_\alpha^{(0)} = \ln L_\alpha$ indeed corresponds to a basal free-energy-like term.

We can straightforwardly generalize this to the multi-gene case.
When gene $\alpha$ is regulated by multiple TF concentrations $g^\zeta$ we can initially write
\begin{align}
 f_\alpha(\lbrace g^\zeta \rbrace) &= \frac{ \prod_\zeta \left(1 + \frac{g^\zeta}{K_{\rm on}^{\alpha\zeta}} \right)^{H^{\alpha\zeta}_{\rm reg}} }
		     { L ~ \prod_\zeta \left(1 + \frac{g^\zeta}{K_{\rm off}^{\alpha\zeta}} \right)^{H^{\alpha\zeta}_{\rm reg}} + \prod_\zeta \left(1 + \frac{g^\zeta}{K_{\rm on}^{\alpha\zeta}} \right)^{H^{\alpha\zeta}_{\rm reg}} }
\label{eqMWC_M1}
\end{align}
where $\zeta$ now loops over all regulatory input species.
Performing the same steps as before for each regulating input separately and regrouping yields
\begin{align}
 f_\alpha(\lbrace g^\zeta \rbrace) = \frac{ 1 }{ 1 + e^{-F_\alpha(\lbrace g^\zeta \rbrace)} } ~,
\end{align}
where the free-energy-like term $F_\alpha$ now reads
\begin{align}
 F_\alpha(\lbrace g^\zeta \rbrace) = \sum_\zeta H^{\alpha\zeta}_{\rm reg} \ln \left( 1 + g^\zeta/K_{\rm on}^{\alpha\zeta} \right) - F^{(0)}_\alpha
 \label{eqFreeEnergyM1}
\end{align}
with $F^{(0)}_\alpha = \ln L$ again.

While the $F$ term above groups all regulatory interactions (including self-regulation) under one sum, 
in our model we want to distinguish maternal regulation from the zygotic interactions, 
and therefore use the following more specific notation
\begin{align}
 f_\alpha(\myvec{g},\myvec{c}) &= \frac{ 1 }{ 1 + e^{-F_\alpha(\myvec{g},\myvec{c})} } 	\nn\\
\end{align}
with
\begin{align}
 F_\alpha(\myvec{g},\myvec{c}) &= \sum_\zeta  H_G^{\alpha\zeta}  \ln \left( 1 + g^\zeta /K_G^{\alpha\zeta} \right) 
			        + \sum_\kappa H_M^{\alpha\kappa} \ln \left( 1 + c^\kappa/K_M^{\alpha\kappa} \right) 
			        - F^{(0)}_\alpha
\end{align}
where $\kappa$ loops over all maternal and $\zeta$ over all zygotic inputs, respectively.
In our standard model with 3 maternal inputs and 4 gap genes expressed downstream of them, $\zeta \in \lbrace 1, 2, 3, 4 \rbrace$ and $\kappa \in \lbrace A, P, T \rbrace$, where the letters stand for the anterior (A), posterior (P) and terminal (T) maternal systems.

\subsection{Combined mRNA and protein model}
\label{secMRNAmodel}
In our original generic stochastic differential equation (SDE) ansatz in Sec.~\ref{secAnsatz} 
we did not take into account the intermediate step of mRNA production.
However, it is known that the translation process can induce additional noise in the output, in particular when protein production occurs in bursts.
In this section we will extend our original framework to effectively incorporate bursty protein production
by adapting the original set of SDEs and the noise covariance terms.

\subsubsection{Incorporating mRNA into the model}
Let us start by extending the set of differential equations in Sec.~\ref{secAnsatz} in order to incorporate the mRNA populations, which we will call $M_i^\alpha$.
In the following we will distinguish mRNA rates and lifetimes from protein rates and lifetimes by the tilde, 
i.e. whenever $\tau_\alpha$ will refer to the protein lifetime, $\tilde\tau_\alpha$ will denote the lifetime of the corresponding mRNA.
The dynamics for a mRNA-protein population pair for gene species $\alpha$ then can be written as 
(abbreviating noise terms as $\mathcal{N}_M$ and $\mathcal{N}_G$, respectively):
\begin{align}
\partial_t M_i^\alpha &= r_\alpha(\myvec{G}_i, \myvec{c}_i) - \frac{1}{\tilde\tau_\alpha} M_i^\alpha + \mathcal{N}_M	\label{eqMRNA+Protein_M}\\
\partial_t G_i^\alpha &= k_\alpha M_i^\alpha - \frac{1}{\tau_\alpha} G_i^\alpha - h_\alpha \sum_\nu \left( G_i^\alpha - G_\nu^\alpha \right) + \mathcal{N}_G
\label{eqMRNA+Protein_P}
\end{align}
Herein, $h_\alpha = D_\alpha / \Delta^2$ is the protein hopping rate between neighboring volumes at distance $\Delta$ 
and $\nu$ runs over all nearest-neighbor volumes, as before.
These equations are still unnormalized, i.e. given in absolute numbers.
Note that the equation for the protein levels $G_i^\alpha$ closely follows the original ansatz.

Let us first look at the means of these equations, denoted by bars over the stochastic quantities:
\begin{align}
 \bar M_i^\alpha &= \tilde\tau_\alpha ~ r_\alpha(\bar{\myvec{G}}_i, \myvec{c}_i)	\\
 \bar G_i^\alpha &= \tau_\alpha k_\alpha \bar M_i^\alpha - \tau_\alpha h_\alpha \sum_\nu \left( \bar{G}_i^\alpha - \bar{G}_\nu^\alpha \right)
\end{align}
The second equation illustrates that we cannot simply define the burst size as $\beta_\alpha$ as
a ratio $\bar G_i^\alpha/\bar M_i^\alpha$, because $\bar G_i^\alpha$ is also affected by diffusion.
We therefore opt for defining the burst size in our model as:
\begin{align}
 \beta_\alpha \equiv k_\alpha \tau_\alpha
\end{align}

\subsubsection{Normalized differential equation for the means of mRNA and protein}
In order to normalize the SDEs introduced in the previous subsection let us define
\begin{align}
 M_{\rm max} &\equiv r_{\rm on}  \tilde\tau		\nn\\
 G_{\rm max} &\equiv k_{\rm max} \tau M_{\rm max} = k_{\rm max}r_{\rm on} \tau\tilde\tau
\end{align}
where $r_{\rm on}$ is the fastest mRNA transcription rate (among all species $\alpha$),
$k_{\rm max}$ is the (fastest) protein translation rate from mRNA,
$\tilde\tau$ is the longest mRNA lifetime,
and $\tau$ the longest protein lifetime of all the species.
This means that $M_{\rm max}$ and $G_{\rm max}$ are the highest possible (per-species) mRNA and protein numbers in the system.
These rates also imply a maximal burst size $\betaMax \equiv k_{\rm max} \tau$;
while the species-specific burst sizes $\beta_\alpha$ in principle may be different,
they will all be smaller than or equal to $\betaMax$, 
and the corresponding ratios $\beta_\alpha / \betaMax$
will appear in the normalized noise covariances further below.

Using the maximal protein copy number $G_{\max}$ for normalization, $\Nmax \equiv G_{\max} = k_{\rm max}r_{\rm on} \tau\tilde\tau$,
and averaging over Eqs.~(\ref{eqMRNA+Protein_M}) and (\ref{eqMRNA+Protein_P})	, we directly obtain the set of coupled ordinary differential equations
describing the time-evolution of the normalized mean expression levels of mRNA and protein populations in our extended model:
\begin{align}
 \partial_t \bar m_i^\alpha &= \frac{r_\alpha(\bar{\myvec{g}}_i, \myvec{c}_i)}{\Nmax} - \frac{1}{\tilde\tau_\alpha} \bar m_i^\alpha
 \label{eqmRNA+Protein_Mmean}\\
 \partial_t \bar g_i^\alpha &= \frac{\beta_\alpha}{\tau_\alpha} \bar m_i^\alpha - \frac{1}{\tau_\alpha} \bar g_i^\alpha - h_\alpha \sum_\nu \left( \bar g_i^\alpha - \bar g_\nu^\alpha \right)
 \label{eqmRNA+Protein_Pmean}
\end{align}
where $\bar m_i^\alpha \equiv \langle M_i^\alpha \rangle / \Nmax$  and $\bar g_i^\alpha \equiv  \langle G_i^\alpha \rangle / \Nmax$.

By further carrying out the normalization of the production function $r_\alpha(\myvec{g}_i, \myvec{c}_i)$ 
we recover almost the same structure as in the original model:
\begin{align}
 \frac{r_\alpha(\myvec{g}_i, \myvec{c}_i)}{\Nmax} &= 
      \frac{1}{\betaMax r_{\rm on}\tilde\tau} \left[ \rb^\alpha + (\rmax^\alpha - \rb^\alpha) f^\alpha(\myvec{g}_i, \myvec{c}_i) \right]	\nn\\
 &=   \frac{1}{\betaMax \tilde\tau} \left[ \rbhat^\alpha + (\rmaxhat^\alpha - \rbhat^\alpha) f^\alpha(\myvec{g}_i, \myvec{c}_i) \right] 
      \equiv \hat r(\myvec{g}_i, \myvec{c}_i)
\end{align}
Note, however, that now the normalized rates $\rbhat^\alpha$ and $\rmaxhat^\alpha$ have a slightly different meaning,
as they refer to the mRNA basal production rate and the mRNA production rate at full induction, respectively;
the additional prefactor $1/\betaMax$ accounts for our choice of normalizing by the maximal protein level.

Eqs.~(\ref{eqMRNA+Protein_M}) and (\ref{eqMRNA+Protein_P}) allow us to forward-integrate all coupled mRNA and protein levels in the system
in order to obtain predictions of the mean expression levels at any desired time point.
Obtaining a similar prediction for the covariance matrix of the coupled system is technically more challenging,
but simplifies considerably when the cases of significantly differing mRNA and protein time scales are considered, as described in the next subsection.

\subsubsection{Approximations for computing noise covariances}
In order to avoid the necessity of repeating the derivations for the covariance matrix with explicit inclusion 
of the additionally arising covariances of the kind $\la \delta G_i^\alpha \delta M_j^\beta \ra$,
we consider the extended model in the regime of separated mRNA and protein time scales,
which allows us to reuse all previous derivations.

If the mRNA dynamics were much faster than the protein dynamics, mRNA fluctuations would be largely averaged out at the protein level
and the mRNA levels would be mainly determined by the protein levels propagated through the regulatory function.
We could then approximate
\begin{align}
 M_i^\alpha \simeq \tilde\tau_\alpha r_\alpha(\myvec{G}_i,\myvec{c}_i)
\end{align}
such that
\begin{align}
\partial_t G_i^\alpha &\simeq k_\alpha \tilde\tau_\alpha ~ r_\alpha(\myvec{G}_i, \myvec{c}_i) - \frac{1}{\tau_\alpha} G_i^\alpha - h_\alpha \sum_\nu \left( G_i^\alpha - G_\nu^\alpha \right) + \mathcal{N}_G
\end{align}
which recovers the original set of equations without explicit treatment of mRNA in Sec.~\ref{secAnsatz}.

In the opposite regime, i.e. $\tilde\tau_\alpha \gg \tau_\alpha$,
which we consider in our model,
we can make another approximation assuming that the mRNA is the relevant species 
and that the protein populations equilibrate on a timescale faster than the mRNA populations are changing.
In this limit, the protein populations effectively ``track'' the mRNA populations and we can make the ansatz:
\begin{align}
  \Geff_i^\alpha \equiv \beta_\alpha M_i^\alpha \equiv k_\alpha \tau_\alpha M_i^\alpha
\end{align}
This allows us to write down an effective SDE for the production and degradation of $\Geff_i^\alpha$
by simply multiplying the corresponding SDE for the mRNA population by the protein burst size $\beta_\alpha$:
\begin{align}
 \partial_t \Geff_i^\alpha &= \beta_\alpha r_\alpha(\myvec{\Geff}_i, \myvec{c}_i) - \frac{1}{\tilde\tau_\alpha} \Geff_i^\alpha + \beta_\alpha \mathcal{N}_{\Geff}
\end{align}
%where we opt for multiplying the noise-covariances by the factor $\tilde\beta \equiv \beta_\alpha$, as described in the case of the simple model.
The equation above does not incorporate protein diffusion yet.
But since $\Geff$ is a very good proxy for the protein population $G$ we can assume
that the $\Geff$ population is altered by diffusion in the same way as the $G$ population,
and that the noise contributions originating from diffusive exchange, which we denote by $\mathcal{D}_{\Geff}$,
can be simply added on top of the other noise sources that affect $\Geff$ (and actually propagate from the mRNA to the protein levels).
We can therefore incorporate diffusion into our ansatz as follows:
\begin{align}
 \partial_t \Geff_i^\alpha &= \beta_\alpha r_\alpha(\myvec{\Geff}_i, \myvec{c}_i) - \frac{1}{\tilde\tau_\alpha} \Geff_i^\alpha 
			      - h_\alpha \sum_\nu \left( \Geff_i^\alpha - \Geff_\nu^\alpha \right) 
			      + \beta_\alpha \mathcal{N}_{\Geff} + \mathcal{D}_{\Geff}
\end{align}
This equation again recovers the structure of our ansatz in Sec.~\ref{secAnsatz}.
The relevant differences are the burst-size prefactors in the production and noise term;
while the former will disappear after normalization, the second prefactor will lead to an additional noise contribution,
as shown further below.
Also note that the relevant integration time scale in the equation above is the mRNA life time $\tilde\tau_\alpha$.

We can now again normalize the approximated SDE above by the normalization constant
$\Nmax = G_{\rm max}$ and defining $\geff_i \equiv \Geff_i/\Nmax$:
\begin{align}
\partial_t \geff_i^\alpha &\simeq \beta_\alpha ~ \hat r_\alpha(\myvec{\geff}_i, \myvec{c}_i) - \frac{1}{\tilde\tau_\alpha} \geff_i^\alpha - h_\alpha \sum_\nu \left( \geff_i^\alpha - \geff_\nu^\alpha \right) + \frac{\beta_\alpha}{\Nmax} \mathcal{N}_{\Geff} + \frac{1}{\Nmax} \mathcal{D}_{\Geff}
\label{eqApproxSDE}
\end{align}
The equation above implies that in the considered limit we can approximate 
the noise powers affecting the protein levels by the noise powers affecting the (absolute) mRNA levels multiplied by prefactor $\beta_\alpha / \Nmax$
and supplemented by the protein-level (normalized) diffusion noise $\mathcal{D}_{\Geff} / \Nmax$,
which has the same structure as in our original model since we normalize by the maximal protein copy number as before.
We calculate the explicit normalized forms of all the modified noise covariances comprised in the extended mRNA-protein model in the subsequent section.

\subsubsection{Normalized noise covariances with mRNA}
\label{secCovsWmRNA}
As in the simpler model without separate treatment of mRNA and protein, for which the noise covariances are treated in Sec.~\ref{secNoisePowers}, 
the noise term $\frac{\beta_\alpha}{\Nmax} \mathcal{N}_{\Geff}$ in Eq.~(\ref{eqApproxSDE}) consists of three contributions
that propagate into or arise at the mRNA level:
(1) the input noise from maternal regulation ($\JNoise_c$),
(2) the input noise from regulation by gap proteins acting as transcription factors to its own or other gap genes' regulatory region ($\JNoise_g$),
and (3) the output noise originating from production and degradation, denoted by ($\ONoise$).
The covariance terms describing these noise contributions are the same ones already used in Sec.~\ref{secNoisePowers},
only that we have to work with the rates and lifetimes relevant to the mRNA populations now,
and account for the additional prefactor $\frac{\beta_\alpha}{\Nmax}$ by multiplying with its square.
For the maternal input noise contributions (cf. Eq.~(\ref{eqInputNoiseMaternal}) in the model without mRNA)
this yields (using $\Nmax = \betaMax \rmax \tilde\tau$):
\begin{align}
 \JNoise_c^\kappa
  &= \frac{\beta_\alpha^2}{\Nmax^2} (\rmax^\alpha - \rb^\alpha)^2 \left[ \left( \frac{\partial f_\alpha}{\partial c^\kappa} \right)^2 \frac{2 c^\kappa}{D_\text{int} l} \Phi_\kappa^\alpha \right]_{c_i^\kappa}	\nn\\
 &= \frac{\beta_\alpha^2}{\Nmax\tilde\tau} \frac{(\rmax^\alpha - \rb^\alpha)^2}{\betaMax^2 \rmax^2} \left[ \left( \frac{\partial f_\alpha}{\partial c^\kappa} \right)^2 \frac{2 c^\kappa \Nmax}{D_\text{int} l \tilde\tau} \Phi_\kappa^\alpha \right]_{c_i^\kappa}	\nn\\
 &= \frac{1}{\Nmax\tilde\tau} (\rmaxhat^\alpha - \rbhat^\alpha)^2 \left(\frac{\beta_\alpha}{\betaMax}\right)^2
    \left[ \left( \frac{\partial f_\alpha}{\partial c^\kappa} \right)^2 2c^\kappa c_0 \Phi_\kappa^\alpha \right]_{c_i^\kappa}
\end{align}
where $c_0 \equiv \Nmax / (D_\text{int}l\tilde\tau)$ now is defined based on the longest mRNA lifetime $\tilde\tau$.
This recovers almost exactly the same formula as before, Eq.~(\ref{eqInputNoiseMaternal}), apart from the prefactor $(\beta_\alpha/\betaMax)^2$,
which vanishes when all burst sizes $\beta_\alpha$ are equal.

A completely analogous calculation yields the corresponding term for the other (zygotic) input noise contributions
(Eq.~(\ref{eqInputNoiseZygotic}) in the model without mRNA):
\begin{align}
 \JNoise_g^\zeta
 &= \frac{1}{\Nmax\tilde\tau} \frac{1}{\gammaMax} (\rmaxhat^\alpha - \rbhat^\alpha)^2 \left(\frac{\beta_\alpha}{\betaMax}\right)^2
    \left[ \left( \frac{\partial f_\alpha}{\partial g^\zeta} \right)^2 2 g^\zeta c_0 \Phi_\zeta \right]_{\gbar_i^\zeta}
\end{align}

The output noise follows the same ansatz as in Sec.~\ref{secNoisePowers}
and is given by the sum of the average mRNA production and degradation rate again multiplied with prefactor $(\beta_\alpha/\Nmax)^2$, 
but care has to put into its correct normalization.
Repeating the original calculation with the required modifications yields
\begin{align}
 \la \left[\ONoise^\alpha_i\right]^2 \ra
 &= \frac{\beta_\alpha^2}{\Nmax^2} \left[ \left( \rb^\alpha + (\rmax^\alpha - \rb^\alpha) f_\alpha(\gbarm_i,\cset_i)\right) + \frac{1}{\tilde\tau_\alpha} \la M_i^\alpha \ra \right]			\nn\\
 &= \frac{\beta_\alpha}{\Nmax}
    \left[ \frac{\beta_\alpha}{\betaMax\rmax\tilde\tau} \left(\rb^\alpha + (\rmax^\alpha - \rb^\alpha) f_\alpha(\gbarm_i,\cset_i)\right) 
    + \frac{1}{\tilde\tau_\alpha} \frac{\beta_\alpha \la M_i^\alpha \ra}{\Nmax} \right]	\nn\\
 &= \beta_\alpha \times \frac{1}{\Nmax}
    \left[ \frac{1}{\tilde\tau} \left(\frac{\beta_\alpha}{\betaMax}\right) \left( \hat\rb^\alpha + (\rmaxhat^\alpha - \hat\rb^\alpha) f_\alpha(\gbarm_i,\cset_i)\right) 
    + \frac{1}{\tilde\tau_\alpha} \gbar_i^\alpha \right]
\end{align}
where in the last line we make use of $\beta_\alpha \la M_i^\alpha \ra / \Nmax = \la G_i^\alpha \ra / G_{\rm max} = \gbar_i^\alpha$.
%The last equation demonstrates: 
While for equal burst sizes $\beta_\alpha$ the prefactor $\beta_\alpha/\betaMax$ preceding the production term again vanishes,
the overall prefactor $\beta_\alpha$ that multiplies the whole output noise always remains.
As known from simpler expression models, this precisely reflects the superpoissonian increase of the output noise due to bursty protein production.

Taken together the normalized noise covariances in the model with mRNA read:
% \fbox{
% \begin{minipage}{0.46\textwidth}
%\newcommand{\NoiseHspace}{-2EM}
\begin{align}
\lefteqn{ \la \left[\NNoise^\alpha_{ii}\right]^2 \ra = \frac{1}{\Nmax} \Bigg\lbrace }	\nn\\
&\quad\quad \frac{\beta_\alpha}{\tilde\tau}  \left(\frac{\beta_\alpha}{\betaMax}\right) 
  \left[ \hat\rb^\alpha + (\rmaxhat^\alpha - \hat\rb^\alpha) f_\alpha(\gbarm_i,\cset_i)\right] + \frac{1}{\tilde\tau_\alpha} \gbar_i^\alpha
&\hspace{\NoiseHspace}\text{``output noise''}	\nn\\
&\quad + \frac{(\rmaxhat^\alpha - \rbhat^\alpha)^2}{\tilde\tau}  \left(\frac{\beta_\alpha}{\betaMax}\right)^2
  \sum_\kappa \left[ \left( \frac{\partial f_\alpha}{\partial c^\kappa} \right)^2 2 c^\kappa c_0 \Phi_\kappa^\alpha \right]_{c_i^\kappa}
&\hspace{\NoiseHspace}\text{``maternal input noise''}	\nn\\
&\quad + \frac{1}{\gammaMax} \frac{(\rmaxhat^\alpha - \rbhat^\alpha)^2}{\tilde\tau}  \left(\frac{\beta_\alpha}{\betaMax}\right)^2
 \sum_\zeta \left[ \left( \frac{\partial f_\alpha}{\partial g^\zeta} \right)^2 2 g^\zeta c_0 \Phi_\zeta^\alpha \right]_{\gbar_i^\zeta}
&\hspace{\NoiseHspace}\text{``zygotic input noise''}	\nn\\
&\quad + 2d \cdot h_\alpha \gbar_i^\alpha + \sum_\nu h_\alpha \gbar_\nu^\alpha	~\Bigg\rbrace
&\hspace{\NoiseHspace}\text{``diffusion noise''}	\\
&&\nn\\
& \la \left[\NNoise^\alpha_{i\nu}\right]^2 \ra = -\frac{h_\alpha}{\Nmax} \left( \gbar_i^\alpha + \gbar_\nu^\alpha \right)
&\hspace{\NoiseHspace}\text{``diffusion noise''}
\end{align}
% \vspace{0.5EX}
% \end{minipage}}

In our optimizations we assume all burst sizes $\beta_\alpha$ to be equal, 
such that the increase of the output noise is the only (but significant) additional noise contribution arising from incorporation of mRNA and bursty protein production.
Note that since all noise covariances derived in this section and Eq.~(\ref{eqApproxSDE}) recover the structure of the corresponding formulae derived in the model without mRNA,
all derivations presented in Sec.~\ref{secCovs} also apply to the extended model, 
such that the computation of the covariances can be carried out in exactly the same way,
with the slight change that protein lifetimes have to be replaced by mRNA lifetimes.

\pagebreak
\section{Numerical computation of means and covariance matrix of gap gene expression}
\label{sec:numerical}
The spatial-stochastic model of gap gene expression derived in the previous sections
establishes two sets of equations for the means and covariances of the gap gene expression levels, respectively,
which we solve numerically.
Note that we need to solve for the means first
because they appear in the definition of the coupled linear system relating the entries of the covariance matrix.

\subsection{Forward integration of the means}
\label{secMeansIntegration}
As already outlined in Sec.~\ref{secMeans}, Eqs.~(\ref{eqMRNA+Protein_M}) and (\ref{eqMRNA+Protein_P}) for the mean expression levels must be integrated numerically. 
We start with zero expression levels at $t=0$ and use the built-in MATLAB ODE solver {\tt ode23t}.
We account for nuclear divisions that result in a doubling of production capacities upon every division
by ramping up the production rate at the start of every new nuclear cycle $c$ via a factor $\phi(c) = 2^c / 2^{14}$,
meaning that the full production rate is only reached at the beginning of nuclear cycle 14 ($\phi(14) = 1$).
The division times (end times of the respective nuclear cycle) used in our model are listed in Table~\ref{tabNucDiv} and follow the timeline described in \cite{Foe1993}.

\begin{table}[ht!]
\centering
\begin{tabular}{|l|r|r|r|r|r|r|r|r|r|r|r|r|r|r|}
\hline
 Nuclear cycle		& 1& 2& 3& 4& 5& 6& 7& 8& 9& 10& 11& 12& 13 \\
\hline
 Division time [min] 	& 12& 20& 28& 36& 44& 52& 60& 68& 76& 84& 94& 110& 126 \\
\hline
 \end{tabular}
 \caption{Nuclear division times / end times of nuclear cycles employed in our model, following the timeline described in \cite{Foe1993}.}
 \label{tabNucDiv}
 \end{table}

\subsection{Numerical computation of the covariance matrix}
\label{secCovMatrixNumerics}
Equation \ref{eqCovarGeneral} defines a coupled linear system for computing the entries of the generalized stationary covariance matrix
that comprises all covariances $C_{ij}^{\alpha\beta}$ across both all the gene species and all positions.
This linear system can itself be represented as a matrix equation
\begin{align}
 \mathcal{M} \mathbf{c} = \mathbf{G}
\end{align}
where $\mathbf{c}$ is a vector containing the entries $C_{ij}^{\alpha\beta}$ in a defined order, 
$\mathcal{M}$ a matrix collecting the coupling terms relating different entries $C_{ij}^{\alpha\beta}$ and $C_{kl}^{\gamma\delta}$ with each other,
and $\mathbf{G}$ a vector containing the respective right-hand side values that do not depend on $C_{ij}^{\alpha\beta}$,
in our case the noise covariance terms.
Thus, by constructing matrix $\mathcal{M}$ and vector $\mathbf{c}$ with a defined bookkeeping (order of $C_{ij}$ entries in $\mathbf{c}$)
and imposing the corresponding right-hand side noise covariances vector $\mathbf{G}$ we can obtain $\mathbf{c}$ by numerial matrix inversion,
\begin{align}
 \mathbf{c} = \mathcal{M}^{-1} \mathbf{G}
\end{align}
and use (map back) its entries for constructing the covariance matrix $\mathcal{C} = (C_{ij}^{\alpha\beta})$.

Note that the matrix $\mathcal{M}$ typically is large and grows quadratically in the system size variables.
Already for a one-dimensional system of typical size,
i.e. for $\Nx = 70$ and $N_{\rm y} = 1$ nuclei in the respective dimensions, and 4 gap genes modeled,
$\mathcal{M}$ is a $78400 \times 78400$ quadratic matrix with more than 6 billion entries ($70 \times 70 \times 4 \times 4 = 78400$).
However, when the short-correlations assumption (Sec.~\ref{secSCA}) is imposed, the vast majority of entries of $\mathcal{M}$ becomes zero.
This sparse matrix can be inverted considerably faster and for the system size parameters above can be carried out in less than one second
of computation time.

%\pagebreak
\section{Calculation of positional information}
\label{secPosInfo}
Our aim is to compute the positional information $I \equiv I(\myvec{g};x)$, which quantifies how much information the local noisy expression levels of the gap genes ($\myvec{g}$) carry about the position at some fixed developmental time when their readout occurs. Based on and validated by our earlier work \cite{Tkacik2015,Dubuis2013_PNAS}, we assume that the local distributions of expression levels, $P(\myvec{g}|x_i)$, can be written as multivariate Gaussian distributions and therefore are completely characterized by
the set of mean expression levels $\gbar(x_i)$ and the local covariance matrices $C^{\alpha\beta}(x_i) = C^{\alpha\beta}_{ii}$ at position $x_i$ predicted by our spatial-stochastic model.
%The validity of this Gaussian approximation has been verified before \cite{Tkacik2015,Dubuis2013_PNAS}.
However, even with this approximation it remains difficult to compute the positional information by direct integration, because  this technically requires integration over the four-dimensional distribution of all possible combinations of local expression levels. This is numerically unfeasible even with Monte-Carlo integration if no further approximations are made \cite{Tkacik2015}.
Therefore, here we opted for an approach which, instead of directly computing the positional information $I(\myvec{g};x)$, computes a lower bound on this quantity from an averaged decoding map (confusion matrix) at the relevant time point ($T_\text{meas})$; the corresponding decoding maps, when averaged, then can be integrated, bounding $I(\myvec{g};x)$ from below \cite{Hledik2019}. The tightest lower bound results from employing \Bayes' optimal decoder \cite{Petkova2019}, which we also opt for here. Additionally, for the case of \Drosophila gap genes, we have previously shown that the decoding-based bound that we employ here is tight to within experimental and numerical errors for the wild-type \cite{tkavcik2015positional}, and moreover predictive about the expression profiles in mutants with knocked-out maternal inputs or gap genes \cite{Petkova2019}.

% With these choices, the individual noisy expression levels used for computing the decoding maps are sampled using the multivariate Gaussian distribution of predicted expression levels, while the posterior distributions of positions inferred from them are computed via \Bayes' rule.

In order to construct the optimal decoder, we first apply \Bayes' rule as follows:
\begin{align}
 P(x^*_j|\myvec{g}) P(\myvec{g}) &= P(\myvec{g}|x^*_j) P(x^*_j)	\nn\\
 \Leftrightarrow\quad\quad\quad P(x^*_j|\myvec{g}) &= \frac{P(\myvec{g}|x^*_j) P(x^*_j)}{P(\myvec{g})} = \frac{P(\myvec{g}|x^*_j)}{\sum_j P(\myvec{g}|x^*_j)}
 \label{eqPosteriorDist}
\end{align}
% \begin{align}
%  P(x_i|\myvec{g}) P(\myvec{g}) &= P(\myvec{g}|x_i) P(x_i)	\nn\\
%  \Leftrightarrow\quad\quad\quad P(x_i|\myvec{g}) &= \frac{P(\myvec{g}|x_i) P(x_i)}{P(\myvec{g})} = \frac{P(\myvec{g}|x_i)}{\sum_i P(\myvec{g}|x_i)}
%  \label{eqPosteriorDist}
% \end{align}
Here the distribution on the left-hand side of the second line is the posterior distribution,
i.e. the distribution over decoded positions $x^*_j$ implied by reading the set of expression levels $\myvec{g}$.
$P(x^*_j) = \frac{1}{\Nx}$  is the prior distribution over positions which in our case is uniform 
(reflecting initial complete ignorance of the nuclei about their position and the nuclei being uniformly spaced along the AP axis)
and therefore cancels in the last formula.
 Equation~(\ref{eqPosteriorDist}) tells us that, for any fixed set of expression levels $\myvec{g}$,
we can obtain the posterior distribution by tabulating the values $P(\myvec{g}|x^*_j)$ for all $x^*_j$ and
normalizing them by their sum over all positions.

For any specific set of local expression levels $\myvec{g}_n(x_i)$, e.g. reflecting the local expression levels at the readout time $T_\text{meas}$
from one individual embryo $n$, we can now construct a decoding map $P_n(x^*_j|x_i)$ as follows:
\begin{align}
 P_n(x^*_j|x_i) = P(x^*_j|\myvec{g}_n(x_i)) = \frac{P(\myvec{g}_n(x_i)|x^*_j)}{\sum_j P(\myvec{g}_n(x_i)|x^*_j)}.
\end{align}
$P_n(x^*_j|x_i)$ establishes the mapping between the actual position $x_i$ and the locally implied positions $x^*_j$
for a single decoding process at all positions along the embryo axis.

Asking for the general performance of the decoding process, i.e. the amount of positional information that the gap genes can encode on average,
we can repeat all the steps leading up to the construction of an individual decoding map $P_n(x^*_j|x_i)$
for a set of $n = 1, .., \Ne$ embryo samples with corresponding expression levels $\myvec{g}_n(x_i)$ sampled from the Gaussian output distributions $P(\myvec{g}|x_i)$,
and then obtain the average decoding map:
\begin{align}
 \langle P \rangle(x^*_j|x_i) \equiv \langle P_n(x^*_j|x_i) \rangle_{\Ne} = \frac{1}{\Ne} \sum_n P_n(x^*_j|x_i)
\end{align}

The decoding capacity can be computed from any (normalized) decoding map $P(x^*|x)$ via
the known formula for mutual information:
\begin{align}
 I(x^*;x) &= \int_0^L \int_0^L P(x^*|x) \log_2 \frac{P(x^*|x)}{P_x(x^*)P_{x^*}(x)} dx^* dx
\end{align}
where by convention the integrand equals zero when $P(x,x^*)=0$ and $P_x$ and $P_{x^*}$ denote
the marginal distributions:
\begin{align}
 P_x(x^*)   &= \int_0^L P(x^*|x) dx 	\\
 P_{x^*}(x) &= \int_0^L P(x^*|x) dx^*
\end{align}
These integrals run over the whole domain of support of $x$ and $x^*$, i.e. the whole embryo length.
In our discrete system the integrals are sums over all nuclear positions,
while the summation is carried out over the average decoding map $\langle P \rangle(x^*_j|x_i)$:
\begin{align}
 I(x^*;x) &= \sum_i \sum_j \langle P \rangle(x^*_j|x_i) \log_2 \frac{\langle P \rangle(x^*_j|x_i)}{P_x(x^*_j)P_{x^*}(x_i)}	\nn\\
 P_x(x^*_j)   &= \sum_i \langle P \rangle(x^*_j|x_i)	\nn\\
 P_{x^*}(x_i) &= \sum_j \langle P \rangle(x^*_j|x_i)
\label{eq:mutinf_sum}
\end{align}
We found empirically that averaging the individual decoding maps over $\Ne = 50$ Monte Carlo embryo ``samples'' (i.e., using $\Nx \times \Ne$ sampled expression vectors $\myvec{g}_n(x_i))$) is sufficient for reproducible computation of $I(x^*;x)$, which serves as an input to the stochastic optimization algorithm described below.

Note that the portrayed way of estimating the positional information closely follows the biological decoding process, where the developing nuclei have to interpret the experienced gap expression levels in a short time frame, thus converting a \emph{single} ``snapshot'' of the noisy pattern into a positional estimate. Thus, while as described above, the computation of $I(x^*;x)$ (as a tractable lower bound on the positional information $I(\myvec{g};x)$) is technically motivated, it is intriguing to think that $I(x^*;x)$ itself is the quantity of primary biological relevance. 

\pagebreak
\section{Optimization procedure}
\label{secOpt}
We use an individually customized simulated annealing algorithm for random optimization of the positional information $I$ 
over the set of regulatory parameters and (in some cases) the gap protein diffusion constant,
followed by a final gradient descent optimization run after random optimization has settled into a minimum.
The optimization procedure starts with a totally random set of parameters uniformely sampled on the prescribed parameter bounds 
(see Table~\ref{tabOptPars} and Sec.~\ref{secOptPars}).
The parameters then are iteratively changed in a random fashion (described in detail below).
Upon each iteration, we solve our spatial-stochastic embryo model for the altered set of parameters,
and (re)compute the positional information $I$ of the pattern as described in Sec.~\ref{secPosInfo}.
The parameter change is always accepted if it leads to an increase of $I$;
if the parameter change decreases $I$, the change is still accepted with a finite probability according to the \Metropolis-\Hastings algorithm,
with an acceptance probability (effective temperature) that is lowered with every subsequent optimization step. 
The detailed cooling protocol is specified further below in Sec.~\ref{secCoolingProtocol}.

\subsection{Parameter changes and acceptance}
For each parameter $n$ we initially define a minimal and maximal bound, $P^{\text{min}}_n$ and $P^{\text{max}}_n$,
which in most optimizations are very generous and cover the full range of biophysically relevant values for the given parameter type
(see Sec.~\ref{secOptPars} for details).
The implied parameter interval $\Delta P_n = P^{\text{max}}_n - P^{\text{min}}_n$ then is subdivided into $N_\delta$ equidistant
subintervals $\delta P_n$.
While in classical approaches the subinterval boundaries together with $P^{\text{min}}_n$ and $P^{\text{max}}_n$ could be used
as a lattice of ($N_\delta + 1$) discrete parameter values on which the optimization can explore the range $\Delta P_n$,
we found that such discretization strongly restricts the set of possible expression patterns,
and therefore opted for the following approach in which all parameter combinations in the bounded hyperparameter space are accessible:
At the start of the optimization the parameter value $P_n^0$ for each optimization parameter $n$ 
is randomly chosen on the interval $\Delta P_n$ with uniform probability ($p(P_n^0) = 1/\Delta P_n$);
this defines the initial parameter vector $\mathbf{\Theta}^0 = (P^0_1, ..., P^0_{N_p})$,
where $N_p$ is the total number of optimized parameters.
At each iteration $i\geq 1$ of the random optimization, first we randomly choose which parameter to change by uniformly sampling 
a random integer number $n_i \in [1, N_p]$.
Second, a new proposed value for parameter $n_i$ is chosen by uniformly sampling it within the $2 \delta P_{n_i}$ interval
around the current value $P_{n_i}^{i-1}$, i.e.
\begin{align}
 r_i &= \mathcal{R}_i([-1,1])	\nn\\
 P_{n_i}^{i} &= P_{n_i}^{i-1} + r_i \delta P_{n_i}
\end{align}
where $\mathcal{R}_i([-1,1])$ denotes a continuous random number uniformly sampled on the interval $[-1,1]$.
If the new value $P_{n_i}^{i}$ reaches beyond the bounds $P^{\text{min}}_n$ or $P^{\text{max}}_n$, it is set (truncated) to the respective boundary value.
The proposed parameter vector $\mathbf\Theta_i^\text{prop}$ for the current optimization iteration then is constructed by replacing 
the $n_i$-th entry of the current vector $\mathbf\Theta_{i-1}$ by $P_{n_i}^{i}$.

We then compute the mean gap gene expression levels and their full covariance matrix for the proposed parameter vector $\mathbf\Theta_i^\text{prop}$
and the corresponding positional information $I_i^\text{prop}$ (see Sec.~\ref{secPosInfo}).
The new value of the objective function $I_i^\text{prop}$
then is used to carry out the ``accept or reject'' test according to the \Metropolis-\Hastings algorithm as follows:
\begin{enumerate}
 \item Compute objective function difference:
       \begin{align}
        \Delta I = I_i^\text{prop} - I_{i-1}	\nn
       \end{align}
 \item Compute the acceptance probability, given current temperature $T=T_i$:
       \begin{align}
        p_\text{acc} = \min(1, \exp(\Delta I / T))	\nn
       \end{align}
 \item \Metropolis-\Hastings test:
       \begin{itemize}
        \item Sample a uniformly distributed random number $\mathcal R \in [0,1]$.
        \item If $\mathcal R \leq p_\text{acc}$, accept the change:
	      \begin{itemize}
	       \item Update parameter vector, $\mathbf{\Theta_i} \leftarrow \mathbf{\Theta}_i^\text{prop}$.
	       \item Update objective function, $I_i \leftarrow I_i^\text{prop}$.
	      \end{itemize}
	\item Else reject:
	      \begin{itemize}
	       \item Parameter vector remains unchanged, $\mathbf{\Theta_i} \leftarrow \mathbf{\Theta}_{i-1}$.
	       \item Objective function value remains unchanged, $I_i \leftarrow I_{i-1}$.
	      \end{itemize}
       \end{itemize}
  \item Decrease temperature according to cooling protocol (see below), $T \leftarrow T_{i+1}$

  \item Iterate, $i\leftarrow i+1$

\end{enumerate}

The random optimization procedure ends when $\Nopt$ optimization steps have been attempted.
By default we set $\Nopt = 250 * \Nx * \Ng$, thus scaling the number of optimization steps
with the system size (number of positions $\Nx$ and number of simulated gap genes $\Ng$).
We found that with our model and optimization procedure, this number of optimization steps is sufficient for finding optimized patterns whose positional information is comparable to or even slightly larger than the $\sim 4.2~\text{bits}$ of the measured gap gene profiles \cite{Dubuis2013_PNAS}.
Finally, the local maximum found by random optimization is taken as the starting point
for a final gradient descent optimization run as described in Sec.~\ref{secGradientDescent}.
% To account for the fact that in the final course of optimization, i.e., close to a local maximum found,
% random parameter changes typically decrease the objective function without being accepted,
% the algorithm retains the parameter set $\mathbf{\Theta}_\text{max}$ that produced the highest
% positional information $I_\text{max}$ during the optimization run,
% and takes this value as a starting point for the final gradient descent optimization described
% in Sec.~\ref{secGradientDescent}.

\subsection{Cooling protocol}
\label{secCoolingProtocol}
The temperature is decreased according to an initially defined exponentially decaying temperature ramp
 as follows:
\begin{align}
T_i = T_\text{max} \exp(-t_i)
\end{align}
where the $t_i$ are $\Nopt$ numbers uniformly spaced on the interval $[0, \log(T_\text{max}/T_\text{min})]$,
such that $T_0 = T_\text{max}$ and $T_{\Nopt}=T_\text{min}$.

After analysing the typical changes in positional information $\Delta I$ upon changing the optimization parameters
we empirically explored different maximal and minimal temperature bounds $T_\text{max}$ and $T_\text{min}$,
and found that the choice of $T_\text{max} = 0.2$ and $T_\text{min} = 0.01$ leads to robust optimization behavior,
i.e. access of local information maxima in parameter space.

\subsection{Final gradient descent}
\label{secGradientDescent}
To ensure that the optimization fully settles into the local maximum $I^*_\text{rand}$ of the objective function (positional information)
found by random optimization, we start an additional gradient-descent optimization run from $I^*_\text{rand}$.
For this we used the built-in MATLAB {\it fmincon} optimizer with negative positional information as an objective function
and $\Nopt/10$ optimization steps, and the same parameter bounds as in random optimization.
In most cases, the final gradient descent leads to only small changes in the parameter values, 
reflecting that random optimization already settled to a point close to the actual local maximum.

%\subsection{Algorithm summary}

\subsection{Exploration of Drosophila-like patterns via constrained optimization}
\label{secConstrainedOpt}
We observed that optimized solutions obtained by starting from random parameter values can be degenerate: we found different gap gene expression patterns that encoded similarly high amounts of positional information. To systematically test whether any of these solutions lies in the vicinity of the observed fruit fly gap gene expression pattern, we employed a recently developed statistical framework which we reported in a separate publication \cite{Mlynarski2021}; we summarize the workflow below.

We first tested whether our model is expressive enough to be able to  generate expression patterns akin to the measured \Drosophila gap gene expression pattern. This was done by minimizing the pattern distance (defined further below) between the measured mean expression profiles and the mean expression profiles generated by our model, {\it without} maximizing the encoded positional information (i.e., by traditional ``fitting''). These tests confirmed that such \Drosophila-like patterns can be generated by our model (with the exception of the posterior Hunchback bump, which we attribute to the likely presence of additional Hb enhancers in the WT system, not included in our model).
We then carried out constrained optimization runs, in which we maximized positional information but supplemented by the pattern distance as a \Lagrange multiplier term, which generates a small ``pulling force'' that biases the stochastic search in parameter space towards expression patterns that are similar to the observed expression pattern.
Note that there is no a priori guarantee that an information-optimizing solution close to the \Drosophila WT exists.
As a last step, we chose the most similar of these optimal, \Drosophila -like patterns and confirmed that they remain a maximum of positional information when the pulling force is taken away again. The key idea behind this workflow  is that while the small ``pulling force'' helps break the degeneracy among information-maximizing solutions to guide the stochastic search towards patterns similar to the \Drosophila wild-type,  the force needs to be small compared to the information term, so that optimization remains  dominated by parameter changes that increase the information. This ``guided optimization'' is very different in the landscape and outcomes compared to pure fitting, which can be recovered in the limit $\lambda\rightarrow \infty$ in Eq.(\ref{utility}) in the function defined below; see also main paper Fig.~2F.

The extended objective function implementing optimization of positional information guided by similarity with the measured expression pattern is defined as
\begin{align}
 \mathcal{L} \equiv I(\myvec{g}; x) - \lambda D(\myvec{g}, \hat{\myvec{g}})
 \label{utility}
\end{align}
where $\myvec{g} = (g_i^\alpha)$ represents represents the pattern being optimized 
and $\hat{\myvec{g}} = (\hat g_i^\alpha)$ the measured pattern,
with expression levels normalized to the range $[0,1]$ in both patterns.
The pattern distance function $D(\myvec{g}, \hat{\myvec{g}})$ is defined as a finite p-norm,\begin{align}
 D(\myvec{g}, \hat{\myvec{g}}) \equiv \underset{\pi}{\min} \left( \frac{1}{N} \sum_\alpha \sum_i |g_i^\alpha - \hat g_i^{\mathcal{P}_\pi(\alpha)}|^p \right)^{1/p}
 \label{eqDistFun}
\end{align}
where $\alpha = 1..\Ng$ runs over all gap gene species, $i = 1..N$ over all positions/volumes in the model,
and index $\pi$ runs over all permutation functions $\mathcal{P}_\pi(\alpha)$ that map a gene species in $\myvec{g}$ to a gene species in $\hat{\myvec{g}}$.
This accounts for the fact that in the optimized pattern the gene identities are not a priori specified,
such that even for identical patterns the distance measure would be nonzero if the distances are not computed on the ``matching'' genes
that minimize the measure.
Interestingly, we found that the pattern distance measure performs better for $p\geq 3$ than for $p=2$, such that we set $p=3$ in all our distance computations.
Moreover, we found that the choice of $\lambda = 6$ for the \Lagrange mulitplier can successfully filter out high-information solutions
that are indeed similar to the measured {\it Drosophila} pattern.

We verified that the extended objective function successfully maximized $I(\myvec{g}; x)$ in two ways.
Firstly, we carried out parameter perturbations away from the optimal parameters which lead to reduction of $I(\myvec{g}; x)$ or negligible changes, as described in SI Fig.~\ref{SI-fig:Optimality}.
Secondly, we computed $I(\myvec{g}; x)$ on ``purely fitted'' solutions in which only $\lambda D(\myvec{g}, \hat{\myvec{g}})$ was minimized (see Fig.~2E of the main text). While the resulting mean expression patterns are similar to the observed pattern, the positional information for these solutions is considerably lower, highlighting that optimization of $I(\myvec{g}; x)$ is crucial for obtaining the parameter set that reproduces the high information content of the measured gap gene pattern.

%\pagebreak
\section{Parameters, simulation geometry and optimization constraints}
We discriminate between parameters that are optimized in our optimization runs as described in Sec.~\ref{secOpt}
and those that are set to fixed values, which include the parameters defining the simulated embryo geometry. The division of parameters into fixed and optimized is not arbitrary: fixed parameters represent biophysical constraints on the system that set the intrinsic levels of noise and thus limit positional information. We will describe the fixed parameters first and then proceed to the optimized parameters.

\subsection{Fixed parameters and geometric constraints}
\label{secFixedPars}
Table~\ref{tabFixedPars} gives an overview of the model parameters that we keep fixed in our optimizations,
although some of them are varied between different ensembles of optimization runs
(such as the number of output / gap genes $\Ng$ or the number of input gradients $\Ni$, see Sec.~\ref{secMutatedOpt} for details).
The geometric parameters are kept fixed in all optimization runs and are chosen to reproduce the geometry
of the Drosophila embryo in nuclear cycle 14.
While our model in principle allows to study arbitrary spatial arrangements of the embryo nuclei,
we restricted our simulatons to two cases:
(1.) A 2d model in which the embryo is abstracted as a cylinder
with $\Nx$ equally spaced nuclei along the axial direction and $\Ny$ equally spaced nuclei along the circumference
with periodic boundary conditions along the $y$-dimension, i.e. the nuclei with indices $y=1$ are coupled 
to the nuclei at $y=\Ny$.
(2.) A reduced 1d model in which the coupling along the $y$-dimension is skipped;
this arrangement is representative of a single ``string'' of nuclei along the half-perimeter
of the ellipsoidal section of the embryo.
To our surprise we found in early simulations that there is little difference
in optimal values of positional information between the 1d and the 2d model,
whereas the computational effort for solving the model is considerably higher in the 2d setting.
Moreover we found that the 1d model already is capable of reproducing the measured features 
of the gap gene system very well.
For the optimization runs presented in this work, we therefore opted for the 1d model in order to boost computational efficiency. Note that a reduction of the simulated lattice of positions has a profound effect on the computational cost, because the numerical computation of the cross-covariances in the system requires the inversion of a matrix that grows quadratically with the number of considered positions (see Sec.~\ref{secCovMatrixNumerics}).
For the same reason, we reduced the spatial size of the system further by excluding the very posterior positions of the embryo where the wild-type embryo does not show any expression of gap genes. This ``posterior cut-off'' amounted to the $N_{\rm c}=5$ posteriormost nuclei.

Our model includes three morphogen input gradients: an anterior gradient (A), representative of Bicoid,
a posterior gradient (P), representative of Nanos, and a terminal system (T), representative of Torso-like, that emerges symetrically
from the opposite poles of the embryo.
We assume an exponential shape for the gradients, and that the posterior gradient is a perfect mirror image of the anterior gradient with respect to mid-embryo.
The length scale $\lambda_{\rm AP}$ is chosen based on the observed properties of the Bicoid gradient, $\lambda_{\rm AP} = 20\%~{\rm EL}$;
the length scale of the terminal system gradients is chosen four times shorter, $\lambda_{\rm T} = 5\%~{\rm EL}$.
Restrictions imposed on the possible regulatory action of the gradients are discussed further below.

While our model is derived such that it allows for different protein and mRNA production rates and lifetimes, 
we make the simplifying assumption that these parameters are equal for all gap gene species, as indicated in Table~\ref{tabFixedPars}.
The same holds for the gap protein burst size and the internal diffusion constant of the transcription factor proteins (both zygotic and maternal).
We further assume that the binding sites of the regulated genes all have the same target size (binding region length) $l$.

We set the value of the inter-nuclear diffusion constant $D$, assumed to be equal for all gap genes,
to $D = 0.5~\mu m^2/s$ by default, and later varied it accross several optimization ensembles (cf. Fig.3E in the main text).

\begin{table}[ht]
\centering
\scalebox{0.9}{
\begin{tabular}{|l|c|c|c|c|}
 \hline
 Name & Symbol & Value  \\ 
 \hline
 \hline
 No. of output genes			& $\Ng$		& 1--5, default = 4	 \\
 No. of maternal input morphogens	& $\Ni$		& 1--3, default = 3	 \\
 No. of nuclei along embryo axis	& $\Nx$		& 70			\\
 Posterior cut-off		& $N_{\rm c}$	& 5			 \\
%  No. of nuclei along circumference	& $N_y$		& 1			& \\
 Corr. length cut-off (SCA)		& $N_{\rm SCA}$	& variable, default = 1	 \\
 Nuclear radius				& $R$		& 3.25~$\mu m$		 \\
 Nuclear volume 			& $\Omega$	& 144~$\mu m^3$		 \\
 Inter-nuclear distance			& $\Delta$	& 8.5~$\mu m$		 \\
%  Resulting embryo half-perimeter	& $L_{\rm P}$	& 603.5~$\mu m$		& \\
 \hline
 Max. gap gene mRNA copy no. (single species)	& $\Mmax$		& 540			 \\
 Max. protein burst size	& $\beta_{\rm m}$	& 12			 \\
 Max. no. of gap proteins (single species)	& $\Nmax$		& 6480			 \\
 Max. input concentration	& $c_{\rm m}$		& 150~nM = 90/$\mu m^3$			 \\
 Input gradient length (anterior and posterior)	& $\lambda_{\rm AP}$	& 0.2~EL	 \\
 Input gradient length (terminal system)	& $\lambda_{\rm T}$	& 0.05~EL	 \\
 \hline
 mRNA life times			& $\tilde\tau_\alpha \equiv \tilde\tau$	& 20~min 		\\
 Protein life times			& $\tau_\alpha \equiv \tau$		& 10~min		 \\
 mRNA prod. rates at full induction	& $\rmax^\alpha \equiv \rmax$		& 27/min		 \\
 Basal mRNA prod. rates			& $\rb^\alpha \equiv \rb$		& 0.027/min		 \\
 Protein burst size			& $\beta_\alpha \equiv \betaMax$	& 12			 \\
 \hline 
 Internal TF diff. const.	& $D_\text{int}$	& 10~$\mu m^2/s$	 \\
 TF binding site length 	& $l$			& 10~nm 		 \\
%  TF prom. state func.		& $\Phi_c$		& 1	& \\
 \hline
\end{tabular}
}
\caption{{\bf Overview of fixed biophysical and geometric parameters.}}
\label{tabFixedPars}
\end{table}

\subsection{Optimization parameters}
\label{secOptPars}
An overview of the model parameters that are optimized for maximizing positional information is shown in Table~\ref{tabOptPars}.
The core set consists of the regulatory parameters that determine the regulation of the gap genes
by the maternal inputs, $H_M^{\alpha\kappa}$ and  $K_M^{\alpha\kappa}$, and their mutual and self-regulation, $H_G^{\alpha\zeta}$ and  $K_G^{\alpha\zeta}$; note that while we separate maternal from zygotic regulation notationally, $K$ and $H$ describe the same quantities (regulation threshold and regulation strength, respectively) in both cases (see Sec.~\ref{secRegFun} for details).
These parameters were optimized on the bounds indicated in the respective column of Table~\ref{tabOptPars},
which were chosen such that all genes have the possibility to be fully expressed or completely inactive throughout the whole system,
i.e. generous enough to guarantee that the expression patterns emerging are not constrained by the parameter bounds 
but by the biophysical laws imposed by our model (specifically, by the structure of the regulatory function and the diffusive coupling).

While keeping the regulatory interactions as generic as possible, we imposed the following additional constraints aiding the optimization performance.
Firstly, optimization looks for the best (if any) activating regulatory parameters for the anterior and posterior morphogen gradients, and for the best (if any) repressive regulatory parameters for the terminal morphogen gradient.
Secondly, optimization looks for the best (if any) repressive interactions between gap genes, and the best (if any) self-activating auto-regulation. 
These two constraints are informed by our prior work~\cite{Walczak2010,Tkacik2012}, which derived optimal topologies for feed-forward and interacting networks, as well as for self-interaction. In this prior work we had analytic control over the optimization scenarios, providing a solid theoretical foundation for the nature of optimal solutions in the more complicated case we analyze numerically here.
For performance reasons, we allow only two of the gap genes to couple to the posterior gradient in the production runs, since that decreases the number of optimization parameters; we verified that this constraint has little influence on the optimal positional information values reached in our standard ensemble described here.
Taken together, these constraints do not change the nature of the optimal solutions that we find, but speed up the search time such that multiple optimization scenarios can be tractably explored. Relaxing these constraints could produce additional solutions that locally maximize positional information, but it will not remove the solutions we do find.

For two optimal ensembles (solid circles Fig.3E in the main text), we also optimized the inter-nuclear diffusion constant $D$ within the bounds indicated in  Table~\ref{tabOptPars}.
Interestingly, the average optimal value found, $D_{\rm opt} = 0.4~\mu m^2/s$, is very close to our initially chosen default value $D = 0.5~\mu m^2/s$ (see also Sec.~\ref{secAlteredDiffConst}).

\begin{table}[ht]
\centering
\scalebox{0.86}{
\begin{tabular}{|l|c|c|c|c|c|}
 \hline
 Name	& Symbol	& No. contained	& Unit	& Default	& Default	\\ 
	&        	& in model 	& 	& bounds	& value		\\ 
 \hline
 \hline
 Maternal regulation thresholds		& $K_M^{\alpha\kappa}$		& $\Ng\times\Ni$	& $1/\mu m^3$ 	& $[0.4, 2000]$		& \\
 Maternal regulation strength		& $H_M^{\alpha\kappa}$		& $\Ng\times\Ni$	& 	  	& $[0.01, 50]$		& \\
 Zygotic regulation thresholds		& $K_G^{\alpha\zeta}$		& $\Ng\times\Ng$	& $\Nmax$ 	& $[5\cdot10^{-7}, 2000]$	& \\
 Zygotic regulation strength		& $H_G^{\alpha\zeta}$		& $\Ng\times\Ng$	& 	  	& $[0.01, 50]$		& \\
 Protein inter-nuclear diffusion const.		& $D$				& 1			& $\mu m^2/s$	& $[0.01, 10]$		& $0.5$\\
 \hline
 \hline
 Total no.				&			& $2\Ng^2 + 2\Ng\Ni + 1$	& &&\\
 \hline
 - for $\Ng = 2$			&			& 21			& &&\\
 - for $\Ng = 3$			&			& 37			& &&\\
 - for $\Ng = 4$			&			& 57			& &&\\
 - for $\Ng = 5$			&			& 81			& &&\\
 \hline
\end{tabular}
}
\caption{
{\bf Overview of optimized biophysical parameters.} The regulatory parameters are always optimized starting from random values uniformly sampled within the indicated parameter bounds. For the analysis summarized in Fig.~3E and F of the main article, we also optimized the inter-nuclear diffusion constant $D$; otherwise the default value was used.
The bounds indicated for the regulatory strength terms $H_M^{\alpha\kappa}$ and $H_G^{\alpha\zeta}$ refer to the absolute value of these quantities, 
i.e. whenever $H_M^{\alpha\kappa}$ or $H_G^{\alpha\zeta}$ were restricted to be negative (representing repressive interactions)  we used the indicated bounds with a negative sign.
By default, the upper bound for the regulatory strength was $H_{\rm max}=50$, but reduced to smaller values in some optimization runs (see Sec.~\ref{secBoundedHmax}).
In the lower part we show how the total number of optimized parameters (including $D$) scales with the number of gap genes simulated, $\Ng$;
these numbers reduce by $2(\Ng-2)$ when the constraint that only two genes can couple to the posterior gradient is imposed, as explained in Sec.~\ref{secOptPars}.
}
\label{tabOptPars}
\end{table}

\subsection{Resource constraints}
One of the most important constraints impacting on the positional information encoded by a noisy gap gene expression pattern
is the total number of molecules that establish the pattern.
If this number could be made arbitrarily high, we could, in principle, remove all intrinsic noise in the system and
attain the maximal positional information possible.
Note that in our discrete model this maximal value is given by the logarithm of the number of nuclei along the embryo axis,
$I_{\rm max} = \log_2 N_x \simeq 6~{\rm bits}$ -- this bound corresponds to a perfect (error-free) identifiability of each nucleus based on a single readout of local gap gene expression patterns. In our model, the constraint on molecular copy numbers is imposed on two levels.

Firstly, we limit the maximal mRNA and protein copy numbers, $\Mmax$ and $\Nmax$, 
that can be reached at full induction of gene expression at each position
by imposing the maximal production rates and burst size indicated in Table~\ref{tabFixedPars}.
These values were chosen such that 
(1.) maximal mRNA copy numbers approximately correspond to experimentally determined values in nuclear cycle 14 \cite{Little2013}, and
(2.) maximal protein copy numbers correspond to current order-of-magnitude estimates ($10^3-10^4$ proteins at maximal induction per nucleus).
With this choice, the gene expression variance in optimized systems turns out to be close close to the experimentally measured variance, which we computed from existing experimental data \cite{Petkova2019}.

Secondly, we impose a constraint on the ``resource utilization'' (RU), which we define to be the fraction of the {\it total} maximal protein number
throughout the whole system (i.e., summing over {\it all} positions) that is available for creating the pattern (see Sec.~\ref{secResourceAlloc} for the corresponding formula).
${\rm RU} < 1$ means that the pattern in which all genes are fully induced at all positions cannot be created,
and the lower RU, the more positions in the embryo must have at least some gap genes below maximal induction.
By default, we set the resource utilization constraint to be equal to the corresponding value in the measured normalized mean expression pattern
of {\it Drosophila}, RU = 0.2, which allows us to compare theoretically optimal solutions under the same molecular cost constraint that the fly embryo experiences.

Note that the experimental value of RU is well below RU = 0.5. RU = 0.5 would lead to shifted ``counter patterns'' in which the genes are expressed at 50\% of the available positions. Such patterns are indeed optimal for a binary code \cite{Hillenbrand2016}, but in wild-type \Drosophila the RU is significantly lower, since the embryo can employ a richer code that is not restricted to binary ``ON''/``OFF'' expression levels and which can productively use intermediate expression levels (also see Sec.~\ref{secAlteredGeneNoAndGradients}). 
This is shown in Fig.~3C in the main text, where we explored higher and lower values of RU in dedicated optimization runs, and how this influences the positional information values reached by optimal solutions. Note that even when the constraint on resource utilization is removed, the positional information values of the optimized patterns still remain $\lesssim 5~{\rm bits}$. The only way to increase positional information beyond this bound would be to allow higher protein copy numbers (i.e., higher mRNA and protein production rates) -- this truly is a fundamental biophysical constraint that can only be removed at a higher time- or metabolic cost.

The resource utilization (RU) constraint is implemented in the optimization runs by simply rejecting simulated patterns that exceed the imposed RU value.

\subsection{Pattern stability constraint}
\label{sec:stability}
When exploring the role of the inter-nuclear diffusion constant $D$ (see Sec.~\ref{secAlteredDiffConst}), we observed that at lower $D$ optimal solutions tend to compensate for the loss of spatial averaging
by producing patterns that have more wiggled spatial profiles--while  at the same time exhibiting a larger amount of non-stationarity of patterns at read-out time $T_{\rm meas}$. As explained in the main text, achieving pattern stability around the read-out time might be essential to the patterning system architecture, motivating us to  explore optimizations with an upper bound on the allowed pattern rate-of-change (RoC).
This constraint was imposed in the same way as the constraint on resource utilization, i.e. by rejecting patterns that exceed the imposed maximal RoC value.

We defined the pattern rate-of-change as the temporal derivative of the normalized gap protein expression levels
evaluated at the read-out time $T_{\rm meas}$ and averaged over all positions and gap gene species, i.e.,:
\begin{align}
 {\rm RoC}(\myvec{g}, T_{\rm meas}) = \la\la \frac{dg_i^\alpha(T_{\rm meas})}{dt} \ra_{\alpha} \ra_i
\end{align}
where $\alpha$ runs over $1..\Ng$ and $i$ over $1..\Nx$.

%\pagebreak
\section{Details of data analysis}

\subsection{Computing positional error profiles}
The local value of the positional error, $\sigma_x(x_i)$, was computed from the first and second moments of the posterior decoding-map distributions. Recall that the average decoding map is $\langle P \rangle(x_j^*|x_i)$, which is defined in the section on positional information, Sec.~\ref{secPosInfo}. The formulae for the moments and the positional error therefore read:
\begin{align}
M_1(x_i) &= \frac{1}{\Nx} \sum_j x_j^* \langle P \rangle(x_j^*|x_i) \\
M_2(x_i) &= \frac{1}{\Nx} \sum_j (x_j^*)^2 \langle P \rangle(x_j^*|x_i) \\
\sigma_x(x_i) &= \sqrt{M_2(x_i) - M_1^2(x_i)}.
\end{align}

\subsection{Determining number of slopes}
In order to numerically determine the number of slopes we first computed (per gene) the numerical derivatives of the normalized expression profiles with respect to space, and then determined local extrema by comparing the sign of successive derivative values $(\bar g^\alpha_i)' \equiv \frac{d\bar g^\alpha}{dx}\vert_{x=x_i} = \frac{\bar g^\alpha_{i+1} - \bar g^\alpha_i}{\Delta}$.
In order to avoid falsely calling small local fluctuations full profile slopes, we restricted the test for derivative slope change to positions with significantly elevated gene expression. Moreover, only absolute derivative values surpassing a required minimum were considered to constitute ``slopes.'' This excluded, e.g., patterns in which the expression level reduced only slightly along the embryo axis and never stretched from very high to very low levels. Since the derivative is a local quantity defined between two successive lattice points, we averaged both the expression levels and derivative values before testing whether they are above the required minimal values. Mathematically, these additional conditions read:
\begin{align}
    (\bar g_{i-1}^\alpha + \bar g_i^\alpha)/2 &\geq \bar g_{\rm min}    \nonumber\\
    (|(\bar g_{i-1}^\alpha)'| + |(\bar g_i^\alpha)'|)/2 &\geq \bar g'_{\rm min}
\end{align}
In practice, $\bar g_{\rm min} = 0.1$ and $\bar g'_{\rm min}=1/EL$ (corresponding to a change from full to zero expression over the whole embryo length) proved suitable choices.
Overall this simple scheme has proven to be very robust in discarding occasional small random bumps in the profiles while calling only larger changes in expression levels significant slopes.

\subsection{Measuring resource allocation}
\label{secResourceAlloc}
For any given expression pattern, we define its ``resource utilization'' as the average expression level across all gene species and positions,
\begin{align}
RU \equiv \frac{1}{\Ng\Nx} \sum_\alpha \sum_i \gbar_i^\alpha \quad ,
\end{align}
where the mean expression levels $\gbar_i$ already are normalized between 0 and 1, as explained in Sec.~\ref{secAnsatz}; the prefactor $\frac{1}{\Ng\Nx}$ ensures that the resource utilization is also normalized between 0 and 1, with 1 corresponding to the situation in which all genes are expressed with the maximal expression rate at all positions.

The partial ``resource utilization per gene'' is defined as
\begin{align}
RU^\alpha \equiv \frac{1}{\Ng\Nx} \sum_i \gbar_i^\alpha \quad ,
\end{align}
where we retain the normalization factor $\frac{1}{\Ng\Nx}$ for comparability with $RU$; notice that the coefficient of variation of this quantity reported for different patterns in the main article (Fig.~3A) is independent of the specific choice of the normalization prefactor.

Similarly, we define the ``resource utilization per position'' as:
\begin{align}
RU_i \equiv \frac{1}{\Ng\Nx} \sum_\alpha \gbar_i^\alpha
\end{align}
Here we locally sum the expression levels of all gene species.

Both $RU^\alpha$ and $RU_i$ define discrete distributions for which we compute the coefficient of variation (CV), by normalizing the standard deviation of the distributions by their respective means (see Fig.~3A and SI Fig.~\ref{SI-fig:AddProp}C).

The resource bias shown in SI Fig.~\ref{SI-fig:AddProp}D was computed according to the formula
\begin{align}
B = \frac{R_A - R_P}{R_A + R_P}
\end{align}
where $R_A$ is the total number of proteins expressed in the anterior half of the embryo, and $R_P$ the remaining total number of proteins expressed in the posterior half, respectively.
For an uneven number of nuclei along the embryo axis, the resources of the central position were assigned to the anterior ($R_A$) and posterior ($R_P$) with weight 1/2, respectively.

\subsection{Determining regulatory interactions}
\label{sec:interactions}
We  visualized  the regulatory interactions of the optimized patterns in two ways.
Firstly, we computed the {\it local} ``regulatory contributions'' by inserting the local concentrations of maternal inputs into the free-energy-like terms that enter the exponent of the MWC regulatory function (see Sec.~\ref{secRegFun}), i.e.,
$H_M^{\alpha\kappa} \log(1 + c_i/K_M^{\alpha\kappa})$ for maternal and
$H_G^{\alpha\zeta} \log(1 + \bar{g}_i/K_G^{\alpha\zeta})$ for zygotic interactions.
This resulted in detailed spatial profiles of the regulatory inputs into each of the simulated genes along the embryo axis. Stacked example profiles for our best \Drosophila-like solution (Fig.~2 of the main article) are shown in SI Fig.~\ref{SI-fig:RegContrib}.

Secondly, in order to obtain an even simpler and stereotyped representation in terms of ``regulatory arrows,'' which ignores spatial dependency of the regulatory contributions, we reconstructed regulatory network ``cartoons'' from the zygotic regulation parameters ($H_G^{\alpha\zeta}$ and $K_G^{\alpha\zeta}$), by evaluating the logarithmic terms (see above) at the highest possible zygotic expression levels (which in our model all are equal to 1).
Terms resulting in values larger than 10 were deemed to be strong interactions depicted by fat arrow symbols in the cartoons; terms resulting in values smaller than 1 were deemed insignificant and therefore not shown in the cartoons.
Networks visualized in this way are shown in SI Fig.~\ref{SI-fig:OptNetworksProfiles} and Fig.~2D of the main article.

%\pagebreak
\section{Additional analyses of the optimal ensemble}
\label{secAddAnalysis}
In addition to quantities discussed in Fig.~3A and B of the main text, we extended our comparison between the optimal ensemble and the random ensemble to further quantities introduced below; comparison results are shown in SI Fig.~\ref{SI-fig:AddProp}.

SI Fig.~\ref{SI-fig:AddProp}A shows the pattern rate-of-change (RoC) distributions for the solutions of the optimal (graded colors) and random (grey) ensembles; for the optimal ensemble, the data shown is the same as used Fig.~3B of the main text. As in the main figure, SI Fig.~\ref{SI-fig:AddProp}A shows that higher positional information (yellow bullets) tends to lead to more stable patterns, i.e., a slower pattern RoC. Interestingly, the RoC is distributed over a similar range in the random ensemble. This is due to the fact that a considerable fraction of the random ensemble consists of solutions with little or no cross- and self-regulation, which tend to have slow dynamics (but also carry little information).

Since we observed that for some solutions the RoC itself may either increase or decrease as the patterns build up, we asked whether such ``RoC slowdown'' or ``RoC speedup'' is correlated with the ability to encode more positional information in the pattern. We defined ``RoC slowdown'' as the ratio between the RoC at 30~min before the readout time $T_{\rm meas}$, i.e. at $t=135~\text{min}$, and the RoC at $T_{\rm meas}=165~\text{min}$; in that respect, a RoC slowdown smaller than one is a RoC speedup. Our RoC estimate is thus a discrete approximation to the second-derivative of the gap gene expression dynamics.

We observe that high positional information tends to correlate with a dynamics slowdown when the patterns progress towards $T_{\rm meas}$; as shown in SI Fig.~\ref{SI-fig:AddProp}B, the optimal patterns tend to reduce their RoC by a factor of 1.6 on average during the last 30~min of patterning.
Comparison to random patterns is not trivial. The random patterns also tend to slow down, as a consequence of a generic anticorrelation between the RoC slowdown and the RoC itself; in other words, patterns with a small RoC at $T_{\rm meas}$ tend to have a larger relative reduction of the RoC before. Since random patterns tend to have a smaller RoC in general, for the reasons mentioned in the last paragraph, they also tend to display a more pronounced slowdown.
However, while for the optimal solutions the RoC slowdown is positively correlated with the positional information of the pattern (${\rm PC}\simeq 0.4$, $p<10^{-13}$), such correlation is very weak (and has an inverted sign) for random patterns (${\rm PC}\simeq -0.12$, $p = 0.026$).

Panel C of SI Fig.~\ref{SI-fig:AddProp} shows, for the standard optimal ensemble and the random ensemble, how uniformly the patterns tend to distribute the available resources (gap gene proteins) along the embryo axis, quantified by the average (across-gene) coefficient of variation (CV) of protein distributions, computed across the length of the embryo. 
Here the median CV is significantly higher (0.46) for the optimized patterns as compared to patterns from the random ensemble (0.13). This reflects the fact that upon optimization the variety of combinations of locally expressed genes increases, meaning that locally more resources have to be devoted in order to obtain a larger number of significant signals (expression levels).
In contrast, randomly generated patterns without significant cross-regulation tend to locally express one, or rarely at most two genes simultaneously, which reduces the variance in the necessary resources.

In SI Fig.~\ref{SI-fig:AddProp}D we analyse how the available protein resources are distributed between the anterior and posterior half of the embryo. The ``resource bias'' is defined such that a value of 0 means equal distribution of resources between the anterior and posterior halves of the embryo, while a value of $\pm 100$ implies a complete attribution of resources to the anterior (+) or posterior (-) half, respectively (see Sec.~\ref{secResourceAlloc} for the corresponding formula).
We observe that optimized solutions, on average, tend to attribute somewhat more resources (bias of $14\%$) to the anterior half, in part because posterior-most 5 nuclei of the system, as described in Sec.~\ref{secFixedPars}, do not contain gap gene patterns. This effectively leads to a  stronger anterior maternal signal and, consequently, optimal solutions tend to allocate more resources there.
%Note, however, that some of the optimized solutions can reach high positional information while devoting up to $\sim 30\%$ more of resources to the posterior.
%A slight over-attribution of resources to the anterior is also observable in the random solutions; the effect is, however, minor and most solutions accumulate around zero resource bias, again reflecting the fact that optimization is indeed required for an optimal usage of available resources under the given external constraints (in this case the maternal input levels).

\section{Optimizations with altered system components}
\label{secMutatedOpt}
In addition to the standard optimized ensemble of Fig.~3 in the main text, we also carried out \emph{ab-initio} optimizations for hypothetical alternative settings in which important components and parameters held fixed in the standard ensemble were changed.
In particular, we varied:
(1.) the inter-nuclear diffusion constant $D$;
(2.) the number of available gap genes expressed downstream of maternal inputs (1--5 genes);
(3.) the number of maternal input gradients regulating the gap genes, i.e. systems with all three maternal gradients present vs. systems in which one or two of these are lacking;
(4.) the maximal regulatory strength for maternal and zygotic regulatory interactions.

In the subsections below we describe these four altered optimiziation settings and their results.

\subsection{Varying gap protein diffusion constant}
\label{secAlteredDiffConst}
In our standard optimization ensemble we used a fixed value of the inter-nuclear diffusion constant of $D = 0.5~\sqMicronPerSec$ for gap gene proteins, which is in the  range reported for Bcd \cite{gregor2007stability}. This choice generates mean expression profiles that agree very well with the experimental data.
We later systematically relaxed this choice and carried out additional optimizations in which (1.) other fixed values of $D = 10^{-2}~\sqMicronPerSec ~\cdots~ 10~\sqMicronPerSec$ were used, and (2.) $D$ itself was considered an additional optimization parameter and optimized for maximizing positional information jointly with the regulatory parameters.
In both cases, the additional optimizations were carried out both without and with a bound on the maximal pattern rate-of-change (RoC).

The corresponding results are summarized in Fig.~3E and Fig.~3F of the main text, where in panel E we show how the mean positional information $I$ and the mean pattern RoC vary with the imposed diffusion constant $D$, while panel F shows example expression patterns for two largely different values of $D$. For the cases in which $D$ itself is optimized, the $I$ and RoC values are plotted at the single value of $D$ which is computed as the average optimal $D$ over optimization runs.

In accordance with previous findings \cite{Sokolowski2015}, the positional information of optimal expression patterns decreases markedly for high diffusion constants, $D\gtrsim 1~\sqMicronPerSec$. In this regime, though spatial averaging is very efficient in removing non-Poissonian  noise components, the gap gene expression patterns become increasingly spread-out and spatially flatter, such that a smaller number of slopes (or combinations thereof) can be accommodated along the embryo axis. This markedly reduces the information content of the ``positional code''; in short, the ability to create sufficiently (but not infinitely!) steep boundaries is essential for good patterning, and high $D$ values prevent the such boundaries from being set up. An example of a suboptimal pattern at large diffusion constant values is  shown in the lower part of Fig.~3F of the main text.

Interestingly, in the opposite regime (and without any bound on the pattern RoC), optima with comparably high positional information ($I\gtrsim4~{\rm bits}$) are found even for very low values of $D\sim10^{-2}~\sqMicronPerSec$. This is due to the fact that in this regime, patterns with very wiggled (high spatial derivatives), irregular shapes can be created, which nominally increases positional information. However, this comes at the cost of decreasing pattern stability, resulting in a high rate-of-change (see red curve in lower part of Fig.~3E of the main text).
Accordingly, when the bound on the pattern RoC is also imposed, the optimal solutions at low $D$ do not reach the plateau of high information any more, and an optimal regime emerges at intermediate values of $D \simeq 0.1~\sqMicronPerSec ~\cdots~ 1~\sqMicronPerSec$.
As expected, this optimal regime is found around the average value of optimized $D_{\rm opt}\simeq 0.4~\sqMicronPerSec$, which happens to be very close to our originally chosen fixed value of $D_{\rm fix} = 0.5~\sqMicronPerSec$.

\subsection{Altered number of gap genes and maternal inputs}
\label{secAlteredGeneNoAndGradients}
An intriguing question to ask is why the \Drosophila system features exactly four gap genes expressed as the first layer of zygotic downstream genes activated by maternal inputs. Could evolution have realized a similar patterning precision with a smaller number, e.g. three, gap genes?  Alternatively, could there be any significant benefit of using more than four gap genes?
Similarly one can ask whether all three of the maternal input systems (anterior gradient, posterior gradient, and terminal system) are necessary for generating patterns with high positional information (and if so, why or why not).

To address these questions, we carried out hundreds of optimization runs with altered numbers of downstream (gap) genes and different combinations of maternal input gradients present in the system (``APT'' = all present, ``AP'' = anterior and posterior gradients only, ``AT'' = anterior gradient and terminal system only, ``A'' = anterior gradient only).

In the runs with altered numbers of gap genes, we disallowed the expression of more than the prescribed number of genes, $N_g < 4$, and in the $N_g = 5$ case we introduced one additional hypothetical gap gene species for simulating the expression of five-gene patterns, allowing the new gene to freely couple to the available morphogen gradients and other gap genes.
In order to compare the optimized ensembles at equal resource utilization, the respective bounds on resource utilization were scaled with the number of gap genes, such that the total amount of expressed molecules was the same for all gene numbers, i.e.: for 3 genes, the resource utilization bound was scaled up by factor 4/3, for 5 genes it was scaled down by factor 4/5, and so forth for the other gap gene numbers. In short, we imagine that there is a fixed total (across gap genes) budget of gap gene expression, which can be split across different numbers of gap gene species: with fewer different gap genes, each can be expressed at more positions in space; with more different gap genes, each must be expressed in a more localized setting. In all these runs, the hard biophysical constraints (i.e., the gap gene product degradation rates and maximal transcription and translation rates) were held fixed.

In the optimization runs with reduced number of maternal inputs the concentrations of the respective gradients ``knocked out'' were simply set to zero, and the optimization parameters pertaining to the coupling of downstream genes to these gradients were taken out of the set of optimization parameters.

Early in the analysis of alternative optimal ensembles we realized that the variety of optimal solutions markedly decreases with decreasing number of available downstream genes, and also with decreasing number of maternal gradients. This, in itself, is an interesting finding.
We therefore carried out a pattern similarity analysis by computing the mutual distances between the optimal patterns found in the respective ensemble (using the distance function defined in Eq.~(\ref{eqDistFun}) of Sec.~\ref{secConstrainedOpt}), and then resorting the distance matrix by hierarchical clustering (via standard MATLAB routines).
%(via the standard MATLAB routine for this task).

We summarize the results of the similarity analyses in SI Fig.~\ref{SI-fig:Sim123genes} and SI Fig.~\ref{SI-fig:Sim45genes}. In these figures, we show the sorted distance matrices for the combinations of gene numbers and maternal input gradients (as indicated in the titles above the matrices). The number of expressed downstream genes increases along the rows. The left column panels all feature the full set of maternal inputs (``APT''), whereas the right column panels show optimizations with the anterior input gradient only (``A''). Selected segments of the distance similarity matrices in which mutual distances are low -- which define ``clusters'' of similar patterns -- are highlighted by red boxes in the matrices; the boxes next to the matrices linked to these clusters show several randomly chosen example patterns from the respective clusters.

We observe the lowest variety of optimal patterns in the case with one downstream gene only (SI Fig.~\ref{SI-fig:Sim123genes}A and B), although that variety is not limited to one single optimal pattern when  all three maternal inputs are present (``1 GENE / APT'', SI Fig.~\ref{SI-fig:Sim123genes}A). We find two optimal shapes, symmetric along the anterior-posterior axis, but, surprisingly, also one solution which prefers to create one bump shallowly increasing towards the posterior and then sharply decreasing again. This can be attributed to the fact that in our system the posterior gradient reaches slightly lower maximal levels due to the ``cut-off'' described in Sec.~\ref{secFixedPars}. 
In contrast, the situation with one gap gene and the anterior input gradient only (``1 GENE / A'', SI Fig.~\ref{SI-fig:Sim123genes}B) is  simpler: here we identify one single cluster of stereotypical solutions, in accordance with our previous theoretical findings \cite{Tkacik2009,Sokolowski2015}.
Notice that while the optimal patterns with one downstream gene nominally are permitted to be fully expressed at 80\% of the positions (to match the resource utilization of the WT), these patterns in fact prefer to use less of the available resources in order to create more varying pattern shapes. 

In sum, stochastic optimization runs for one gap gene converge on a single solution (for one morphogen input) or a pair of qualitatively different solutions (for three morphogens inputs), showing that optimizing positional information in this setup strongly constraints the space of optimal expression patterns.  

Solutions optimized with two gap genes show the increase in the variety of optimal solutions, but we can still clearly identify structural similarities (SI Fig.~\ref{SI-fig:Sim123genes}C and D). Optimal solutions for two gap genes responding to a single maternal morphogen form a small number of discrete clusters (SI Fig.~\ref{SI-fig:Sim123genes}D, ``2 GENES / A''), which are furthermore qualitatively consistent with our previous work~\cite{Tkacik2009,Walczak2010}; the stochastic optimization repeatedly finds multiple solutions belonging to these stereotyped clusters. Amongst all these solutions, the available resources are predominantly assigned to the anterior half of the embryo.

%----

For optimal solutions with two gap genes responding to three maternal morphogens, in addition to almost symmetric solutions in which the two genes are expressed in opposite halves of the embryo, we also found asymmetric solutions in which the resources are assigned predominantly either to the anterior or posterior of the embryo (see example patterns shown in SI Fig.~\ref{SI-fig:Sim123genes}C). Notice that in these solutions one of the two gap genes follows a profile that constitutes an optimal solution for the optimizations with one gene (SI Fig.~\ref{SI-fig:Sim123genes}C), while the other gap gene profile is ``added on top.''

The observed trend of increasing solution diversity continues as the number of available gap genes is increased beyond two. With three gap genes and all three maternal morphogens, there are more clusters, each containing more degenerate solutions, and the optimal patterns within each cluster are also less similar to each other (SI Fig.~\ref{SI-fig:Sim123genes}E). Patterns remain more stereotypic and similar to each other, though, within the clusters in the optimal ensemble with 3 gap genes and anterior maternal morphogen only (SI Fig.~\ref{SI-fig:Sim123genes}F).

With 4 and 5 gap genes and all three maternal gradients we still can sometimes find clusters of structurally similar solutions, but as a rule the optimal solutions now are dissimilar (SI Fig.~\ref{SI-fig:Sim45genes}A and C). Once again the clusters of similar solutions are larger with one maternal gradient only, and the solutions within these clusters also are more similar to each other (SI Fig.~\ref{SI-fig:Sim45genes}B and D). Interestingly, some of these solutions do not make use of all available genes but prefer to assign all available resources to a smaller set of genes (i.e., the pattern consists only of nonzero expression for 3 out of 4, or 4 out of 5, available genes).

Overall, the similarity between the patterns decreases less when going from 3 to 4 genes as compared to going from 2 to 3 genes.
This is best seen in the comparison of the pattern distance histograms across the ensembles with different gene numbers, SI Fig.~\ref{SI-fig:Sim45genes}E and F (upper panels). Here the ensembles with 1 or 2 gap genes extend out towards very low pattern distances, while starting from 3 genes a characteristic distance distribution forms at larger pattern distances.
The data for 5 genes was omitted from these plots because it overlaps almost perfectly with the data for 4 genes.
Below the distance histograms we also show the corresponding histograms of positional information encoded by the optimized patterns.
Here the largest jumps occur between 1 and 2, and between 2 and 3 gap genes.
3-gap-gene patterns can reach surprisingly high values of positional information (maximal value $I_3^{\rm max} = 4.56~{\rm bits}$),
but 4-gap-gene patterns still reach about 0.2~bits more ($I_4^{\rm max} = 4.77~{\rm bits}$); the increase between 4- and 5-gene patterns is smaller ($I_5^{\rm max} = 4.84~{\rm bits}$).
Note that while the nominal increases of these maximal $I$-values may seem minuscule, due to the logarithmic nature of the information-theoretic quantities, these information values correspond to an increase from 23 to 27 (and 28, respectively) perfectly distinguishable states (positions).

\subsection{Bounding maximal regulatory strength}
\label{secBoundedHmax}
We also asked about how the maximal allowed strength of regulatory interactions affects the nature of optimal solutions. By default, we set the bounds on the the maximal regulatory interaction strength to be very generous, $H_{\rm max} = 50$. As we found in our earlier work for a single output gene under the control of a single maternal input gradient (but using the same type of regulatory function, see Sec.~\ref{secRegFun}), this maximal value is high enough to produce a sharp, step-like transition of the activation function from full to zero expression within two successive nuclei along the anterior-posterior axis \cite{Sokolowski2015}.

In order to assess the influence of this generous choice, we also generated optimization ensembles with lower values of $H_{\rm max} \in [3, 4, 5, 10]$.
Moreover, we extended this analysis to two additional hypothetical restrictions on the possible regulatory interactions: (1.) regulation with different $H_{\rm max}$ and anterior gradient only (``A only''); (2.) ``feed-forward regulation only'' (``FF only'') at different $H_{\rm max}$, meaning that zygotic regulatory interactions, i.e. mutual regulation and self-regulation, are not possible at all, while maternal regulation is constrained by lowering $H_{\rm max}$.
For the highest value $H_{\rm max} = 50$, we also generated the ``FF only'' optimal ensembles with two (anterior + posterior = ``AP'', and anterior + terminal = ``AT'') gradients.

Fig.~4B and C of the main text summarize how these extra constraints impact the positional information encoded by optimized patterns.
While in all three hypothetical settings the positional information decreases with decreasing maximal strength of regulatory interactions $H_{\rm max}$, mutual and self-regulation can partly compensate for this decrease compared to the ``FF only'' case. This is particularly pertinent at very low $H_{\rm max}=3$. Qualitatively, the same conclusion holds with one morphogen input gradient only (``A only''): so long as $H_{\rm max}$ if very high, the positional information difference between optimal patterns generated only by a single anterior morphogen gradient and the full set of three morphogen gradients is smaller than at lower regulatory strengths. Taken together, at very high $H_{\rm max}$, solutions under different scenarios appear more degenerate, but as the constraint on $H_{\rm max}$ kicks in, the importance of having gap gene self- and cross-regulation, as well as responding to the complete set of three (vs only a single) morphogen becomes evident.

In the ``FF only'' optimal ensembles, the diversity of optimal patterns is strongly reduced compared to the optimal ensembles with full regulation; this diversity decreases further when three maternal morphogen inputs are replaced by a single anterior morphogen input. This is summarized in SI Fig.~\ref{SI-fig:SimFFonly}, which shows the pattern distance matrices sorted by hierarchical clustering, along with several chosen example patterns from identified clusters (following the same logic as described in Sec.~\ref{secAlteredGeneNoAndGradients}).
With all three maternal inputs (``APT'') we observe that one class of optimal patterns assigns resources predominantly to the opposing ends of the embryo, while ``avoiding'' the central region. With less than three maternal inputs, the patterns become increasingly asymmetric and devote more resources to the anterior. These solutions highlight that without the ability of gap genes to cross-regulate, optimization drives the patterns towards locations where maternal input concentrations are high, as to minimize the impacts of input noise.

\clearpage
\def\url#1{}
\bibliographystyle{sn-nature}
\bibliography{si_nourl}

%%%%%%%%%%%%%%%
%%% FIGURES %%%
%%%%%%%%%%%%%%%

\newcommand{\SimFigWidth}{0.85\linewidth}

%%%%%%%%%%%%%%%%%%%
%%% SI Figure 1 %%%
%%%%%%%%%%%%%%%%%%%
\begin{figure*}[ht!]
\centering
\includegraphics[width=\textwidth]{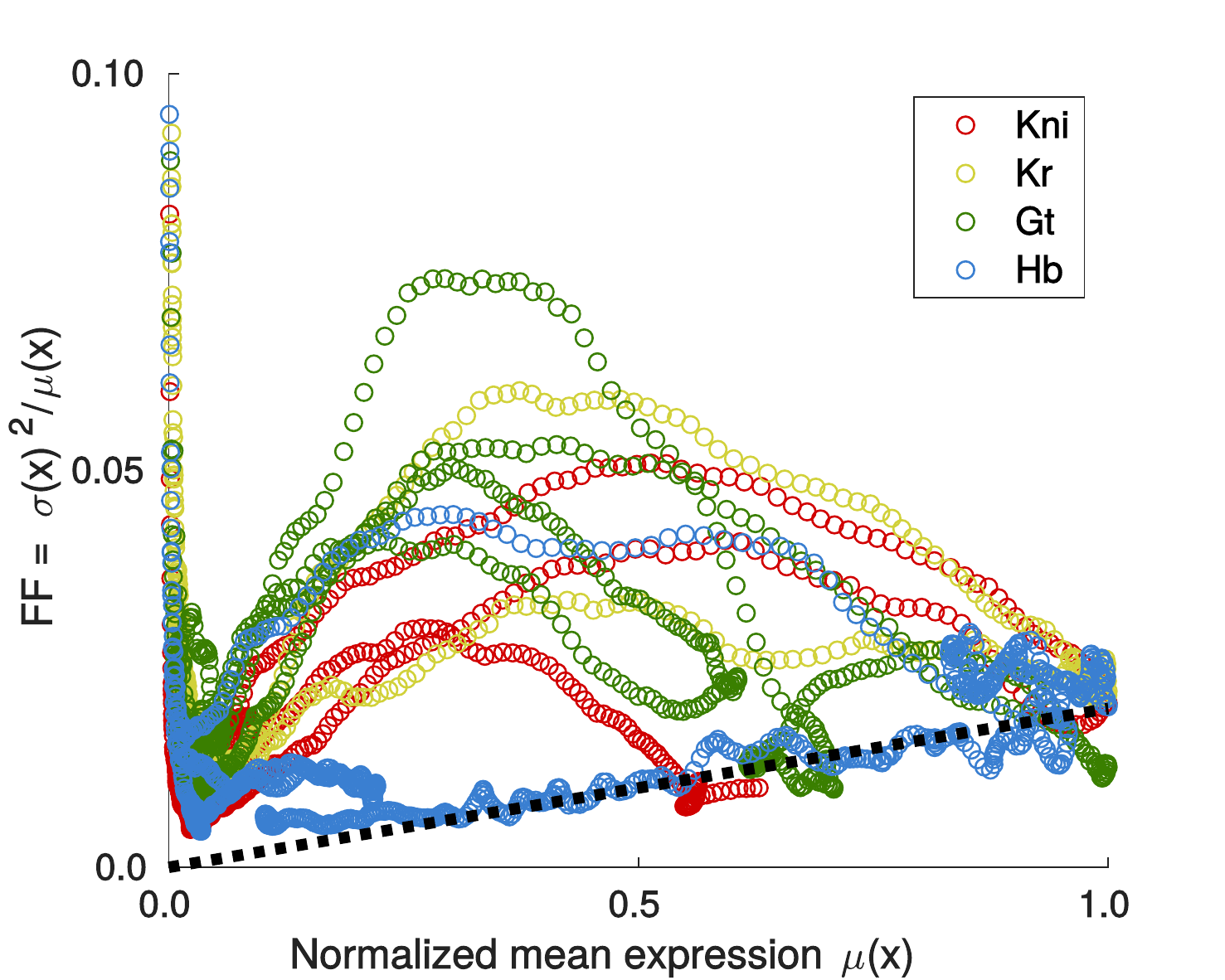}
\caption[Estimating extrinsic noise from data.]
{
{\bf Estimating extrinsic noise from data.}
Here we plot the Fano Factor (FF), i.e. the local {\it variance} of the normalized expression levels along the anterior-posterior ($x$) axis of the embryo divided by the local {\it mean} expression levels, $\sigma^2(x)/\mu(x)$, against these mean levels $\mu(x)$, for the gap gene expression levels recorded by Petkova et al. \cite{Petkova2019}.
In addition to the expected variance increase at intermediate expression due to noise propagation, the plot reveals a variance component which appears to scale quadratically with the mean, $FF(x) = \sigma^2(x) / \mu(x) \sim \mu(x)$, i.e. $\sigma^2(x) \sim \mu^2(x)$; note that this relationship is independent of the normalization factors (actual protein copy numbers) of the genes, as both $\sigma^2(x)$ and $\mu^2(x)$ are normalized by the same values.
We attribute this variance component seen in the experimental data to extrinsic noise (between-embryo variations affecting all expression levels simultaneously), which cannot be explicitly predicted by our model. We therefore added extrinsic noise as an additional phenomenological contribution to our other computed noise sources. Mathematically, this was implemented   by adding a quadratic term $m (\bar{g}^\alpha_i)^2$, where $m$ is a constant, to the local variances $\sigma^2_{i,\alpha}$ predicted by our model (see Sections \ref{secCovsWmRNA} and \ref{secCovMatrixNumerics}), where $\bar{g}^\alpha_i$ is the predicted local mean expression level of gene $\alpha$ (see Sec.~\ref{secMeansIntegration}). The constant $m=0.02$ is estimated from data as shown in the figure, where it represents the linear ``noise floor'' approximately shared by all gap genes.
}
\label{SI-fig:QuadNoiseTerm}
\end{figure*}

%%%%%%%%%%%%%%%%%%%
%%% SI Figure 2 %%%
%%%%%%%%%%%%%%%%%%%
\begin{figure*}[ht!]
\centering
\includegraphics[width=0.8\textwidth]{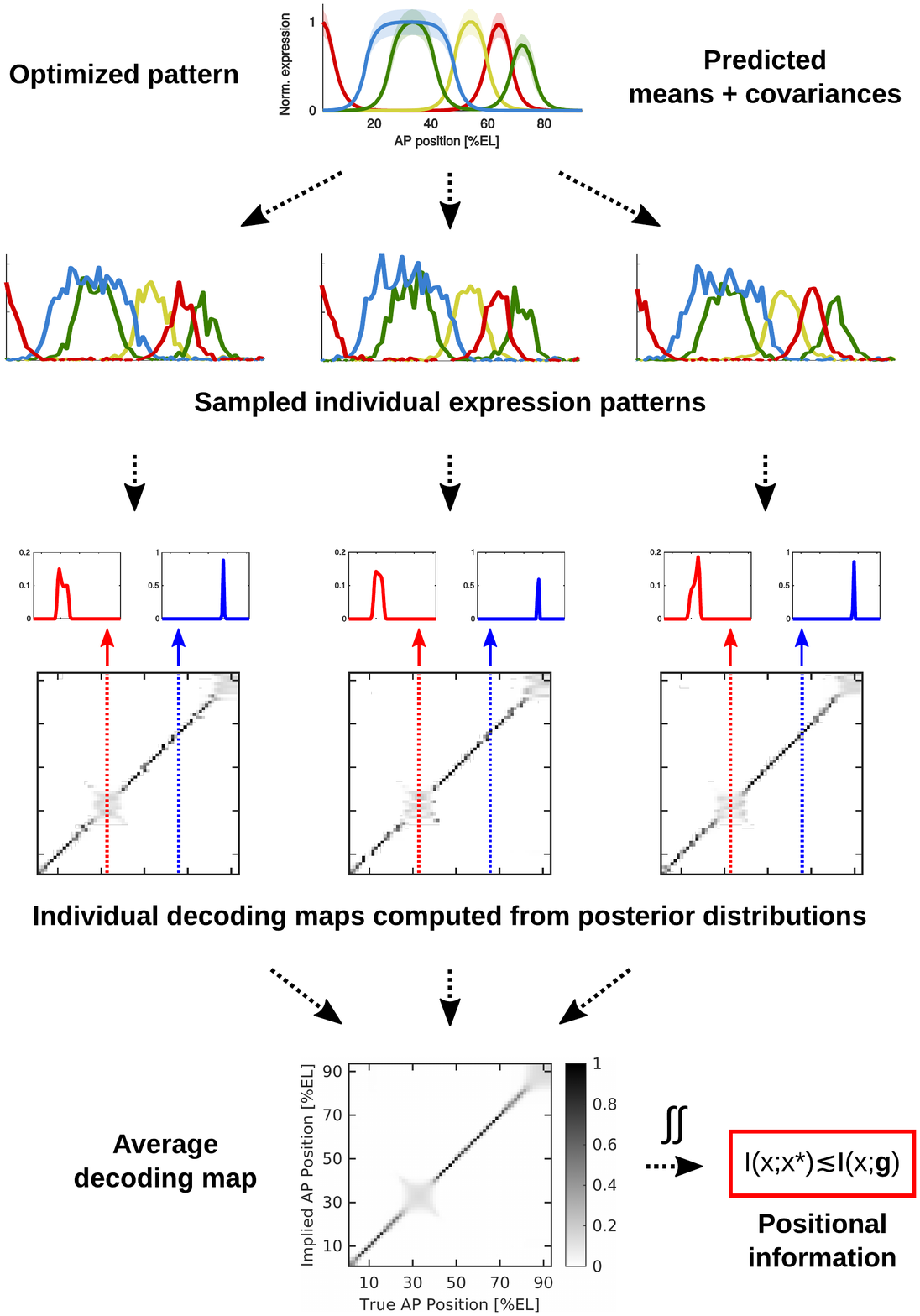}
\caption[Approximate computation of positional information via optimal decoding.]
{
{\bf Approximate computation of positional information via optimal decoding.} See Sec.~\ref{secPosInfo} for details.
A schematic of the steps used to estimate positional information $I(\myvec{g};x)$ based on the mean expression profiles and the system-wide covariance matrix computed from the model as described in Sec.~\ref{sec:numerical}.
First, we draw an ensemble of $\Ne$ expression profiles $\myvec{g}_n(x_i)$, $n=1,\dots,\Ne$, where each $n$ indexes an expression profile in an individual simulated embryo at the readout time $T$, in accordance with the predicted covariance levels and mean profile shapes.
From each individual embryo we then construct the posterior distributions $P(x^*_j|\myvec{g}_n(x_i))$ of {\it implied} (decoded) positions $x^*_j$ given the noisy local expression levels $\myvec{g}_n(x_i)$  using \Bayes' rule, and assemble the individual decoding maps $P_n(x^*_j|x_i)$ from these posteriors.
Finally, individual decoding maps are averaged to obtain the average decoding map, $\langle P \rangle(x^*_j|x_i) \equiv \langle P_n(x^*_j|x_i) \rangle_{n}$, which subsequently is integrated using the standard definition of mutual information, Eq.~(\ref{eq:mutinf_sum}), yielding the (decoding) lower bound on mutual information~\cite{Petkova2019}.
}
\label{SI-fig:InfoComp}
\end{figure*}

%%%%%%%%%%%%%%%%%%%
%%% SI Figure 3 %%%
%%%%%%%%%%%%%%%%%%%
\begin{figure*}[ht!]
\centering
\includegraphics[width=0.85\textwidth]{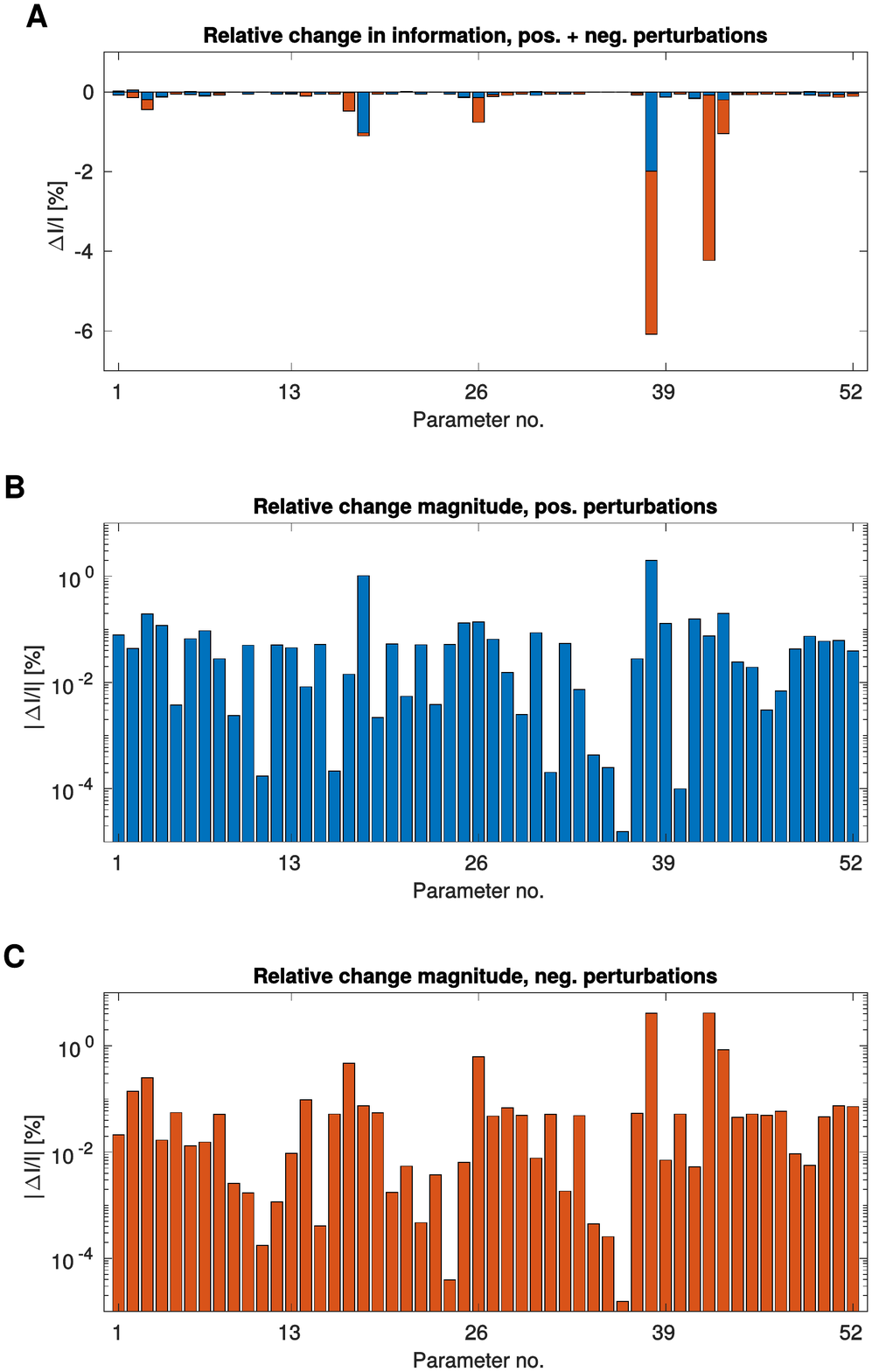}
\caption[Optimality check for the Drosophila-like solution.]
{
{\bf Optimality check for the Drosophila-like solution.} Here we analyze the optimal solution shown in Fig.~2 of the main paper by perturbing its parameters.
{\bf (A)} Bars show the relative change in positional information, $\Delta I / I$, in percent, upon perturbing individual parameters (ordered and indexed by 1--52 along the x-axis), for a positive (blue) and negative (red) perturbation by $0.1\%$ of the considered parameter range (see Sec.~\ref{secOptPars}), respectively. Changes of the same sign are stacked along the corresponding axes. The vast majority of perturbations results in reduction of $I$ by up to $6\%$ of the unperturbed value. A few perturbations slightly increase $I$ by less than $0.04\%$ of the unperturbed value, which is much lower than the typical stochastic fluctuations encountered when computing $I$ (described in detail in Sec.~\ref{secPosInfo}).
Panels {\bf (B)} and {\bf (C)} show the {\it absolute} values (magnitudes) of the bars in (A) in matching colors on a logarithmic scale.
}
\label{SI-fig:Optimality}
\end{figure*}

%%%%%%%%%%%%%%%%%%%
%%% SI Figure 4 %%%
%%%%%%%%%%%%%%%%%%%
\begin{figure*}[ht!]
\centering
\includegraphics[width=0.9\textwidth]{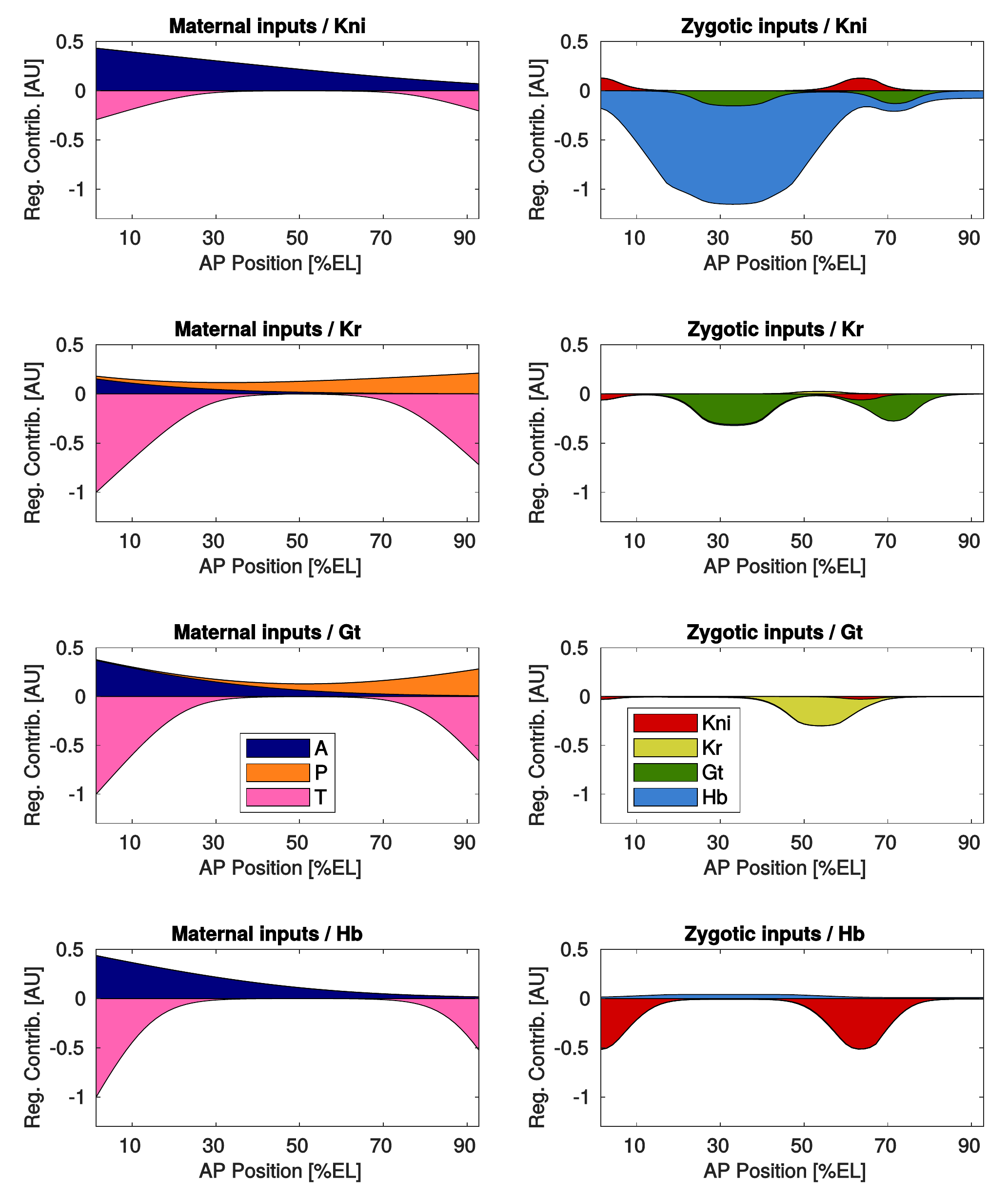}
\caption[Detailed map of regulatory contributions for the Drosophila-like solution.]
{
{\bf Detailed map of regulatory contributions for the Drosophila-like solution.} See Sec.~\ref{sec:interactions} for details.
The panels show the maternal (left column) and zygotic (right column) regulatory contributions to the four gap genes (indicated in the plot titles, $\alpha=$ Kni, Kr, Gt, Hb from top to bottom), as a function of the anterior-posterior (AP) position, for the regulatory network of the optimized solution shown in Fig.~2 of the main text.
Regulatory contributions were computed by inserting the local mean concentration values of the regulating species (i.e., morphogens and gap genes) into the respective terms summed in the exponent of the MWC regulatory function for gap gene $\alpha$, i.e.
$H_M^{\alpha\kappa} \log(1 + c_i^\kappa/K_M^{\alpha\kappa})$ for maternal inputs $c^\kappa$ ($\kappa=$ A, P, T, see color legend), and
$H_G^{\alpha\zeta} \log(1 + g_i^\zeta/K_G^{\alpha\zeta})$ for zygotic inputs $g^\zeta$ ($\zeta=$ Kni, Kr, Gt, Hb, see color legend),
where $i$ denotes the nuclear position along the AP axis.
Color legends shown in the third row apply to all panels. Regulatory contributions have been normalized to a maximum of 1 (across maternal and zygotic inputs), for each regulated gene separately. Positive contributions are activating; negative contributions are repressive.
}
\label{SI-fig:RegContrib}
\end{figure*}

%%%%%%%%%%%%%%%%%%%
%%% SI Figure 5 %%%
%%%%%%%%%%%%%%%%%%%
\begin{figure*}[ht!]
\centering
\includegraphics[width=\textwidth]{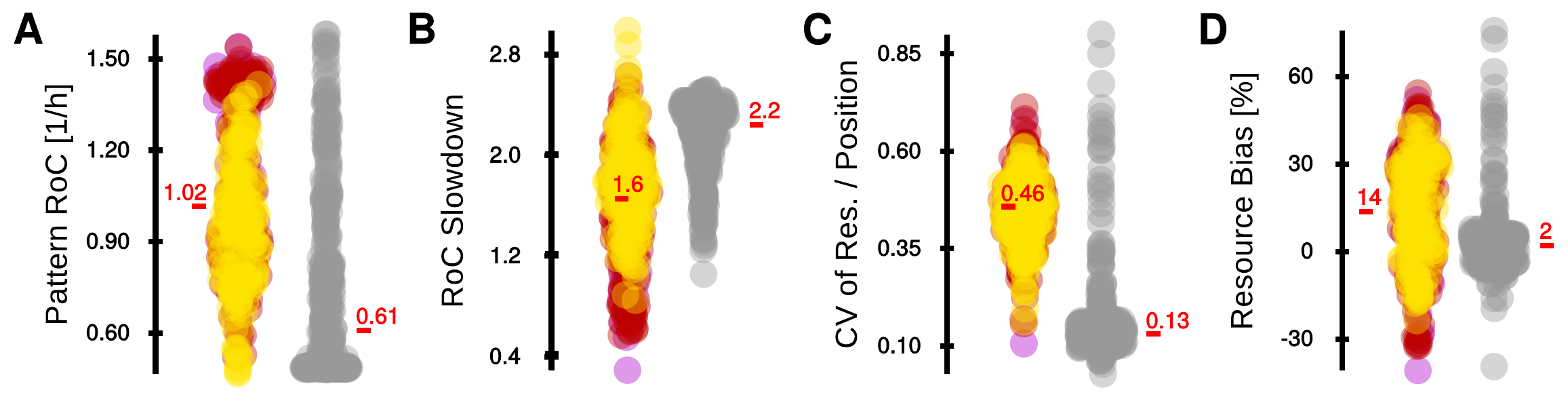}
\caption[Additional patterning phenotypes in the optimal and random ensembles.]
{
{\bf Additional patterning phenotypes in the optimal and random ensembles.} See Sec.~\ref{secAddAnalysis} for details.
Each dot represents either a solution from the optimal ensemble (color, solutions from “WT RU” in Fig.~3C of the main text, N = 324) or a draw from the random ensemble (gray, including only solutions $> 0.5$ bit that are at or below WT resource utilization, delineated by dashed yellow lines in Fig.~3C of the main text, N = 319). 
Violet, red, and yellow indicate the lowest, middle, highest third of the information interval.
{\bf (A)} Higher positional information in the optimal ensemble correlates with lower pattern rate-of-change (RoC) phenotype (cf. Sec.~\ref{sec:stability}). Points from the optimal ensemble are replotted from Fig.~3B of the main text.
Solutions from the random ensemble spread out almost uniformly along the range of observed RoC values, but also display an enrichment of solutions at very low RoC values, which strongly biases the median towards low RoC as well. Random patterns tend to have a smaller RoC because they typically lack cross- and self-regulation.
{\bf (B)} While solutions optimized for positional information can display both an acceleration (speedup, values $<1$) or deceleration (slowdown, values $>1$) of the pattern rate-of-change (RoC); the highest-info solutions (yellow) tend to display the strongest pattern slowdown (${\rm PC}\simeq 0.4$, $p<10^{-13}$).
Note that the slowdown of random patterns is more pronounced than for optimal patterns. This is due to the fact that, across all observed ensembles, patterns with a low RoC also tend to have a stronger slowdown, and random patterns on average have a low RoC (see panel A).
However, for random patterns the slowdown is (very weakly if at all) \emph{negatively} correlated with increased positional information (${\rm PC}\simeq -0.12$, $p = 0.026$), in stark contrast to the strong \emph{positive} correlation for the optimized ensemble. RoC slowdown is defined as the ratio between the RoC at $T_{\rm meas}-30~{\rm min}$ and the RoC at $T_{\rm meas}$.
{\bf (C)}
Optimal solutions display a higher degree of variability (coefficient of variation, CV) in resources per position (cf. Sec.~\ref{secResourceAlloc}) along the AP axis than random solutions, highlighting the fact that optimization promotes a rich code with different numbers, combinations, and levels of locally expressed genes. In comparison, patterns in the random ensemble often do not even spatially modulate some of the gap genes.
{\bf (D)} Optimal solutions allocate more resources to the anterior of the system. The panel shows the ``resource bias'' in percent of the total available resources, where the value of $100\%$ means that all resources are allocated to the anterior half of the system (cf. Sec.~\ref{secResourceAlloc}). While pattern optimization can lead to over-allocation both in the anterior or posterior, high information correlates with anterior resource bias (median = $14~\%$). Note that this value is close to the resource bias of the experimentally measured pattern, which (under the same measure) amounts to $12~\%$. In contrast, patterns from the random ensemble cluster around zero resource bias.
}
\label{SI-fig:AddProp}
\end{figure*}

%%%%%%%%%%%%%%%%%%%
%%% SI Figure 6 %%%
%%%%%%%%%%%%%%%%%%%
\begin{figure*}[ht!]
\centering
\includegraphics[width=\textwidth]{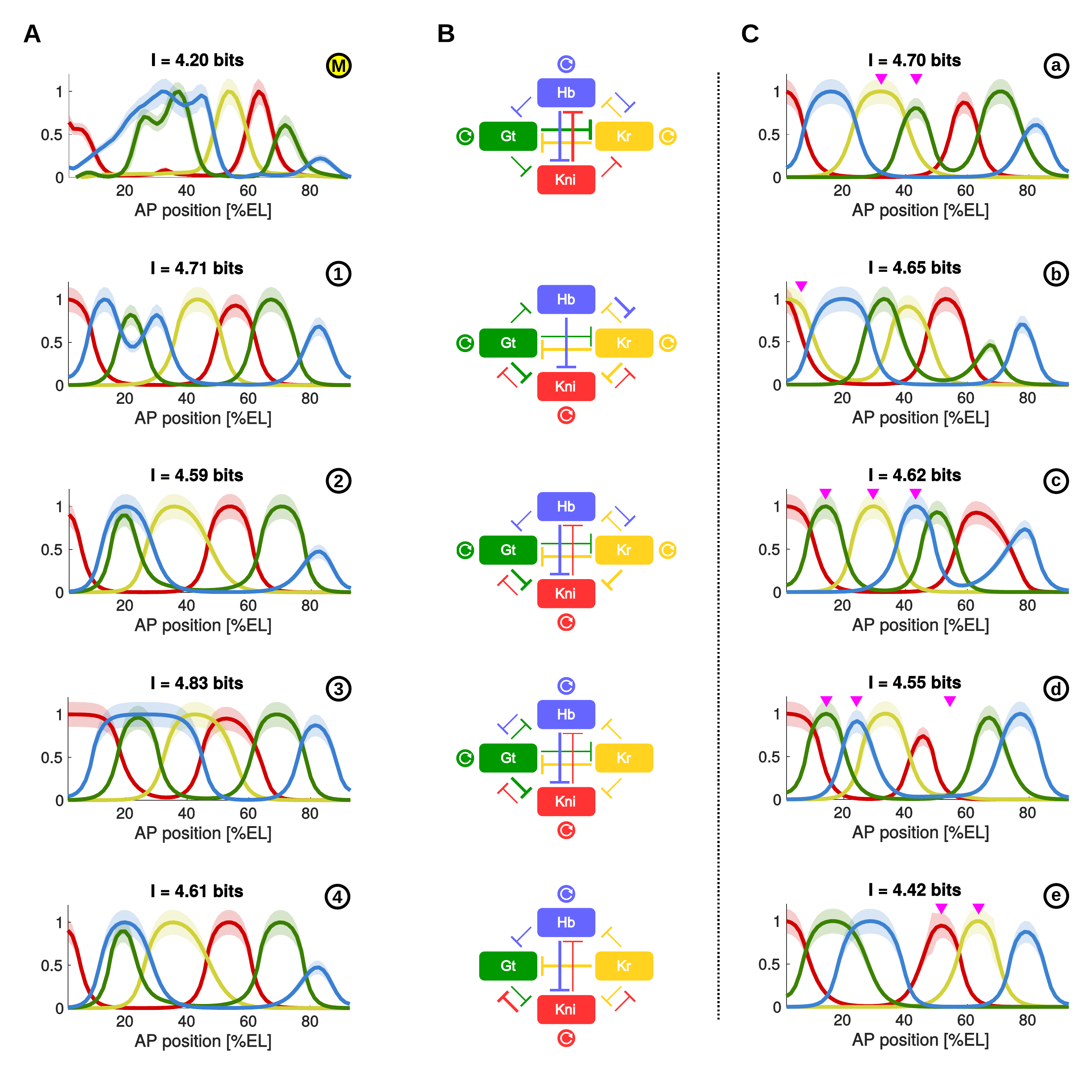}
\caption[\emph{Ab initio} optimized solutions similar to the Drosophila WT expression patterns.]
{
{\bf \emph{Ab initio} optimized solutions similar to the Drosophila WT expression patterns.} Selected optimal solutions shown here (except the top row) are obtained purely \emph{ab initio}, by maximizing positional information with zero  ``pulling force'' towards the \Drosophila WT pattern, i.e., $\lambda=0$ in Eq.~(\ref{utility}).
{\bf (A)} Mean expression profiles with standard deviations corresponding to solutions (1), (2), (3), (4) enumerated in Fig.~3D of the main text. The measured solution is shown in the first row for comparison and is denoted by the yellow ``M'' symbol.
{\bf (B)} Regulatory network cartoons summarizing significant regulatory interactions among the gap genes forming the patterns shown in (A) (cf. Sec~\ref{sec:interactions}). Strong interactions are shown with bolder symbols than weak interactions. The circles next to the genes show significant self-activation. The topmost network shows the literature-based reconstruction, identical to the Fig.~2D (right) of the main text~\cite{Jaeger2010}.
{\bf (C)} Examples of optimal solutions with expression domains swapped or additionally present as compared to the experimentally observed order of expression domains in \Drosophila. From top to bottom we can identify the following cases: (a) Kr and anterior Gt domains swapped; (b) additional anterior Kr stripe; (c) Kr and anterior Hb domains swapped; (d) most anterior domains squeezed into the anterior half; (e) Kr and posterior Kni domains swapped.
}
\label{SI-fig:OptNetworksProfiles}
\end{figure*}

%%%%%%%%%%%%%%%%%%%
%%% SI Figure 7 %%%
%%%%%%%%%%%%%%%%%%%
\begin{figure*}[ht!]
\centering
\includegraphics[width=\textwidth]{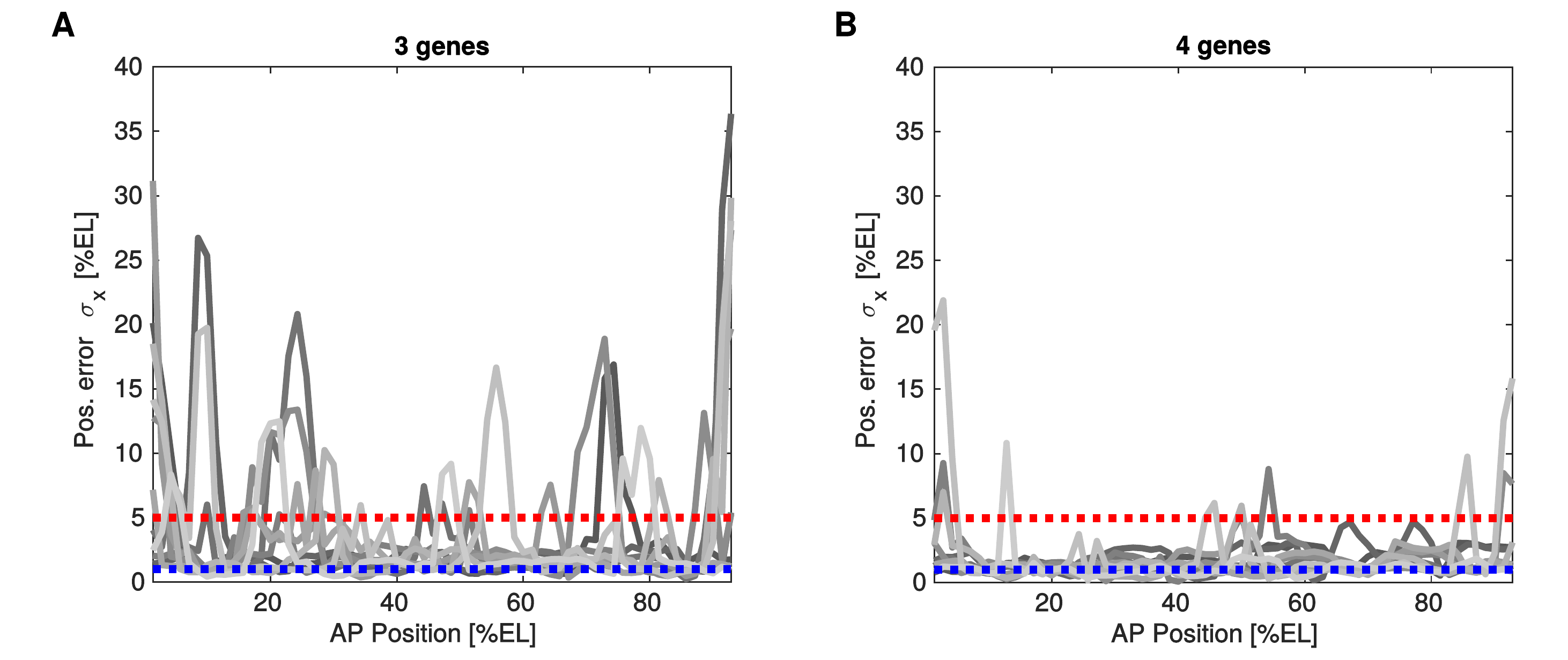}
\caption[Optimal solutions with 3 gap genes have localized positional defects.]
{
{\bf Optimal solutions with 3 gap genes have localized positional defects.}
Panels {\bf (A)} and {\bf (B)} show positional error ($\sigma_x$; SD of the posterior distribution $P_n(x*|x)$) profiles along the anterior-posterior (AP) embryo axis, for N = 10 example optimal solutions obtained in optimizations with 3 gap genes {\bf (A)} and with 4 gap genes {\bf (B)}, respectively, in units of embryo length percentage ($\%{\rm EL}$). The solutions were chosen uniformly from the optimal ensembles sorted by increasing positional information $I$, with lighter grey shades representing higher-info solutions. The blue and red dashed lines mark positional error levels of $1~\%{\rm EL}$ and $5~\%{\rm EL}$, respectively, for reference.
While median and maximal $I$ values differ, respectively, by $0.06~{\rm bits}$ and $0.22~{\rm bits}$ between the 4- and 3-gene ensembles (cf. Fig.~4A of the main text), which is a significant but fractionally small difference, individual 3-gene solutions have markedly more positions in the embryo where the local positional error exceeds $5~\%{\rm EL}$, which we call ``positional defects'': at those positions, placement of any downstream positional marker would be very inaccurate. In contrast,  4-gene solutions feature much more uniform positional error profiles with defects limited in number, magnitude, and preponderance typically at the extremes of the patterning axes. See Ref~\cite{Dubuis2013_PNAS} for the theoretical claim that optimal patterning implies not only a low average positional error, but \emph{uniformly low} positional error, which optimal 4-gap-gene patterns can, but optimal 3-gap-gene patterns cannot, achieve reproducibly.
}
\label{SI-fig:PosErrProfiles}
\end{figure*}

%%%%%%%%%%%%%%%%%%%
%%% SI Figure 8 %%%
%%%%%%%%%%%%%%%%%%%
\begin{figure*}[ht!]
\centering
\includegraphics[width=\SimFigWidth]{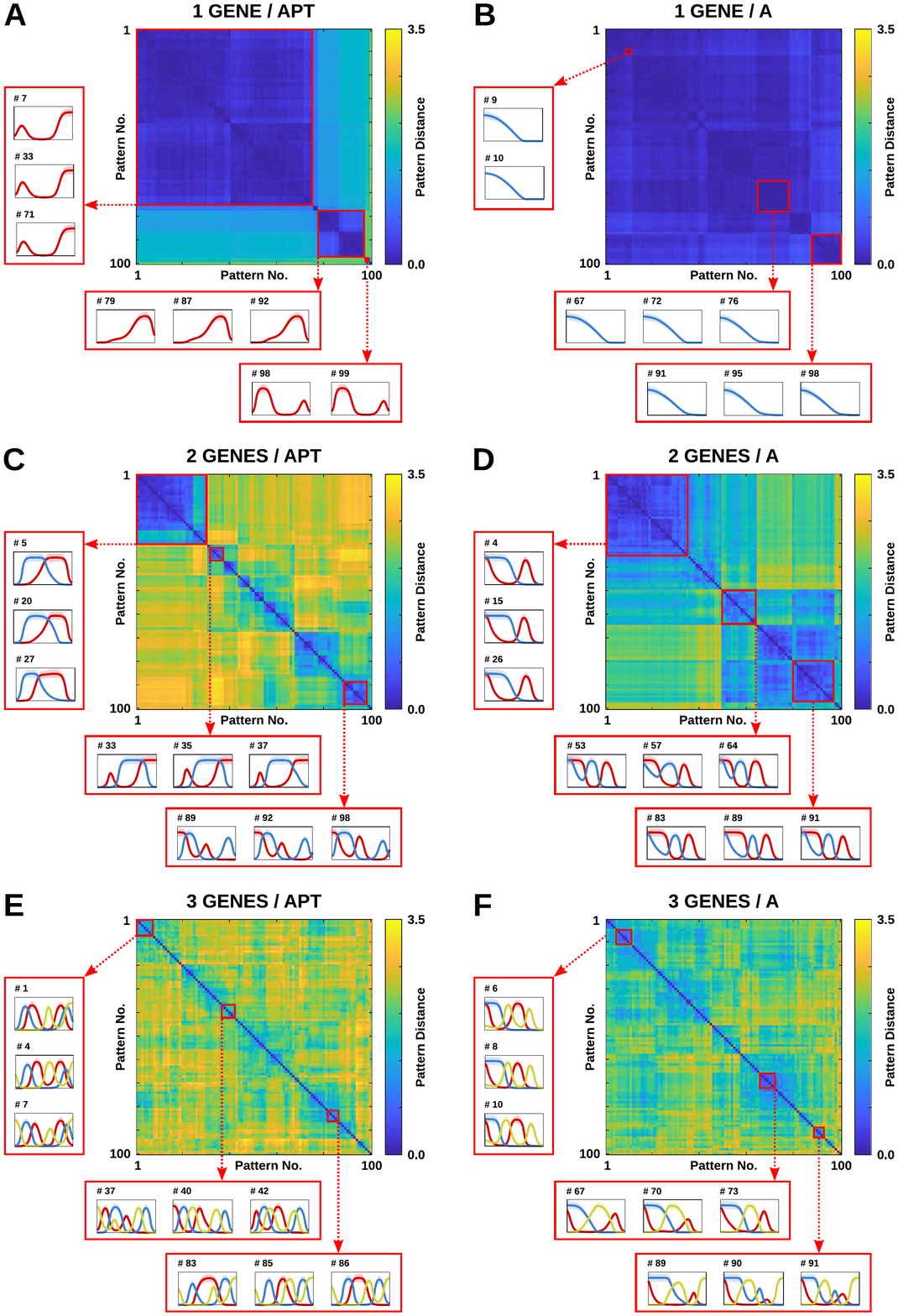}
\caption[Altered number of gap genes and maternal morphogen gradients.]
{
{\bf Altered number of gap genes and maternal morphogen gradients.} For details see Sec.~\ref{secMutatedOpt} and Fig.~4A,B of the main text.
Each panel shows the resorted (hierarchically clustered) distance matrix between 100 optimal solutions optimized in systems with one {\bf (A, B)}, two {\bf (C, D)}, or three {\bf (E, F)} zygotically expressed gap genes, activated either by all three {\bf (A, C, E)}, or only by the anterior maternal gradient {\bf (B, D, F)}. Bluish hues indicate similar patterns (small distance), yellowish hues indicate different patterns (large distance); distance is defined in Eq.~(\ref{eqDistFun}). The number of gap genes and maternal morphogen combination is indicated in the each plot title. 
Red boxes display randomly chosen example patterns from the respective clusters, as highlighted in the distance matrix.
}
\label{SI-fig:Sim123genes}
\end{figure*}

%%%%%%%%%%%%%%%%%%%
%%% SI Figure 9 %%%
%%%%%%%%%%%%%%%%%%%
\begin{figure*}[ht!]
\centering
\includegraphics[width=\SimFigWidth]{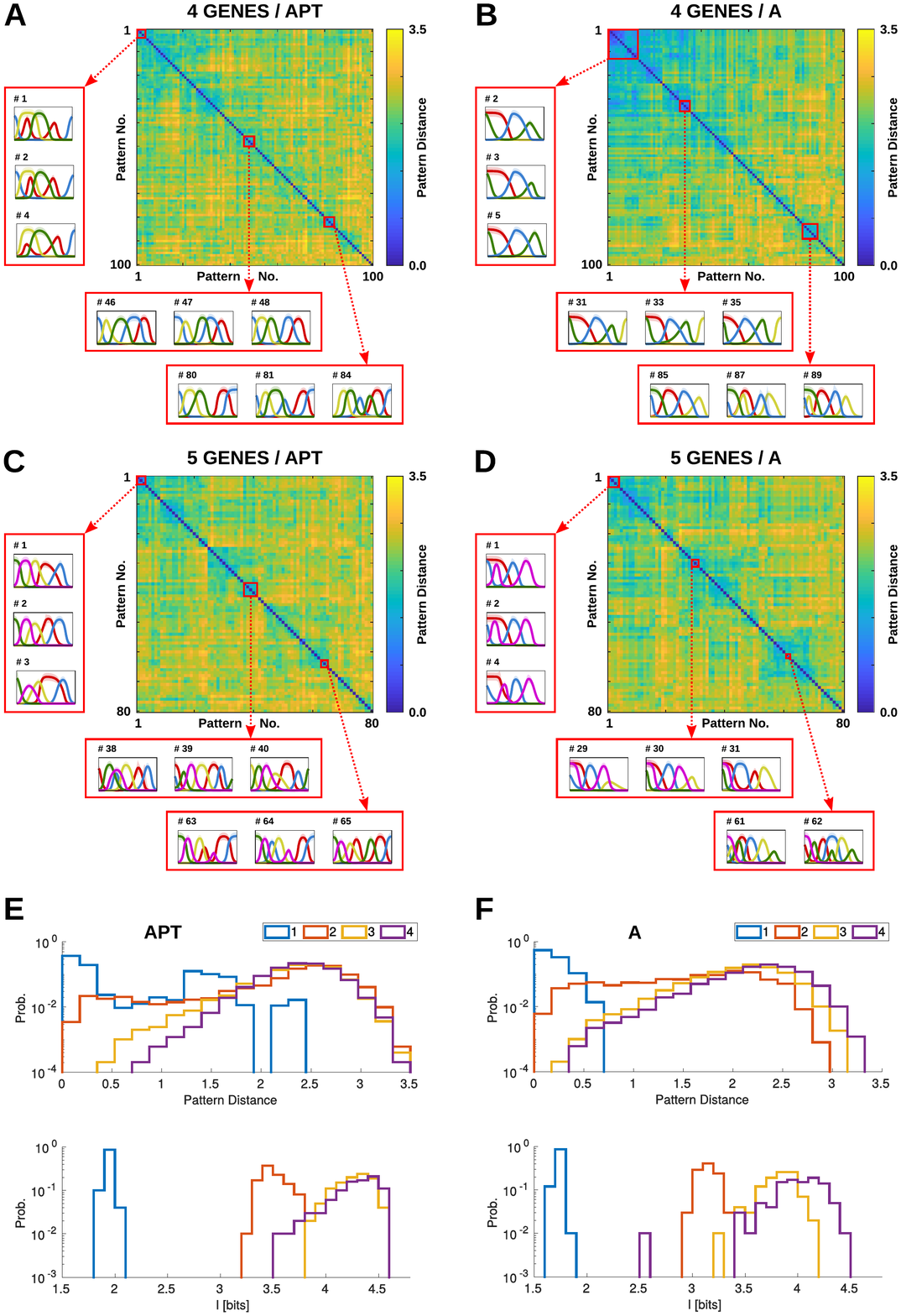}
\caption[Altered number of gap genes and maternal morphogen gradients, cont'd.]
{
{\bf Altered number of gap genes and maternal morphogen gradients, cont'd.}
For details see Sec.~\ref{secMutatedOpt} and Fig.~4A,B of the main text.
Panels {\bf (A--D)} show the resorted (hierarchically clustered) distance matrix between 100 optimal solutions optimized in systems with four {\bf (A, B)} and five {\bf (C, D)} zygotically expressed gap genes activated either by all three {\bf (A, C)} or only the anterior maternal gradient {\bf (B, D)}.
Bluish hues indicate similar patterns (small distance), yellowish hues indicate different patterns (large distance); distance is defined in Eq.~(\ref{eqDistFun}).
The number of gap genes and maternal morphogen combination is indicated in the each plot title. 
Red boxes display randomly chosen example patterns from the respective clusters, as highlighted in the distance matrix.
Panels {\bf (E)} and {\bf (F)} show a comparison of the distance (upper plots) and positional information (lower plots) histograms, computed for the optimal ensembles with 1--4 gap genes driven by all three {\bf (E)}, or only the anterior {\bf (F)}, maternal morphogen gradient(s), respectively.
}
\label{SI-fig:Sim45genes}
\end{figure*}

%%%%%%%%%%%%%%%%%%%
%%% SI Figure 10 %%%
%%%%%%%%%%%%%%%%%%%
\begin{figure*}[ht!]
\centering
\includegraphics[width=\SimFigWidth]{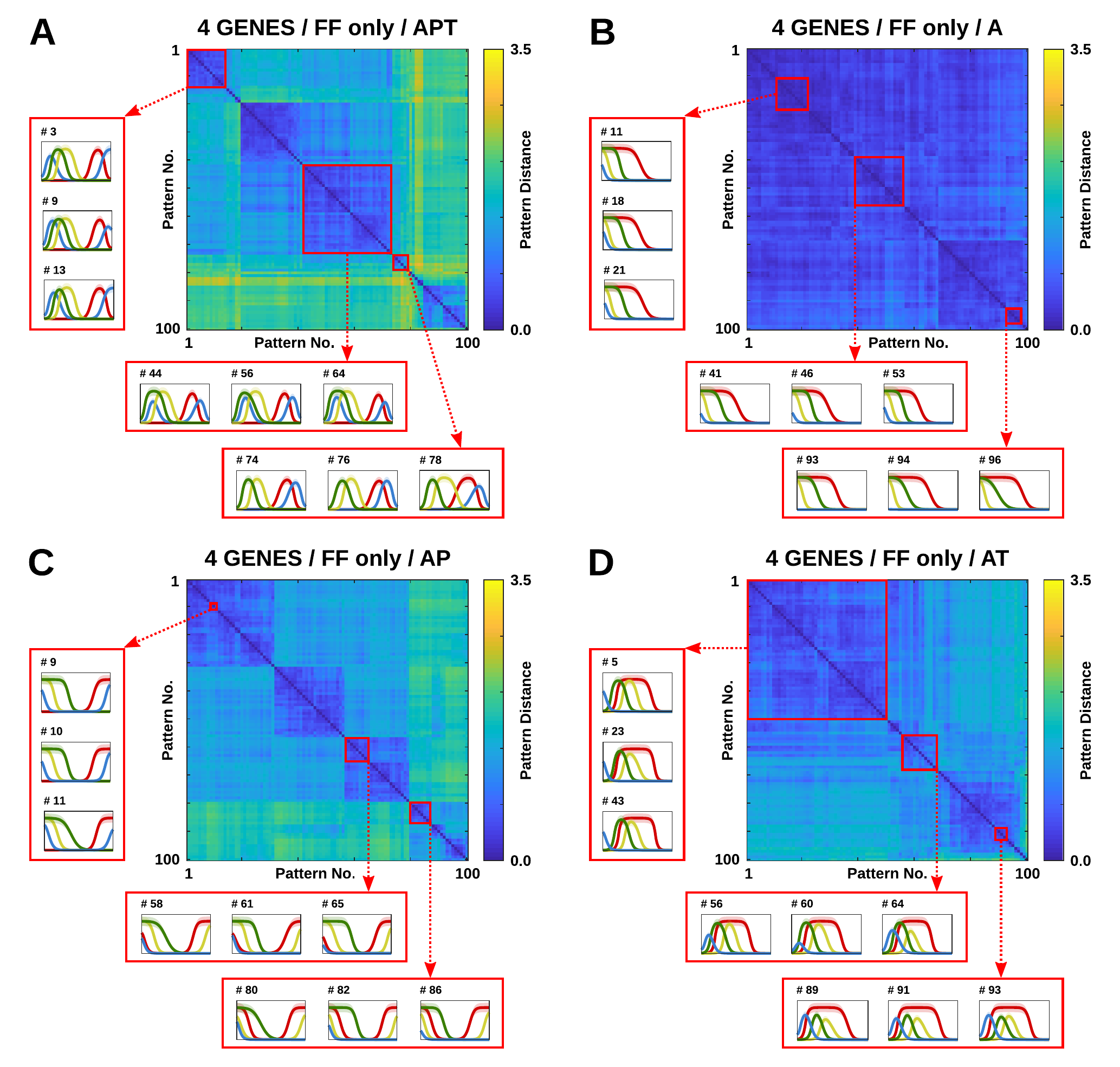}
\caption[Feed-forward regulation by maternal morphogen gradients only.]
{
{\bf Feed-forward regulation by maternal morphogen gradients only.} For details see Sec.~\ref{secMutatedOpt} and Fig.~4B of the main text.
Panels {\bf (A--D)} show the resorted (hierarchically clustered)  distance matrix between 100 optimal solutions optimized in systems with four zygotically expressed gap genes regulated by maternal inputs only, without self- or cross-regulation, subject to the following different combinations of maternal gradients:
{\bf (A)} all three maternal input gradients (anterior, posterior and terminal system, ``APT''); 
{\bf (B)} anterior gradient only (``A'');
{\bf (C)} anterior and posterior gradients (``AP'');
{\bf (D)} anterior gradient and terminal system (``AT'').
Bluish hues indicate similar patterns (small distance), yellowish hues indicate different patterns (large distance); distance is defined in Eq.~(\ref{eqDistFun}).
The number of gap genes and maternal morphogen combination is indicated in the each plot title. 
Red boxes display randomly chosen example patterns from the respective clusters, as highlighted in the distance matrix.
}
\label{SI-fig:SimFFonly}
\end{figure*}

%% file: content/abstract-summary.tex
%Optimization has long guided our thinking about biological systems, driven by the speculation that evolutionary adaptation has maximized crucial functions. 
Many biological systems approach physical limits to their performance, motivating the idea that their behavior and underlying mechanisms could be determined by such optimality. Nevertheless, optimization as a predictive principle has only been applied in very simplified contexts. Here, in contrast, we explore a mechanistically-detailed class of models for the gap gene network of the \emph{Drosophila} embryo, and determine its 50+ parameters by optimizing the information that gene expression levels convey about nuclear positions, subject to physical constraints on the number of available molecules.  Optimal networks recapitulate the architecture and spatial gene expression profiles of the real organism.  Our framework makes precise the many tradeoffs involved in maximizing functional performance, and allows us to explore alternative networks to address the questions of necessity vs contingency.  Multiple solutions to the optimization problem may be realized in closely related organisms.
% Nevertheless, no rigorous optimization-derived \emph{ab initio} prediction of a living system in a mechanistic setting exists. 
%predict the regulatory network, dozens of its parameters, and the resulting spatial expression profiles for the \emph{Drosophila} gap genes by numerically maximizing the positional information they encode under biophysical constraints. 
%We uncover seemingly redundant interacting regulatory mechanisms in this system that give rise to multiple patterning solutions with high positional information, delineating possible evolutionary outcomes. By systematically optimizing alternative scenarios deficient for morphogens, gap genes, and regulatory mechanisms, we quantitatively characterize their ultimate impact on patterning performance, thereby establishing their functional roles and evolutionary necessity.

%% file: content/main-content.tex
\MyDropcap{O}ptimization is the mathematical language of choice for a number of fundamental problems in physical and statistical sciences. Stochastic optimization likewise constitutes the foundation of evolutionary theory, where selection continually improves organismal fitness by favoring adaptive traits~\cite{parker1990optimality,walsh2018evolution}. 
This evolutionary force pushes against quantifiable physical constraints and there are many examples where the organisms we see today operate very close to the physical limit: photon counting in vision~\cite{rieke1998}, diffraction limited imaging in insect eyes~\cite{barlow1952}, molecule counting in bacterial chemotaxis~\cite{berg1977}, and more.  Experimental evidence for optimal performance can be promoted to an optimization principle from which one can derive non--trivial predictions about the functional behavior and underlying mechanisms, sometimes with no free parameters~\cite{bialek2012biophysics,tkavcik2016information}.  Attempts at such ambitious \emph{ab initio} predictions include the optimization of coding efficiency in  visual and auditory sensory processing~\cite{karklin2011efficient,olshausen1996emergence,smith2006efficient,mlynarski2019ecological};  growth rates in metabolic networks~\cite{ibarra2002escherichia}; matter flux in transport networks~\cite{tero2010rules,katifori2010damage}; information transmission in regulatory  networks~\cite{tkavcik2008information}; and  the design of molecular machines and assemblies~\cite{savir2010cross}.
%\footnote{We emphasize that this list is illustrative rather than exhaustive.}

%None of these successes, however, extend to 
We are unaware of any successful optimization predictions for complex, multi-component biological systems whose interactions are described in molecular detail.
%a reasonable level of interacting molecular components. 
Whether {\em any} first principles prediction is even possible at this level remains unclear. As a consequence, we cannot determine whether the existence of a particular gene, genetic interaction or regulatory logic is an evolutionary necessity or merely a historical contingency~\cite{blount2018contingency}. This difficulty is not resolved by genetic tests for necessity, since these cannot rule out alternative evolutionary histories that would have unfolded without (or with modified) molecular components.

%%% Original position of Fig. 1 in PNAS template %%%

Here we address these issues during the early stages of development in the \emph{Drosophila} embryo~\cite{nusslein1980mutations}. About two hours post fertilization, the four major gap genes \emph{hunchback}, \emph{Kr\"uppel}, \emph{giant}, and \emph{knirps} are expressed in an elaborate 
spatiotemporal pattern  along the anterior--posterior (AP) axis of the embryo~\cite{jaeger2011gap}. The gap genes regulate one another, forming a network that responds to the anterior (Bicoid), posterior (Nanos), and terminal (Torso-like) maternal morphogen gradients~\cite{nusslein1980mutations,briscoe2015morphogen}.  The states of the gap gene network in turn drive the expression of pair rule genes in striped patterns that presage the segmented body plan of the fully developed organism~\cite{lawrence1992making}.
At readout time, about $40$ minutes into nuclear cycle 14 (NC14), the local gap gene expression levels peak and encode $4.3\pm 0.1~{\rm bits}$ of positional information~\cite{wolpert1969positional, dubuis2013positional,tkavcik2015positional}.  This information is necessary and sufficient for the specification of downstream pair--rule  expression stripes and other positional markers with a positional error as small as $\sim 1\%$ of the embryo length (EL)~\cite{petkova2019optimal}, roughly corresponding to the spacing between nuclei. Multiple lines of evidence further suggest that the flow of positional information through this system -- comprising both its encoding into gap gene profiles and its readout by the pair--rule genes -- is nearly optimal~\cite{dubuis2013positional,petkova2019optimal,tkavcik2021many}. These empirical observations lead us to the hypothesis that the gap gene network itself may be derivable from an optimization principle.

%and this analysis also provides evidence for optimality of information flow in the gap gene network.  First, despite the complex spatial patterns of gap gene expression, the positional errors are nearly uniform along the AP axis, as predicted if the distribution of maternal inputs is matched to the signal and noise characteristics of the network; quantitatively, the $4.3\pm 0.1~{\rm bits}$ of positional information~\cite{wolpert1969positional} encoded by the gap genes is $98.4\pm 0.3\%$ of the maximum that could be achieved by further adjusting this match~\cite{dubuis2013positional}.  Second, not only does the positional error of $\sim 1\%$ match the spacing between cells, the precision of pair rule gene stripes, and other downstream events, the optimal decoding algorithm correctly predicts, with no free parameters, distortions of the striped patterns in mutant embryos lacking one or two of the maternal inputs~\cite{petkova2019optimal}.  These observations on the optimality of inputs and readouts motivates us to ask about optimization of the gap gene network itself.

\begin{figure}[ht!] % [ht!] for PNAS style
\centering
\includegraphics[width=\FigOneWidth]{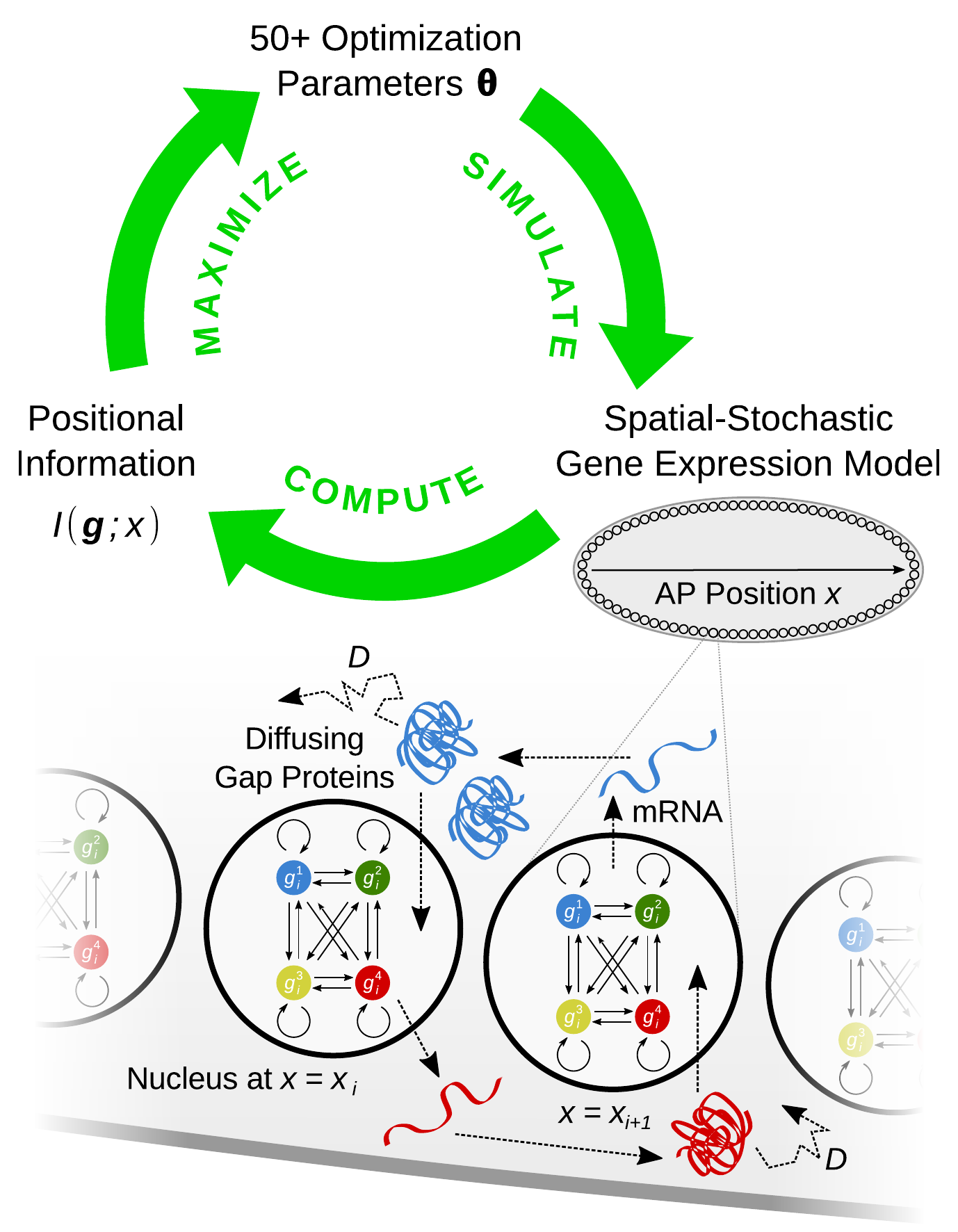}
\caption{{\bf Deriving a genetic regulatory network from an optimization principle.}
%{\bf The logic of an optimization-based approach to regulatory network prediction.} 
We simulate patterning during early fly development in a biophysically realistic, spatial--stochastic gap gene expression model (bottom; see Box 1) that accounts for the stochastic gene expression dynamics in individual nuclei along the anterior-posterior (AP) axis of the embryo. Regulatory interactions among four gap genes (arrows between colored circles in each nucleus), their response to three maternal morphogen gradients, and spatial coupling between neighboring nuclei are parameterized by a set of over $50$ parameters $\boldsymbol{\theta}$. For each parameter set, we numerically simulate the resulting noisy gap gene expression patterns, compute the system's positional information $I(\mathbf{g}; x)$, and adjust $\boldsymbol{\theta}$ using stochastic optimization to iteratively maximize the encoded $I$ (top). For details, see Box~1 and SI Sec.~1--5.}
\label{fig:1}
\end{figure}
%\vspace*{0.5EM}

Quantitative experiments, genetic manipulations, and attempts to fit mathematical models of the gap gene network to data have uncovered a wealth of detail about this system
~\cite{sanchez2001logical,jaeger2004dynamic,jaeger2004dynamic2,perkins2006reverse,manu2009canalization,manu2009canalization2,ashyraliyev2009gene,verd2017dynamic,verd2018damped,seyboldt2022latent}.  These facts are, in part, what an optimization theory for the gap gene network should explain.  But there are also major conceptual questions:  Is behavior of the network more constrained by its evolutionary history~\cite{smith1985developmental} or by the developmental constraints and physical limits that arise from the limited numbers of mRNA~\cite{little2013precise,zoller2018diverse} and protein~\cite{gregor2007probing} molecules?
Are all three maternal morphogens and four gap genes necessary? Most importantly for our discussion, are the interactions among gap genes and the resulting expression patterns coincidental, or determined by some underlying theoretical principle~\cite{tkavcik2021many}?  In simpler terms, can we derive the behavior of the gap gene network, rather than fitting its parameters to data?

\ifthenelse{\isodd{\IncludeTheBox}}
{
% Will include the box when \IncludeTheBox is set to 1
\begin{figure*}[hp!]
\input{content/box}
\captionsetup{labelformat=empty}
\end{figure*}
}
{
% Do not include box / do nothing here
}

\MySection{Optimization in a realistic context}

To answer the questions outlined above, we have formulated a detailed and realistic spatial--stochastic model of patterning that encompasses gap gene regulation by maternal morphogens; gap gene cross--regulation; discrete nuclei, including their divisions; transcription, translation, and degradation processes;  and diffusion of gap gene products (Fig.~\ref{fig:1}, Box~1, SI Sec.~1).  Within this class of models, we search for those that generate gene expression patterns encoding maximal positional information given limits on the number of molecules that can be synthesized.  %Here we give a preliminary account of our work, with subsequent analyses to be reported in a longer paper.

We considered the three maternal morphogens as well as maximal gap gene transcription, translation, and degradation rates to be physical constraints fixed to their measured or estimated values (Box 1, SI Table 2). This leaves more than $50$ parameters which govern how gap genes integrate transcriptional regulatory signals from other gap genes and from their morphogen inputs; we refer to all these parameters together as $\boldsymbol{\theta}$ (SI Table 3).  As an example, for each gene regulated by another, there is a parameter that measures the concentration at which the regulator exerts half--maximal activating or repressive effect on its target, and another parameter that measures the strength of this regulatory interaction.  Different points in this 50+ dimensional space describe a wide spectrum of regulatory networks and their diverse expression patterns, most of which are nothing like the real fly embryo but nonetheless are {\em possible} networks given the known component parts.  For any set of parameters we simulate the time evolution of our model, evaluating the mean spatial pattern of expression for all four gap genes as well as the gap gene (co)variability at every nuclear location along the AP axis. These calculations, carried out in the Langevin formalism, are complex yet numerically tractable (SI Sec.~2); they properly account for maternal morphogen gradient variability and intrinsic biochemical stochasticity.

\begin{figure*}[ht!]
\centering
\includegraphics[width=1\linewidth]{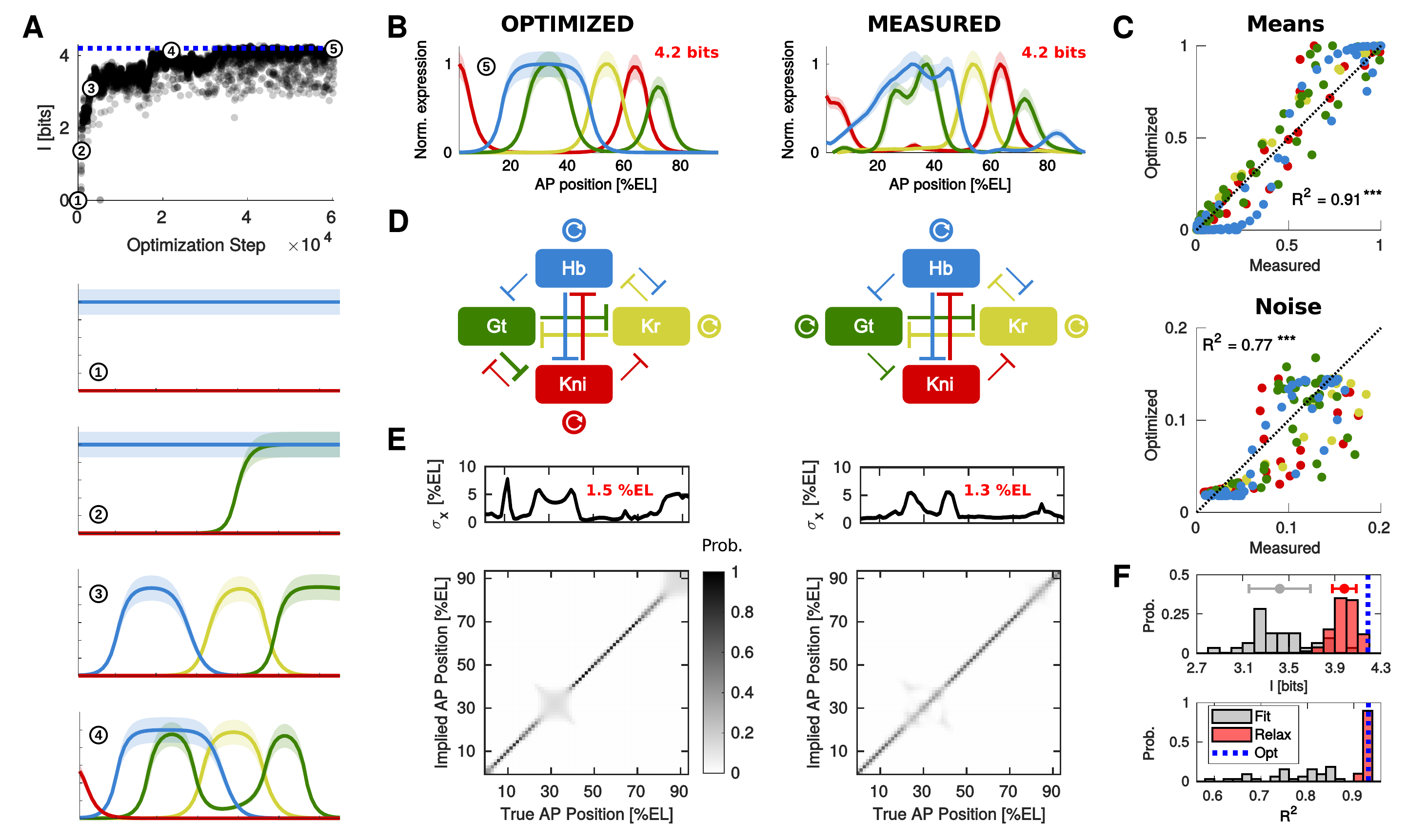}
\caption{
{\bf Networks that maximize information transmission recapitulate the measured gap gene expression patterns and the regulatory network interactions.} 
{\bf (A)} Positional information increases during a single optimization run, starting with the homogeneous profile at $0~{\rm bits}$ (1), proceeding through more complex spatial patterns (2-4), to the final solution (5, pattern in panel B) that reaches $\sim 4.2~{\rm bits}$ (dashed blue line). 
{\bf (B)} Predicted optimal (left) vs.  measured gap gene expression pattern (right;~\cite{dubuis2013accurate}), 40 min into NC14 (blue = \emph{hunchback}/Hb; green = \emph{giant}/Gt; yellow = \emph{Kr\"uppel}/Kr; red = \emph{knirps}/Kni; shade = standard deviation in gene expression). Positional information estimate from data is consistent with that reported in~\cite{dubuis2013positional}. 
{\bf (C)} Measured vs. predicted mean expression (top) and variability (bottom) are highly correlated (color code as in B; Pearson $p<10^{-3}$).  
{\bf (D)} Predicted gap gene regulatory network (left; blunt arrows = repression; circular arrows = self--activation) vs. literature--based reconstruction (right;~\cite{jaeger2011gap}). Interaction strength is depicted by the width of the arrows.
{\bf (E)} Predicted (left) vs. measured (right) decoding map (bottom) shows a nearly unambiguous code (diagonal band) with $\sim 1.5\%$ median positional error and few outlier positions (top inset)~\cite{petkova2019optimal}.
{\bf (F)} Fitting the model to mean WT gap gene expression profiles yields lower positional information (upper histogram; gray bars = distribution over replicate fits starting with random parameters, N = 32; red bars = distribution over replicate fits ``relaxed'' from the optimized solution, N = 250; horizontal error bars show mean $\pm$ SD), compared to the optimized solution (blue dashed line). The optimal solution better recapitulates the mean WT pattern compared to model fitting with random initial parameters; ``relaxation'' from the optimal solution does not significantly improve the fit (lower histogram showing goodness-of-fit as $R^2$; colors as above). 
}
\label{fig:2}
\end{figure*}

Positional information $I(\mathbf{g};x)$ can be formalized as the mutual information between the set of gap gene expression levels $\mathbf{g}\equiv \{g_1,g_2,g_3,g_4\}$ and the AP coordinate $x$~\cite{dubuis2013positional,tkavcik2015positional,tkavcik2016information,tkavcik2021many}. This quantity can be computed from the means and covariances of gap gene expression, which are the results of our model simulation at fixed $\boldsymbol{\theta}$ (see Box~1, SI Sec.~3; SI Fig.~2). 
If the gap gene system indeed has been strongly selected to maximize positional information at some readout time $T$, then the real network should be near the optimal setting of parameters,  $\boldsymbol{\theta}^* = \mathrm{argmax}_{\boldsymbol{\theta}} I\left(\mathbf{g}(T);x\right)$.  This problem is well posed because there are physical limits: the maximal rates of molecular synthesis combine with degradation rates to limit the maximum number of molecules for each species, setting the scale of the noise which in turn limits information transmission.  We have previously solved simplified versions of this optimization problem on small subnetworks, which inform the regulatory topology choices we make in this work~\cite{tkavcik2009optimizing,walczak2010optimizing,tkavcik2012optimizing,sokolowski2015optimizing,hillenbrand2016beyond,sokolowski2016extending}, but understanding the whole network at the level where comparisons with data are  possible required a new computational strategy (Box~1).  This larger scale numerical approach, combining simulation and optimization (Fig.~\ref{fig:1}),  provides a route to derive the first \emph{ab initio} prediction for a gene network in a realistic context. 

%Constraints in our case arise naturally, since fixing maximal production and degradation rates directly limits maximal copy numbers of gap gene products per nucleus, thereby placing a lower bound on the intrinsic noise in the system and an upper bound on the positional information; the embryo can utilize even less molecules in total than allowed by maximal production rates, as quantified by the ``resource utilization'' (RU) bound that we can  vary (see Box~1). Following up on promising empirical hints of optimality~\cite{tkavcik2008information,petkova2019optimal} and building on our understanding of optimization in simpler models~\cite{tkavcik2009optimizing,walczak2010optimizing,tkavcik2012optimizing,sokolowski2015optimizing,hillenbrand2016beyond,sokolowski2016extending}, we chose large-scale optimization as the route to derive the first \emph{ab initio} prediction for a gene network in a realistic context. 

%\FloatBarrier

\MySection{Comparing optimal networks with the real network}

We used a custom simulated annealing code to optimize the gap gene system for positional information~(Fig.~\ref{fig:2}A). We first biased the search towards solutions that might exist in the proximity of the wild--type (WT) \emph{Drosophila} gap gene expression pattern using a recently developed statistical methodology~\cite{mlynarski2021statistical}, and then removed the bias to be sure that we have found a true optimum (SI Sec.~4; SI Fig.~3).
Figure~\ref{fig:2}B compares the gap gene expression profiles generated by the optimized network to data. The match in mean expression profiles is very good~(Fig.~\ref{fig:2}C), although not perfect.  Mismatches---e.g., double--peaked anterior giant domain, posterior--most hunchback bump, linear anterior ramp of hunchback---likely trace their origin to the fact that the class of models we consider still is a bit too simple; even if we fit the parameters of the model to the data we cannot resolve these discrepancies.
%in particular our model can only approximate, but not properly account for, gene expression that is driven additively by multiple cis--regulatory elements (CREs). 
The predicted gap gene variability similarly recapitulates the measured behavior.

\begin{figure*}[ht!]
\centering
\includegraphics[width=1\linewidth]{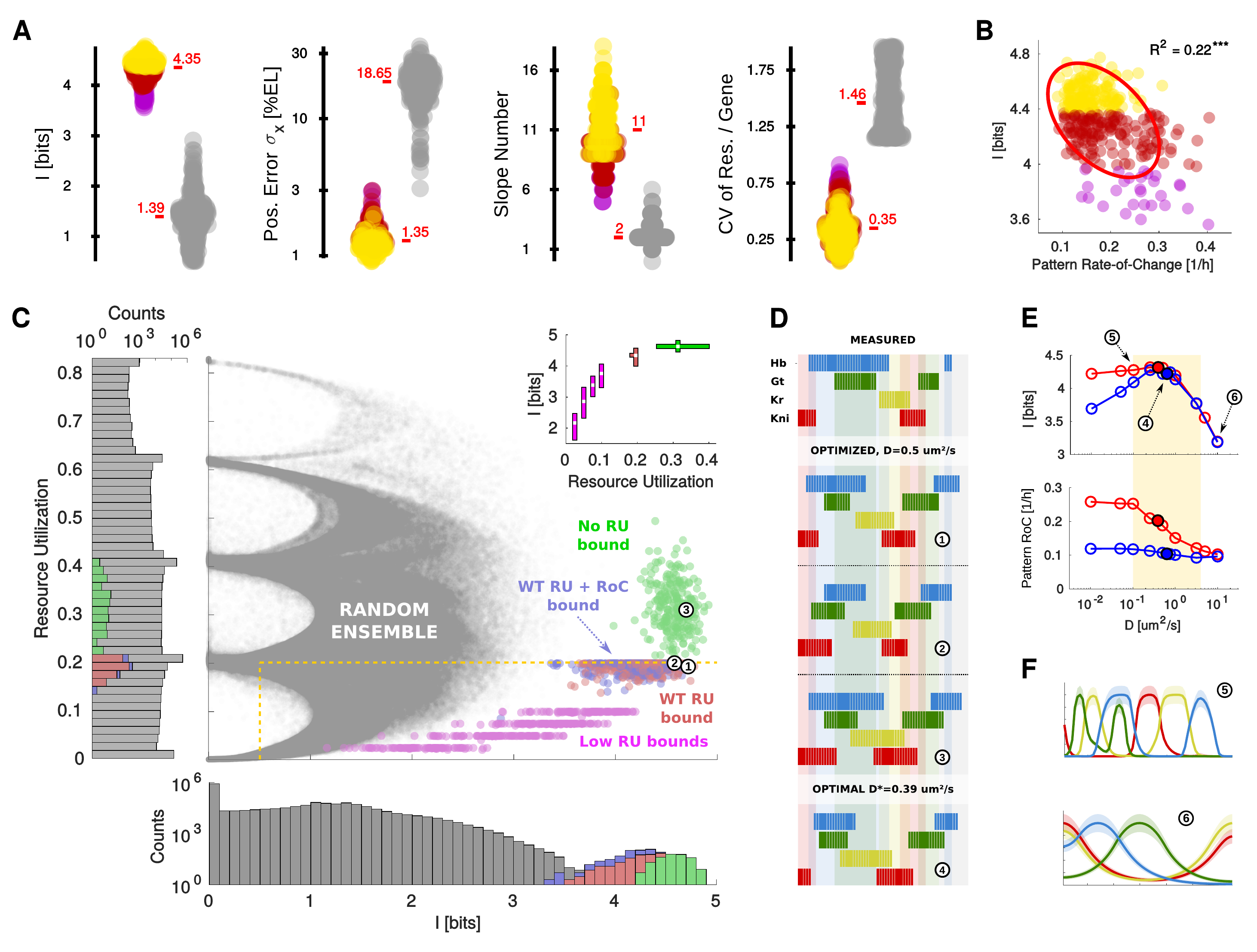}
\caption{
{\bf Optimal and random gap gene network ensembles.} 
{\bf (A)} Patterning phenotypes for optimal ensemble (color, solutions from ``WT RU'' in panel C,; N = 324) vs. random ensemble (gray, including only solutions $> 0.5$ bit that are at or below WT resource utilization, delineated by dashed yellow lines in panel C; random choice of N = 319 solutions meeting this criterion) reveal that high positional information (leftmost; violet, red, yellow indicate lowest, middle, highest third of the information interval) correlates with low positional error, high number of gap gene ``slopes'', and a more uniform utilization of resources across the four gap genes (red numbers = ensemble medians). 
 {\bf (B)} Within the optimal ensemble, higher information correlates with higher dynamical stability, i.e., lower pattern rate--of--change (each dot = one optimal solution; red ellipse = 1 SD contour in the I vs. RoC plane; color code and optimal ensemble as in A; N = 324). 
 {\bf (C)} Random (gray, N = 1580129) and various optimal ensembles (red = resource utilization bounded by \emph{Drosophila} WT denoted by dashed horizontal yellow line, N = 350; magenta = progressively smaller resource utilization, N = 100, 126, 100, 100; blue = WT resource utilization plus a bound on pattern rate--of--change, N = 607; green = no resource utilization or rate--of--change bounds, N = 239) depicted in the information vs. resource utilization plane (each dot = unique parameter combination). Histograms in the margins show the raw counts of evaluated parameter combinations. Inset: Information vs. resource utilization (median, 0.1--0.9-quantile intervals over ensembles in the main panel shown as central white squares and ribbons, respectively). 
 {\bf (D)} %\textcolor{red}{[Maybe you could show graphs as in Fig. 2B? I don't think you need the observed pattern again, and you needn't mix changing $D$ into this panel!]} 
 Example optimal solutions (1--3) from panel C optimized at fixed gap product diffusion ($D=0.5\;\mathrm{\mu m^2/s}$), and an example solution (4) where $D$ was also optimized from the ensemble in panel E, qualitatively match WT gap gene expression domains (top) and the regulatory network architecture. 
 {\bf (E)} Mean positional information (top) and pattern rate--of--change (bottom) as a function of gap the gene diffusion constant $D$ (empty circles = mean across optimal ensembles), capped at WT resource utilization (red; N = 100, 99, 100, 100, 350, 100, 99, 100, 100) or with an additional rate--of--change constraint (blue; N = 100, 100, 100, 303, 300, 200, 100, 100, 100). Solid circles = mean values for the case where $D$ itself is also optimized (N = 100); yellow shade = broad range of $D$ consistent with literature reports. 
 {\bf (F)} Two example solutions optimized at lower-than-optimal (top) and higher-than-optimal (bottom) diffusion constant values. 
 }
\label{fig:3}
\end{figure*}

The mechanistic nature of our model allows us to rationalize how the optimal pattern emerges.
% from a combination of multiple  factors. 
For example, the precision of the system output, manifested in the low  variability ($\sim 10\%$) of gap gene expression levels at fixed position, is  achieved through a combination of temporal averaging 
%(accumulation of relatively long-lived gap gene mRNA) 
and spatial averaging via diffusion, which substantially reduces noise components transmitted from upstream regulators and morphogens~\cite{erdmann2009role,sokolowski2012mutual,sokolowski2015optimizing}. The spatial patterns of expression in the optimal solution are shaped significantly by mutual repression and self--activation, closely mimicking  what had been inferred about the structure of the network from genetic interventions~(Fig.~\ref{fig:2}D). Optimization correctly predicts strong mutual repression between \emph{hunchback} and \emph{knirps}, between \emph{giant} and \emph{Kr\" uppel}, as well as most weak repressive interactions and self--activation of \emph{hunchback}~\cite{sokolowski2012mutual} (see SI~Fig.~4 for detailed analysis of regulatory interactions). Together, these  factors combine to encode positional information nearly unambiguously, with a median positional error of $\sim 1.5\%$~(Fig.~\ref{fig:2}E); even the elevated positional uncertainty around the cephalic furrow and in the far posterior is consistent between the optimal prediction and the real embryo~\cite{petkova2019optimal}. 

\emph{Ab initio} optimization performed here makes minimal use of empirical data to derive a wide range of predictions, in stark contrast to traditional model fitting~\cite{mlynarski2021statistical}. This has three further important consequences. First, when fitting, objective functions are purely statistical (e.g., maximum likelihood, mean-square-error, etc.), lacking any biological interpretation. In contrast, positional information used in optimization constitutes a meaningful and independently measurable phenotype of the patterning system. For example, our optimal solution (Fig.~\ref{fig:2}A,B) reaches $4.2~{\rm bits}$, to be compared with $4.2-4.3~{\rm bits}$ estimated  directly from data~\cite{dubuis2013positional,tkavcik2015positional}. Second, if fitting is performed instead of optimization, e.g., by minimizing the mean-square-error of the predicted mean gap gene expression, the best fits underestimate the positional information~(Fig.~\ref{fig:2}F, top). This is because fitting fails to take into account the functional consequences of noise and pattern variability. Furthermore, the goodness-of-fit landscape is very rugged, causing many fits starting from random initial parameters to get trapped in suboptimal solutions; information optimization regularizes this ruggedness and leads to solutions whose goodness-of-fit cannot be significantly improved further~(Fig.~\ref{fig:2}F, bottom).  In other words, the network that we predict through optimization cannot be brought significantly closer to the real network even by fitting 50+ parameters to the data.
Third, optimization can identify  locally optimal solutions that are qualitatively different from the  gene expression patterns observed in the embryo but functionally near degenerate.

Multiple optimization runs indeed produce diverse solutions that locally maximize positional information  while not exceeding the resource utilization of the wild type pattern. Together, these solutions constitute the \emph{optimal ensemble}.   A natural comparison is provided by the \emph{random ensemble}, where parameters $\boldsymbol{\theta}$ are drawn independently and uniformly from broad but realistic intervals. The vast majority of the random ensemble forms no patterns ($I\approx 0$ bits); we thus focus our comparisons in Fig.~\ref{fig:3} on a non-trivial subset of the random ensemble.   Optimization for positional information automatically leads to significantly lower positional error (Fig.~\ref{fig:3}A), higher number of boundaries where gap gene expression switches from low to high or vice--versa (``slopes''), more uniform utilization of resources across gap genes, as well a slight but significant over--allocation of resources in the anterior, 
%where morphogens are more precise,  
as can be seen in data as well~(see SI~Fig.~5 for further patterning phenotypes). Within the optimal ensemble, solutions with higher information tend to be more dynamically stable at readout time~(Fig.~\ref{fig:3}B), which we quantify by pattern rate--of--change (RoC), i.e., mean temporal derivative of gap gene expression~\cite{krotov2014morphogenesis}. Low RoC is relevant since pair--rule genes appear to read out gap gene expression via fixed decoding rules~\cite{jaeger2004dynamic,petkova2019optimal}, implying that temporally varying solutions could cause larger spatial drifts in pair--rule stripes.
% and thus lead to misspecified cells.

Networks in the random ensemble that transmit large amounts of information are exceedingly rare: the probability of drawing a network with positional information of $4~{\rm bits}$ or more by chance is $\ll 10^{-6}$~(Fig.~\ref{fig:3}C). In contrast, optimization strongly and robustly enriches for solutions above $4~{\rm bits}$~(Fig.~\ref{fig:3}C).  In our optimization we have constrained the maximum numbers of molecules, and the real embryo uses $\sim 20\%$ ($\mathrm{RU}=0.2$) of this maximum, on average.  This  resource utilization appears necessary for high-information solutions, whereas permitting more utilization within the same maximal rate limits does not noticeably increase positional information.  In fact, among $>10^3$ optimization runs we never found a solution exceeding $5~{\rm bits}$, indicating that such information values likely cannot be accessed within realistic  constraints.

The random and the optimal ensemble are closely related to the evolutionary concepts of the \emph{neutral} and the \emph{selected} phenotype distributions~\cite{hledik2022accumulation}. The random ensemble delineates what is physically possible in absence of selection for function, while the optimal ensemble delineates solutions that maximize function within fixed physical constraints. How closely natural selection \emph{could} approach this optimality (as quantified by the selected phenotype distribution), or indeed \emph{has} approached it (via the actual WT pattern), depends on selection strength and its history,  genetic load, linkage disequilibrium, and other limitations that are of negligible concern to \emph{in silico} optimization. Successful prediction of the pattern in Fig.~\ref{fig:2}B implies that selection was sufficiently strong to overcome such limitations and push the gap gene system beyond evolutionary tinkering~\cite{jacob1982possible} towards optimality~\cite{bialek2012biophysics,sella2005application,mlynarski2021statistical}. Even in strictly \emph{ab initio} runs with zero bias towards the WT pattern we repeatedly find solutions that closely reproduce the overall size and placement of expression domains in \emph{Drosophila}~(Fig.~\ref{fig:3}D, SI~Fig.~6), the encoded positional information, as well as the regulatory interactions. Tantalizing early experimental work suggests that dipteran species related to \emph{Drosophila} may feature a broadly consistent gap gene domain arrangement whose expression domains are, however, shifted~\cite{crombach2014evolution,wotton2015megaselia} or swapped~\cite{goltsev2004different}, as we find in our optimal ensembles~(SI~Fig.~6).

Taken together, our results paint a nuanced picture of the ``necessity vs. contingency'' dichotomy.  In the 50+ dimensional parameter space of possible networks, there is a highly non--random, locally optimal solution which produces expression patterns very similar to what we see in real fly embryos, but there are many other local optima that transmit about the same amount of positional information; all of these solutions are rare in the random ensemble. Careful analysis of optimal ensembles suggests that three or more interacting gap genes are necessary for a wide spectrum of optimal solutions~(SI~Figs.~8--10). It is an open question whether network architectures that enable such a wide spectrum could themselves be evolutionarily favored due to higher evolvability~\cite{payne2019causes}. It is also an open question whether alternative optima quantitatively recapitulate gap gene patterns seen in other dipterans or whether the degeneracy is removed by selection for additional phenotypes beyond positional information. 

\MySection{Alternatives to the real network}

Our theory provides a framework within which we can explore tradeoffs beyond the structure of the gap gene network.  In our standard optimal ensemble (``WT RU'' in Fig.~\ref{fig:3}C), we have taken the effective diffusion constant of gap gene products to be a fixed physical property of the cytoplasm, $D = 0.5~\mathrm{\mu m^2/s}$, in line with existing measurements~\cite{gregor2007probing}.  But we can view $D$ as one more parameter to be optimized, and remarkably we find that there is a broad optimum at the experimentally estimated value~(Fig.~\ref{fig:3}E). Larger diffusion constants lead to a precipitous drop in information even when all other parameters $\boldsymbol{\theta}$ are re-optimized, because high diffusion smooths gap gene profiles to the extent that adjacent nuclei can no longer be distinguished reliably~(Fig.~\ref{fig:3}F, bottom).  On the other hand, slower diffusion does not serve as effectively to average over local super--Poisson noise sources; the optimization algorithm compensates by finding parameters that generate more and steeper transitions between high and low expression levels (Fig.~\ref{fig:3}F, top), but even these unrealistic patterns do not transmit quite as much information. Thus, a single parameter displaced away from its optimum causes significant decreases in positional information; to lessen the impact the optimization algorithm adjusts other network parameters, driving the predicted patterns of  gene expression away from what we see in the real embryo.
% as well as large changes in predicted expression patterns that rapidly deviate from the WT.

\begin{figure*}[ht!]
\centering
\includegraphics[width=1\linewidth]{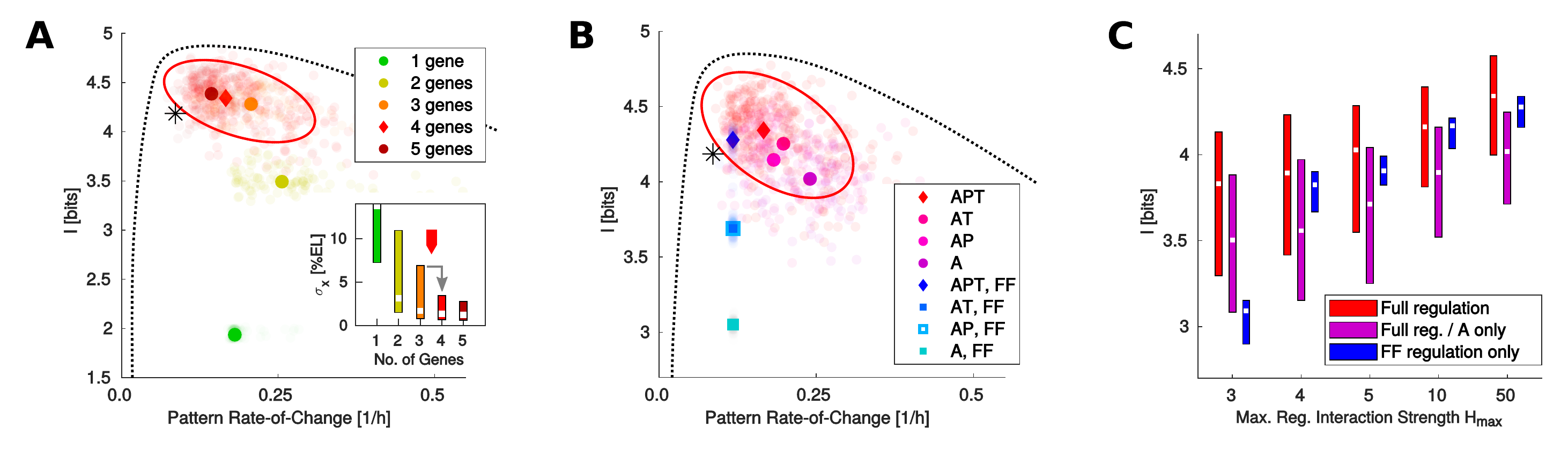}
\caption{
{\bf Necessity and sufficiency of gap gene regulatory network mechanisms.}
{\bf (A)} Optimal ensembles (transparent symbols = individual optimal solutions; solid symbols = ensemble medians) for networks with $1,\,2,\,\cdots{},\,5$ gap genes (legend colors; N = 100, 100, 100, 350, 83) optimized at the WT resource utilization (for reference, red diamond + red ellipse at 1 SD = WT--like optimal ensemble from Fig.~\ref{fig:3}B). Solutions delineate the accessibility frontier (dotted black line for visual guidance) in positional information (I) vs. pattern rate--of--change (RoC)  plane. 
{\bf (Inset)} While the median positional error (white squares) plateaus for optimal networks with three gap genes or more, the variability in positional error (ribbons denote 0.1- and 0.9-quantiles across AP positions in individual embryos) significantly shrinks only with 4 gap genes or more (red arrow).
{\bf (B)} Optimal ensembles for networks responding to different subsets of the three morphogens (A = anterior; P = posterior; T = terminal; red/purple circle symbols = ensemble median; red dots, diamond, ellipse = WT--like ensemble as in A). Optimal ensembles for  purely feed forward networks (``FF only'', i.e., no gap gene self-- or cross--regulation) denoted in bluish hues (legend). Sample numbers (order as in legend): N = 350, 100, 93, 100, 100, 100, 100, 100.
{\bf (C)} Positional information in optimal ensembles with gap gene self-- and cross--regulation (red = with APT morphogens, purple = with A morphogen only) or without self-- and cross--regulation (blue = ``FF only'' networks with APT morphogens), for different maximal regulatory strength, $H_\text{max}$. White squares denote median over optimal ensembles in A, B; ribbons denote the corresponding 0.1- and 0.9- quantile ranges. Lower $H_\text{max}$ values potentiate the positional information advantage of optimal networks with self--regulation, cross--regulation, and three maternal morphogens, compared to architectures deficient in these mechanisms.
N = 100, 100, 99, 100, 350 (full reg.); N = 100 for all $H_\text{max}$ (full reg. / A only and FF reg. only).
}
\label{fig:4}
\end{figure*}

We next address the question of evolutionary necessity and sufficiency. To this end, we make structural changes to the network and then re-optimize all of its parameters to explore  ``alternative evolutionary histories'' that could have unfolded with changed molecular components or mechanisms. As an example, Figure~\ref{fig:4}A characterizes solutions obtained using $1,\,2,\,\cdots{},\,5$ gap genes, subject to the same {\em total} resource utilization as the WT, plotting the positional information vs. the rate at which expression patterns are changing at readout time.  Networks that transmit $4~{\rm bits}$ or more---as in the real embryo---are completely inaccessible using only one or two gap genes, even though these networks are allowed to utilize the same total number of molecules as in the optimal four gene networks above. Such networks also cannot produce a wide spectrum of optimal solutions~(SI~Fig.~8,~9). With three gap genes the optimized networks grow significantly more diverse and can transmit a total information comparable to what is seen experimentally, but detailed analysis reveals that three--gene networks all have local defects where the positional error spikes above $5-10\%$ of the embryo length~(SI~Fig.~7), in contrast to the much more uniform distribution of precision along the length of the real embryo~\cite{dubuis2013positional}; we can quantify this by looking at the variations in the positional error along the AP axis (Fig.~\ref{fig:4}A, inset).  This failure of the three gene networks arises because they cannot realize  a sufficient number of slopes or switches between high and low expression levels. Four gap genes thus are necessary to ensure that high positional information translates into defect--free patterning not just on average, but uniformly across the entire AP axis of every embryo~\cite{dubuis2013positional}. The marginal benefit of the putative fifth gap gene appears small and may not be sufficient to establish the required additional regulatory mechanisms or to maintain them at mutation--selection balance~\cite{gerland2002selection}.

We can explore, in the same spirit, the role of the multiple maternal morphogens. In the fly embryo, the anterior (A, Bicoid), posterior (P, Nanos), and terminal (T, Torso-like) systems jointly regulate gap gene expression~\cite{petkova2019optimal}.   In our model, we can remove one or two of these inputs and re--optimize all the parameters of the gap gene network, and find that there are moderate yet statistically significant losses in both positional information  and stability (Fig.~\ref{fig:4}B). The impact of primary morphogen deletions is limited because the optimization algorithm adjusts the gap gene cross--regulation parameters to restore informative spatial patterns. This ability, however, disappears entirely if gap gene cross--regulation is not permitted and the gap gene network is feed forward (FF) only~(light gray arrows in Box figure, Fig.~\ref{fig:4}B); in the absence of cross-regulation,  removal of each primary morphogen system is associated with a large decrease in positional information.

Figure~\ref{fig:4}B suggests that stable, high information patterns could be generated by utilizing all three maternal morphogens even without the ability of gap genes to regulate one another. But in the absence of cross--regulation, the time scale for variations in the pattern is determined solely by the intrinsic lifetime of the most stable species (mRNA).   In contrast, gap gene interactions allow for the emergence of longer time scales which both slow the variations and can reduce noise by temporal averaging~\cite{tkavcik2012optimizing}; possible evidence for these effects has been discussed previously~\cite{krotov2014morphogenesis}.  Evolutionarily, adding gap gene cross--regulation creates variability in the rate--of--change phenotype that could  additionally be selected for. Indeed, the WT--like solution of Fig.~\ref{fig:2}B falls close to the accessibility frontier of Fig.~\ref{fig:4}B, suggesting such a preference.

Lastly, we varied the maximal allowed strength of regulatory interactions, $H_\text{max}$ (see Box~1), in our model. This parameter determines how strongly each individual input, either a morphogen or a gap gene acting via self-- or cross--regulation, can impact the expression of a target gap gene. 
% can impact the expression of a target gap gene.  
In simple microscopic models, this parameter measures the number of transcription factor molecules that bind cooperatively to their target sites as they regulate a single gene, and correlates with the steepness (or sensitivity) of the resulting induction curve.   Optimizations presented so far used $H_\text{max}=50$, sufficiently high not to impose any functional constraint. As $H_\text{max}$ is lowered and the constraint kicks in, the optimal feed forward  solution of Fig.~\ref{fig:4}B (dark blue) suffers large drops in encoded positional information~(Fig.~\ref{fig:4}C); optimal feed forward architectures are thus heavily reliant on levels of effective cooperativity that appear unrealistic.  Further, one might have been tempted to interpret Fig.~\ref{fig:4}B by saying that cross--regulation and multiple input morphogens provide alternative or even redundant paths to high information transmission, but we see that this degeneracy is progressively lifted when we limit the effective cooperativity to more realistic levels. From an evolutionary perspective, gap gene cross--regulation therefore is favourable for two reasons: first, it generates temporally stable phenotypes at the accessibility frontier (as in Fig.~\ref{fig:4}B); and second, it permits high information solutions also in networks where the strength of individual regulatory interactions is limited (as in Fig.~\ref{fig:4}C). 

\MySection{\DiscussionName}

The idea that living systems can approach fundamental physical limits to their performance, and hence optimality, goes back at least to explorations of the diffraction limit in insect eyes and the ability of the human visual system to count single photons~\cite{bialek2012biophysics}.   The specific idea that biological systems optimize information transmission emerged shortly after Shannon's formulation of information theory, in the context of neural coding~\cite{barlow_59,tkavcik2016information}.  Despite this long history, most optimality analyses in biological systems have been carried out in very simplified contexts, using functional models with a small number of parameters.  Here we have instantiated these ideas in a much more realistic context, using mechanistic models for genetic regulatory networks that permit direct interpretation in terms of  molecular mechanisms and interactions. 

We focused on the \emph{Drosophila} gap gene system, one of the paradigms for developmental biology and for physical precision measurements in living systems~\cite{gregor2014embryo}. Our work extends previous mathematical models of this system~\cite{schroeder2004transcriptional,sanchez2001logical,jaeger2004dynamic,jaeger2004dynamic2,perkins2006reverse,segal2008predicting,manu2009canalization,manu2009canalization2,ashyraliyev2009gene,bieler2011whole,duque2014simulations,verd2017dynamic,verd2018damped}, as well as attempts to predict it \emph{ab initio}~\cite{franccois2007deriving,franccois2010predicting,rothschild2016predicting}. In contrast to previously studied models, we systematically incorporate the unavoidable physical sources of noise, highlighting how patterning precision can emerge from noisy signals by a synergistic combination of multiple  mechanisms.  Crucially, we do not  \emph{fit} the parameters of the model to data, but rather \emph{derive} them \emph{ab initio} via optimization. In contrast to previous prediction attempts, our constraints and comparisons to data are not stylized, but fully quantitative and commensurate with the precision of the corresponding experiments.

We have found networks that maximize positional information with a limited number of molecules, and there is at least one local optimum quantitatively matching a large range of observations in the wild--type \emph{Drosophila} system: its spatial patterns of expression and variability, the resulting decoding map, the molecular architecture of the network, as well as subtler biases in spatial resource utilization. Our optimization framework furthermore provides a platform for exploring the necessity and sufficiency of various network components that ensure maximal information transmission. Using this framework to deliver on our introductory questions, we have established that four gap genes appear necessary for defect--free patterning and that the apparent redundancy between the three maternal morphogens and gap gene cross--regulation is lifted under reasonable constraints on the strength of regulatory interactions. 

Numerical optimization  clearly is not evolutionary adaptation, yet its results nevertheless provide perspective on evolutionary questions. Discussions about the interplay of evolutionary optimization and developmental constraints, necessity versus contingency, and limits to selection have a venerable history~\cite{wilkins2002evolution,smith1985developmental}.
% and are nearly synonymous with the field of evolutionary biology itself~\cite{wilkins2002evolution,smith1985developmental}.
Rather than discussing these questions in qualitative terms, here we explored the  role of physical  constraints  and tradeoffs quantitatively, in the context of an expressive mechanistic model, using the powerful concepts of the random and the optimal ensembles. In the words of Jacob~\cite{jacob1982possible}, the random ensemble delineates the space of the ``possible.'' Within this space, our optimization principle acts as a proxy for strong selection for high positional information, thereby identifying a much more restricted optimal ensemble. It is surprising that this principle alone is sufficient to ensure that the optimal ensemble contains a solution very close to Jacob's ``actual'', the \emph{Drosophila} gap gene network that we observe and measure.

%se ensembles delineate the space of the ``possible,'' the space that we can and should understand based on physical laws. Only within this space did the evolutionary past, with all of its idiosyncrasies, realize the ``actual'' -- the \emph{Drosophila} gap gene network that we can observe and measure.

%% file: content/box.tex
\begin{minipage}[t]{1\linewidth}
\begin{tcolorbox}[
sharp corners=all,
colback=yellow!7.5,
%colframe=blue!75,
size=tight,
boxrule=0.5mm,
left=3mm,right=3mm,
top=3mm,bottom=3mm
]
\footnotesize
\begin{multicols}{2}
\noindent {\bf Box 1. Model description and assumptions.}\vspace{0.2cm}\\
%\\
\noindent \input{content/box-content}

\end{multicols}

%\centering
\input{content/box-figure}

% \includegraphics[width=\linewidth]{figs/Fig_Box1_v7.pdf}
% \caption*{{\bf Spatial-stochastic model for gap gene expression.} The gap gene regulatory network (center; colored circles = gap genes; grayscale circles = maternal morphogens) in each nucleus transforms maternal inputs with known spatial profiles (left) into a gap gene expression pattern at readout time $T$ (right; solid lines = computed mean expression; shade = computed standard deviation). Each interaction in the network stands either for the feed forward (FF) regulation of a gap gene by a morphogen input (light gray arrows),  or for the regulation of a gap gene by other gap genes or by itself (cross-- and self--regulation, dark gray arrows), and is described by two parameters (concentration threshold $K$ and regulatory strength $H$; several parameter pairs are shown, corresponding to nearby thicker arrows). All parameters denoted by regulatory arrows are jointly optimized to maximize positional information $I(\mathbf{g};x)$.}
% \label{fig:box}

\end{tcolorbox}
\end{minipage}

%% file: content/main-acknow.tex
We thank Nicholas H. Barton for his comments on the manuscript, Benjamin Zoller for inspiring discussions, and Aleksandra Walczak and Curtis Callan for  early discussions that shaped this work. Special thanks to Eric F. Wieschaus for many persistently inspiring conversations. This work was supported in part by the Human Frontiers Science Program, the Austrian Science Fund (FWF P28844), U.S. National Science Foundation, through the Center for the Physics of Biological Function (PHY-1734030);  by National Institutes of Health Grants R01GM097275, U01DA047730, and U01DK127429; by the Simons Foundation; and by the John Simon Guggenheim Memorial Foundation.